\documentclass[12pt]{report}

\usepackage{sty/DL_thesis}  

\usepackage{notoccite} 
\usepackage{algorithm} 
\usepackage{algorithmic} 
\usepackage{amsmath} 
\usepackage{mathrsfs} 
\usepackage{amssymb} 
\usepackage{booktabs} 
\usepackage{boxedminipage} 
\usepackage{breqn} 
\usepackage{changepage} 
\usepackage{cite} 
\usepackage{enumitem} 
\usepackage{fancyhdr}
\usepackage{flafter} 
\usepackage[flushleft]{threeparttable} 
\usepackage{geometry} 
\usepackage{tabularx} 
\usepackage[table]{xcolor} 
\usepackage{graphicx} 
\graphicspath{{figures/}}
\usepackage[export]{adjustbox} 
\usepackage{epstopdf} 
\usepackage{longtable} 
\usepackage{ltablex} 
\usepackage[maxfloats=36]{morefloats} 
\usepackage{moresize} 
\usepackage{multirow} 
\usepackage[list=true,listformat=simple]{subcaption} 
\usepackage{outlines} 
\usepackage{rotate} 
\usepackage{rotating} 
\usepackage{scrextend} 
\usepackage{pdflscape} 
\usepackage[doublespacing]{setspace} 
\usepackage{siunitx} 
\usepackage[all]{nowidow} 
\usepackage{indentfirst}

\usepackage[colorinlistoftodos,textwidth=25mm,shadow,backgroundcolor=green]{todonotes} 
\usepackage{marginnote} 

\usepackage{verbatim} 
\usepackage{wrapfig} 
\usepackage{xspace} 
\usepackage[bookmarks=true,backref=page, hyperfigures=true, pdfpagelabels=false]{hyperref}
\usepackage{bookmark} 
\usepackage[capitalize]{cleveref}
\creflabelformat{equation}{#2\textup{#1}#3}
\usepackage[utf8]{inputenc}
\usepackage[nogroupskip,toc,acronym]{glossaries} 

\usepackage{lineno} 
\usepackage{cite}

\makeatletter
\renewcommand{\@todonotes@drawMarginNoteWithLine}{%
 \begin{tikzpicture}[remember picture, overlay, baseline=-0.75ex]%
  \node [coordinate] (inText) {};%
 \end{tikzpicture}%
 \marginnote[{
  \@todonotes@drawMarginNote%
  \@todonotes@drawLineToLeftMargin%
  }]{
  \@todonotes@drawMarginNote%
  \@todonotes@drawLineToRightMargin%
 }%
}
\makeatother

\newgeometry{letterpaper, margin=1in} 

\newcommand{\qed}{\mbox{$\ \Box$}}
\newcommand{\lqed}{\qed\vspace{.1in}}

\newcommand{\myendfig}{\vskip -0.14in \end{figure} } 
\newcommand{\myendtab}{\vskip -0.01in \end{table} }

\newtheorem{ex}{Example}[chapter]
\newtheorem{theor}{Theorem}[chapter]
\newtheorem{mtd}{Method}[chapter]
\newtheorem{mydef}{Definition}[chapter]
\newtheorem{mylem}[theor]{Lemma}

\newenvironment{proof-of}[1]{\noindent {\bf Proof of #1:}}{\hfill \lqed}

\AuthorFirstName{Yingnan}
\AuthorLastName{Xu}

\ThesisTitle{Global Fits}{ in the }{Supersymmetric Georgi-Machacek Model}
\GraduateDepartment{Physics}

\FirstDegree%
\FirstDegreeType{B.A.}
\FirstDegreeMajor{Physics, Mathematics}
\FirstDegreeUniversity{Oberlin College}

\SecondDegree%
\SecondDegreeType{M.S.}
\SecondDegreeMajor{Physics}
\SecondDegreeUniversity{Southern Methodist University}

\ThesisDefenseDateYear{2024}
\ThesisDefenseDateMonth{April}
\ThesisDefenseDateDay{30}

\GraduationDateYear{2024}
\GraduationDateMonth{May}
\GraduationDateDay{11}

\SMUCollege{Dedman College}

\AdvisorFullName{Dr. Roberto Vega, Advisor}
\AdvisorTitle{Southern Methodist University}

\CommitteeMemberA{Dr. Pavel Nadolsky, Committee Chair}
\CommitteeMemberTitleA{Southern Methodist University}

\CommitteeMemberB{Dr. Joel Meyers, Committee Member}
\CommitteeMemberTitleB{Southern Methodist University}

\CommitteeMemberC{Dr. Stephen Sekula, Committee Member}
\CommitteeMemberTitleC{Queen's University and SNOLAB}

\hypersetup{
 bookmarksopen=true, 
 bookmarksnumbered=true, 
 breaklinks=true, 
 filecolor=magenta, 
 citecolor=blue, 
 colorlinks=true, 
 linkcolor=blue, 
 linktocpage=true, 
 pdfnewwindow=true, 
 plainpages=false, 
 pdffitwindow=false, 
 pdfmenubar=true, 
 pdfstartview={FitH}, 
 pdftoolbar=true, 
 pdfauthor={Author}, 
 pdfcreator={Author}, 
 pdfkeywords={Keyword1} {Keyword2} {Keyword3} {Keyword4} {Author}, 
 pdfproducer={Producer}, 
 pdfsubject={Subject}, 
 pdftitle={Title of Dissertation}, 
 unicode=false, 
 urlcolor=cyan 
}

\makeglossaries
\newglossaryentry{LHC}{
 text=LHC,
 long=Large Hadron Collider,
 name={\glsentrylong{LHC} (\glsentrytext{LHC})},
 first={\glsentryname{LHC}},
 sort={large hadron collider},
 description={Large Hadron Collider}
}

\newglossaryentry{thesis}{
 name=thesis,
 description={https://xkcd.com/1403/}
}

\begin{document}



\ApprovalTitlePages%

\Copyright%


\begin{Acknowledgment}
 I would like to express my deepest gratitude and appreciation to everyone who has helped me during my Ph.D. program.

First and foremost, I am profoundly thankful to my advisor, Dr. Roberto Vega. This dissertation would have been impossible without his guidance and mentorship. Dr. Vega has not only taught me particle physics and research skills but also many aspects of life. His expertise and constructive feedback have been instrumental in shaping my academic taste and research.

I am also indebted to the members of my dissertation committee, Dr. Nadolsky, Dr. Meyers, and Dr. Sekula, for their scientific insights and critiques. Their commitment to academic excellence has greatly enriched the quality of my work.

I acknowledge the Physics department at SMU for providing me with the opportunity to accomplish my Ph.D. degree and teaching assistantship for living expenses in the past six years. I would like to thank every faculty member who has ever taught me in class or provided invaluable help during my research in this department. They are Dr. Stephen Sekula, Dr. Joel Meyers, Dr. Fredrik Olness, Dr. Randall Scalise, Dr. Ryszard Stroynowski, Dr. Pavel Nadolsky, and Dr. Jingbo Ye. I also thank Lacey Breaux and Michele Hill for their administrative work.

Thanks to Mr. Junfeng Bi, my father-in-law, for assisting me in understanding and writing computer codes that are crucial to completing my research project. We had extensive discussions on the packages I used in data analysis and modification methods to suit my needs in this project. Besides his guidance in \texttt{Fortran} and \texttt{C++} languages, I am also truly grateful for the care and love he gives to my daughter Blair, along with my mother-in-law. Without their help in raising Blair, I would not have been able to focus on my work and complete my degree on time.

I would also like to thank my parents and my elder sister, who supported me in pursuing academic and career goals in the US, both emotionally and financially. Their encouragement and understanding have always been my pillar of strength, and I am truly grateful for the sacrifices they have made.

Last but not least, I would like to express my inexpressible love and gratitude to my wife Catherine. She gave up her career for almost three years and came to the U.S. to accompany me. Her company and support are extremely valuable to me. She is an excellent partner and a truly great mother.
\end{Acknowledgment}

\newgeometry{margin=1in}
\begin{Abstract}
 In this dissertation, we study a supersymmetric extension of the Standard Model with Higgs triplets in the scalar sector. We begin with a review of the Standard Model (SM), particularly the electroweak sector and the Higgs mechanism. In the SM, the Higgs mechanism requires the presence of a complex Higgs doublet to break the electroweak symmetry and endow particles with a mass; this process is called Spontaneous Symmetry Breaking (SSB). Although this is the simplest possibility, higher scalar representations may also contribute to the electroweak breaking process (EWSB).  The extent to which these higher representations contribute to EWSB is constrained by precise measurements of the $\rho$ parameter. The model must predict $\rho\approx 1$ at tree level. It is a fortuitous circumstance that simple doublet representations satisfy this requirement exactly. The underlying reason is that models with doublets satisfy an accidental custodial symmetry.  Therefore, one can add any number of scalar doublets and still satisfy this experimental constraint. For higher representations, it is a bit trickier to maintain the custodial symmetry. We study in this work a supersymmetric model that incorporates triplet representations, satisfies the custodial symmetry, and predicts $\rho\approx 1$ at tree level. The non-supersymmetric Georgi-Machacek (GM) model is one example of a custodial invariant model of SSB with Higgs triplets. However, the GM model has a fine-tuning problem beyond that of the SM. The solution to both issues is the Supersymmetric Custodial Triplet Model (SCTM). The supersymmetric Georgi-Machacek model arises as a low energy limit of the SCTM model. It is this model that we study here. We make use of public code, \texttt{GMcalc} and \texttt{Higgstools}, to perform global fits to the parameters of this model and obtain strong limits on the triplet vacuum expectation values, mixing angles, mass differences between the new heavy exotic Higgs bosons, as well as their decay width, at the $95\%$ confidence level. For these new hypothetical scalars, we identify the dominant decay channels and extract bounds on their branching ratios from the global fits. We also examine the possible presence of a $95$ GeV Higgs Boson in the SGM model.

\end{Abstract}

\glsunsetall%
\tableofcontents%
\listoffigures%
\listoftables%
\glsresetall%

\newgeometry{margin=2in}
\Dedication{
\textit{This thesis is dedicated to my wife, Catherine, and my daughter, Blair, \\
for their endless love and continuing support.}
}
\newgeometry{margin=1in}%

\begin{thesis}

 \chapter*{Notations and Conventions}\label{chapter:preface}
\addcontentsline{toc}{chapter}{\protect\numberline{}Notations and Conventions}
In this dissertation, we take the natural unit system
\begin{equation}
\hbar=c=1 .
\end{equation}
All physical quantities in natural units are related to the dimension of mass. In particular, \begin{equation}
[L]=[T]=[M]^{-1}, \quad[p]=[E]=[M].
\end{equation}
The dimension of scalars, spinors and vectors are \begin{equation}
[\phi]=[M], \quad[\psi]=[M]^{3 / 2}, \quad[V]=M.
\end{equation}
The Minkowski metric we choose is \begin{equation}
\eta_{\mu \nu}=\eta^{\mu \nu}=\operatorname{diag}(1,-1,-1,-1).
\end{equation}
The Greek indices run over 0, 1, 2, 3 and Roman indices run over 1, 2, 3. We use Einstein's convention where repeated indices are summed. The inner product is 
\begin{equation}
x\cdot y=\eta_{\mu\nu} x^{\mu} y^{\nu}=x^0y^0-\vec{x}\cdot\vec{y}.
\end{equation}
The Pauli matrices are defined as \begin{equation}
\tau^1=\left(\begin{array}{ll}
0 & 1 \\
1 & 0
\end{array}\right), \quad \tau^2=\left(\begin{array}{cc}
0 & -i \\
i & 0
\end{array}\right), \quad \tau^3=\left(\begin{array}{cc}
1 & 0 \\
0 & -1
\end{array}\right)
\end{equation}
Pauli matrices are related to the generators of the $SU(2)$ group in the $s=1/2$ representation, \begin{equation}
T_i=\frac{\tau_i}{2}, \quad \left[T^i, T^j\right]=i \epsilon^{i j k} T^k.
\end{equation}
We choose the chiral basis for gamma matrices, where \begin{equation}
\gamma^0=\left(\begin{array}{cc}
0 & I_2 \\
I_2 & 0
\end{array}\right), \quad \gamma^k=\left(\begin{array}{cc}
0 & \sigma^k \\
-\sigma^k & 0
\end{array}\right), \quad \gamma^5=\left(\begin{array}{cc}
-I_2 & 0 \\
0 & I_2
\end{array}\right).
\end{equation}
The projection operators are \begin{equation}
P_{\mathrm{L},\mathrm{R}}=\frac{1}{2}\left(1\pm\gamma^5\right).
\end{equation}

 \begin{singlespace}
\chapter{Introduction}\label{chapter:introduction}
\end{singlespace}

By now, we know that there exist four types of interactions among matter: the electro-magnetic force that binds the electrons and the nuclei in atoms; the weak interaction, which is responsible for the radioactive decay of heavy atoms; the strong interaction, that binds quarks into hadrons; and gravity, which is a pulling force that holds planets, stars, and galaxies together at cosmological scale. Physicists have been trying to construct a grand unified theory that unites the four interactions for a long time but have so far failed. To date the most successful theoretical framework in particle physics is the Standard Model (SM), in which the electromagnetic, weak and strong interactions are unified in the framework of a gauge theory. The first gauge theory was formulated by Maxwell in classical electrodynamics, where the symmetry group is an abelian $U(1)$. Non-abelian gauge theory was first postulated by Chen Ning Yang and Robert Mills\cite{PhysRev.96.191}. These non-abelian gauge theories did not attract much attention because all the particles were predicted to be massless. It was not until the Higgs mechanism was incorporated into the $U(1)_Y\times SU(2)_L$ gauge theories by Weinberg and Salam \cite{Weinberg:1967tq,Salam:1968rm} that this theory gained attention. We will review the Higgs mechanism, or equivalently electroweak symmetry breaking (EWSB), in chapter~\ref{chapter:SM Higgs}. The SM also incorporates a so-called color symmetry which is associated with the strong interactions; the symmetry group in this case is $SU(3)_C$\cite{GELLMANN1964214}. Here we will not be concerned with this aspect of the SM. 

The SM is a self-consistent, renormalizable quantum field theory. Despite its success, it still does not come close to explaining all the observed phenomena in our universe. For example, the fourth interaction, namely gravity, is explained by General Relativity. However, a consistent quantum theory of gravity has yet to be developed and it is not clear if it ever will be. Besides that, there are a plethora of challenges that cannot be tackled with the SM in modern physics. Here we briefly discuss some of them. 

\begin{itemize}
  \item \textbf{Naturalness problem:} Naturalness is a prejudice that dimensionless parameters of a theory should be of order one. For example, the masses of the fundamental particles in a theory should not differ from each other by large factors. The SM does not satisfy this expectation. In the SM, some parameters vary by several orders of magnitude. Therefore, it is deemed unnatural.
  \item \textbf{Hierarchy problem:} The hierarchy problem is related to the naturalness problem in that one expects the masses of the fundamental particles to lie close to the energy scale at which the model is applicable. In the SM this scale is of the order of 1 TeV. The mass spectrum of the particles in the SM are naturally restricted to be below 1 TeV by the underlying gauge symmetries. The mass of the Higgs particle, or any other scalar particle, is an exception. Scalar mass terms are perfectly consistent with gauge symmetries and are therefore not protected from large radiative corrections. If there is a large energy scale in the theory, the Higgs mass will tend to be on the order of that scale. This hierarchy problem arises only in the context of envisioning the SM as a low energy limit of a grand unified theory. Presumably, in this more complete theory, many of the 19 free parameters in the SM would be fixed. The scale of this theory is expected to be on order of $10^{15}$ GeV and we would expect the Higgs mass to be of this order. The Higgs mass, however, has been measured to be around 125 GeV.
  \item \textbf{Ad hoc parameters:} The SM requires 19 unrelated and arbitrary free parameters whose values are not fixed by the theory. The discovery of neutrino oscillations further expands the number of parameters not fixed by the model. The number of parameters can be increased by 9 if neutrinos have mass. None of these parameters is predicted by the SM. Ideally one would expect that a fundamental theory has at most one unfixed parameter.
  \item \textbf{Generation problem:} Quarks and leptons in the SM are divided into 3 generations. The main difference between generations appears to be their mass. The SM does not explain this hierarchy nor why there should be exactly three generations, or even if there are more generations. The observed universe is essentially composed of just first-generation particles. 
  \item \textbf{Inconsistencies with $\Lambda$CDM model:} The SM fails to coincide with $\Lambda$CDM model, which tries to explain the origin of our universe. In the $\Lambda$CDM model, the isotropy and homogeneity of the visible universe requires a super-rapidly inflationary period after the Big Bang. The SM does not contain any information about the inflaton. Also, the theorized vacuum zero-point energy from the SM is of 50 to 120 orders of magnitude larger than the observed vacuum energy from cosmology\cite{adler1995vacuum}. Perhaps the most glaring shortcoming of the SM in Cosmology is that it does not explain dark matter or dark energy.
\end{itemize}

The model considered here could possibily solve the heirarchy problem and could explain the origin of Dark Matter. The simplest extension of the SM Higgs sector is to add any number scalar doublets. Scalar doublets naturally preserve the custodial $SU(2)$ symmetry, which is responsible for the result that $\rho_{\text{tree}}=M_W^2/M_{Z}^2 \cos ^2 \theta_{\text{W}}=1$ in the SM. It is possible to add higher scalar representations of the SM gauge group. However, to preserve the custodial symmetry and maintain agreement with $\rho=1.00038 \pm 0.00020$ in electroweak precision measurement\cite{ParticleDataGroup:2022pth}, one must be careful. The first model with an electroweak triplet in the Higgs sector was proposed by Gelmini and Roncadelli\cite{Gelmini:1980re}. By adding an extra Higgs boson that strongly couples to neutrinos but very weakly to other fermions, this model provides a Seesaw Mechanism for generating Majorana neutrino masses. However, adding a single triplet spoils the custodial symmetry and the model would require fine-tuning to the triplet v.e.v to maintain $\rho\approx1$ and thus is deemed unnatural. A small triplet v.e.v. also limits the discovery of such a Higgs boson significantly. In 1985, Howard Georgi and Marie Machacek \cite{Georgi:1985nv,Gunion:1989we} proposed an extended Higgs sector model, the Georgi-Machacek model, where the Higgs sector consists of a real $SU(2)_L$ triplet field $\xi$ with $Y=0$ and a complex triplet field $\chi$ with $Y=1$, in addition to the SM-like Higgs doublet field $\Phi$. The vacuum expectation values of $\xi$ and $\chi$ fields can be arranged so that a custodial $SU(2)_C$ symmetry is maintained. We will refer to this model as the GM model.

The GM model suffers from the same naturalness and hierarchy problems as the SM. In addition, it turns out that radiative corrections to $\rho\approx1$ in this model are infinite and thus the model has to be fine-tuned in this respect also \cite{Gunion:1990dt}. It is therefore natural to look at the supersymmetric extension of the GM model. 

Supersymmetry, a symmetry that relates integer-spin to half-integer-spin particles, and is consistent with Poincare invariance, was found to alleviate the naturalness and hierarchy problem to some extent. The simplest consistent version of Supersymmetry, the Minimal Supersymmetric Model (MSSM), however, requires the doubling of the particle spectrum. It also requires an additional Higgs scalar doublet and half-integer-spin partners for all the old and new scalar particles. The number of parameters grows to over 100.

As in the MSSM, the Supersymmetric Custodial Higgs Triplet Model (SCTM) doubles the GM spectrum\cite{Cort:2013foa,Garcia-Pepin:2014yfa,Delgado:2015bwa}. Furthermore, these additional scalars are paired with additional fermions, i.e. higgsinos, leading to a more complicated mass spectrum. $\rho\approx1$ is still maintained since the global $SU(2)_L\times SU(2)_R$ symmetry breaks down to the custodial $SU(2)_C$ symmetry after EWSB. 

Vega et al. \cite{Vega:2017gkk} showed that the GM model can arise as the low energy limit of the SCTM. In this scenario, half of the extra scalars in the model become very massive and decouple, and the other half remain at low mass and reproduce the scalar sector of the GM model. In this so-called decoupling limit, the SCTM scalar potential reduces in form to that of a restricted GM model. In this restricted GM model, the number of free parameters in the potential is reduced because supersymmetry imposes relations between the parameters of the original potential. In the present work, we conduct a phenomenological study of the decoupling limit of the SCTM; we call this the SGM model. We perform global fits for the free parameters in SGM model, using the latest available experimental data, for the first time. For comparison purposes, we do this for the GM model as well. The global fits are done using open-source software: \texttt{GMCalc}\cite{Hartling:2014xma}, a calculator for GM parameters; and \texttt{HiggsTools}\cite{Bahl:2022igd}, a toolbox for comparing BSM (beyond the Standard Model) models to the latest experimental data sets.

This dissertation is organized as follows. We briefly review the Standard Model in Chapter~\ref{chapter:SM Higgs} and its triplet extensions of the Higgs sector in Chapter~\ref{chapter:GM}. In Chapter~\ref{chapter:SCTM} we introduce the Supersymmetric Custodial Triplet Model and identify the Supersymmetric Georgi-Machacek Model as the decoupling limit of the SCTM model. In Chapter~\ref{chapter:package} we introduce the two open source packages we used in the global fits. The theoretical constraints of the SGM model on the parameter space are discussed. We also consider all available experiment data on exotic Higgs boson direct searches. Scan results are shown in Chapter~\ref{chapter:Results}. The possible presence of a Higgs Boson with a mass about 95 GeV within both the GM model and SGM model are examined in Chapter~\ref{chapter:95GeV}. A summary and discussion of future work conclude the dissertation. Some statistical concepts are reviewed in Appendix~\ref{appendix:A}. Experimental data related to GM and SGM scalars are listed in Appendix~\ref{appendix:exp}. 

 \begin{singlespace}
\chapter{The Standard Model}\label{chapter:SM Higgs}
\end{singlespace}

The framework of the SM is that of a quantum field theory. To each fundamental particle there corresponds a quantum field, and the time evolution and interactions of these particle states are governed by a Lagrangian constructed out of the quantum fields. The interactions can be understood as results of symmetries of the Lagrangian. Symmetries refer to transformations that leave the form of the Lagrangian and the equations of motion unchanged. The symmetries which involve spacetime coordinate transformations are governed by the Poincare group. Symmetries, such as those of the Poincare group, are called global symmetries because they are defined by parameters which are independent of spacetime coordinates. There are also internal (local) symmetries whose parameters do depend on spacetime coordinates; they are called gauge symmetries. The mediators of the fundamental forces are related to the local gauge symmetries. To each generator of the gauge group there corresponds a mediator.  Therefore, the gauge symmetries define the dynamics of the system. The gauge mediators are particles of spin 1, known as vector bosons. The matter particles are all of spin $1/2$. The Higgs particle, which is of spin $0$, is in a class of its own. We will briefly review the Standard Model, especially the Higgs sector in this chapter following \textit{Quantum Field Theory and the Standard Model} from Schwartz \cite{Schwartz:2014sze} and \textit{The Standard Model: A Primer} from Burgess and Moore\cite{Burgess:2006hbd}.

\section{The Poincare Group}

The Poincare symmetry is the full symmetry of special relativity, which includes translations in spacetime, rotations in space and boosts, for which the generators are denoted by $P_i$, $J_i$ and $K_i$, respectively. Here, boosts refer to Lorentz transformations that connect two uniformly moving objects in spacetime. The latter two, rotations and boosts, make up the Lorentz symmetry. The Lorentz symmetry is a global, exact, and unbroken symmetry in the SM. As a result, all elementary particles in the SM form representations of the Lorentz group. The full Lorentz symmetry group $O(1,3)$ is a subgroup of the Poincare group, which contains all isometries in Minkowski spacetime. The elements in the Lorentz group are called Lorentz transformations, denoted by $\Lambda^\mu_{\ \nu}$, with $\det \Lambda =\pm1$. We say a Lorentz transformation is proper if $\det\Lambda=1$ and improper if $\det\Lambda=-1$. 
Two fundamental improper Lorentz transformations are parity and time reversal.

The parity transformation, $\mathsf{P}$, is a reflection about the origin in the three dimensional spatial coordinates,
\begin{equation}
\mathsf{P}=\left(\begin{array}{cccc}
1 & 0 & 0 & 0 \\
0 & -1 & 0 & 0 \\
0 & 0 & -1 & 0 \\
0 & 0 & 0 & -1
\end{array}\right).
\end{equation}
The time-reversal transformation, $\mathsf{T}$, is a reflection about the origin for the time axis, 
\begin{equation}
\mathsf{T}=\left(\begin{array}{cccc}
-1 & 0 & 0 & 0 \\
0 & 1 & 0 & 0 \\
0 & 0 & 1 & 0 \\
0 & 0 & 0 & 1
\end{array}\right).
\end{equation}
A Lorentz transformation is orthochronous if $\Lambda^0{ }_0 \geq+1$ and non-orthochronous if $\Lambda^0_{ \ 0} \leq-1$. The time-reversal operator is a non-orthochronous Lorentz transformation. Furthermore, while the proper orthochronous Lorentz transformations are parametrized by a set of six continuous real numbers, the parity and time reversal symmetries are discrete symmetries. We state without proof that the most general Lorentz transformation can always be expressed as a direct product of a proper orthochronous Lorentz transformation times the parity and time reversal transformations. It is convenient then to treat the  proper orthochronous Lorentz transformation separately. This is also motivated by the fact that parity $\mathsf{P}$ and time reversal $\mathsf{T}$ are violated.

The fundamental particle states must form representations of the proper orthochronous Lorentz group, i.e. the restricted Lorentz group $SO^+(1,3)$. 
The Lie algebra of this group is, \begin{equation}
\begin{aligned}
{\left[J_i, J_j\right] } & =i \epsilon_{i j k} J_k, \\
{\left[J_i, K_j\right] } & =i \epsilon_{i j k} K_k, \\
{\left[K_i, K_j\right] } & =-i \epsilon_{i j k} J_k.
\end{aligned}
\end{equation}
If we define linear combinations of generators $J_i$ and $K_i$ to form new generators $J^+$ and $J^-$, where
\begin{equation}
J_i^{+} \equiv \frac{1}{2}\left(J_i+i K_i\right), \quad J_i^{-} \equiv \frac{1}{2}\left(J_i-i K_i\right).
\end{equation}
These new generators satisfy the Lie algebra \begin{equation}
\begin{aligned}
  \left[J_i^{+}, J_j^{+}\right]&=i \epsilon_{i j k} J_k^{+}, \\
  \left[J_i^{-}, J_j^{-}\right]&=i \epsilon_{i j k} J_k^{-}, \\
  \left[J_i^{+}, J_j^{-}\right]&=0.
\end{aligned}
\end{equation}
This implies 
\begin{equation}
SO(1,3)\cong SU(2)\times SU(2).
\end{equation}
In mathematical jargon we say that the universal cover of $SO^+(1,3)$ is isormorphic to $\mathrm{SL}(2, \mathbf{C})$.
More simply, representations of the Lorentz group can be constructed as a direct product of irreducible representations of $SU(2)$.

\section{The Gauge Groups}

Under the Lorentz group elementary particles are classified into three categories by their spin properties: the spin-$1/2$ fermions, the spin-0 scalar bosons and the spin-1 vector bosons. In the SM the only scalar particle is the Higgs boson; it is added ad hoc for the purpose of generating the electroweak symmetry breaking and endowing the gauge bosons with mass. The spin-1 vector bosons are the mediators of fundamental interactions. They are associated with each of the generators of the gauge group. The three types of fundamental interactions described by gauge theories in the SM are: the electromagnetic, weak and strong interactions. The gauge group of the SM is $U(1)_Y\times SU(2)_L\times SU(3)_C$. The fundamental particles form representations of these three groups. The subscript denotes the quantum numbers associated with each symmetry. The hypercharge $Y$ is associated with the abelian $U(1)$ symmetry, while the $L$ indicates that only the left-handed fermions carry nonzero charges under the weak symmetry group $SU(2)$, and the $C$ denotes the quantum number associated with the color charge of the strong interactions.  For this work we will focus exclusively on the electroweak interactions and its associated gauge group, $SU(2)_L\times U(1)_Y$.

Since the group $SU(2)$ has three generators and the group $U(1)$ one,
there are four vector bosons associated with the $SU(2)_L\times U(1)_Y$ gauge group. We designate these as $W^a_\mu(x)$, $a=1,2,3$, and $B_\mu(x)$ respectively. The $W$'s form a triplet representation under $SU(2)_L$ and have zero hypercharge. Likewise the $B_\mu$ carries $Y$ hypercharge and has zero weak-charge. As we will see shortly these gauge bosons are not the physical vector bosons observed in nature, they represent massless degrees of freedom. It is through the mechanism of spontaneous symmetry breaking, $SU(2)_L\times U_Y(1) \longrightarrow U(1)_Q$, that these four vector boson massless degrees of freedom are transformed into the four physical gauge bosons, the massless photon $A_\mu$, and the massive $W_\mu^\pm$-, and $Z_\mu$-bosons observed in nature.

\begin{figure}[hbt!]
 \centering
 \includegraphics[scale=0.22]{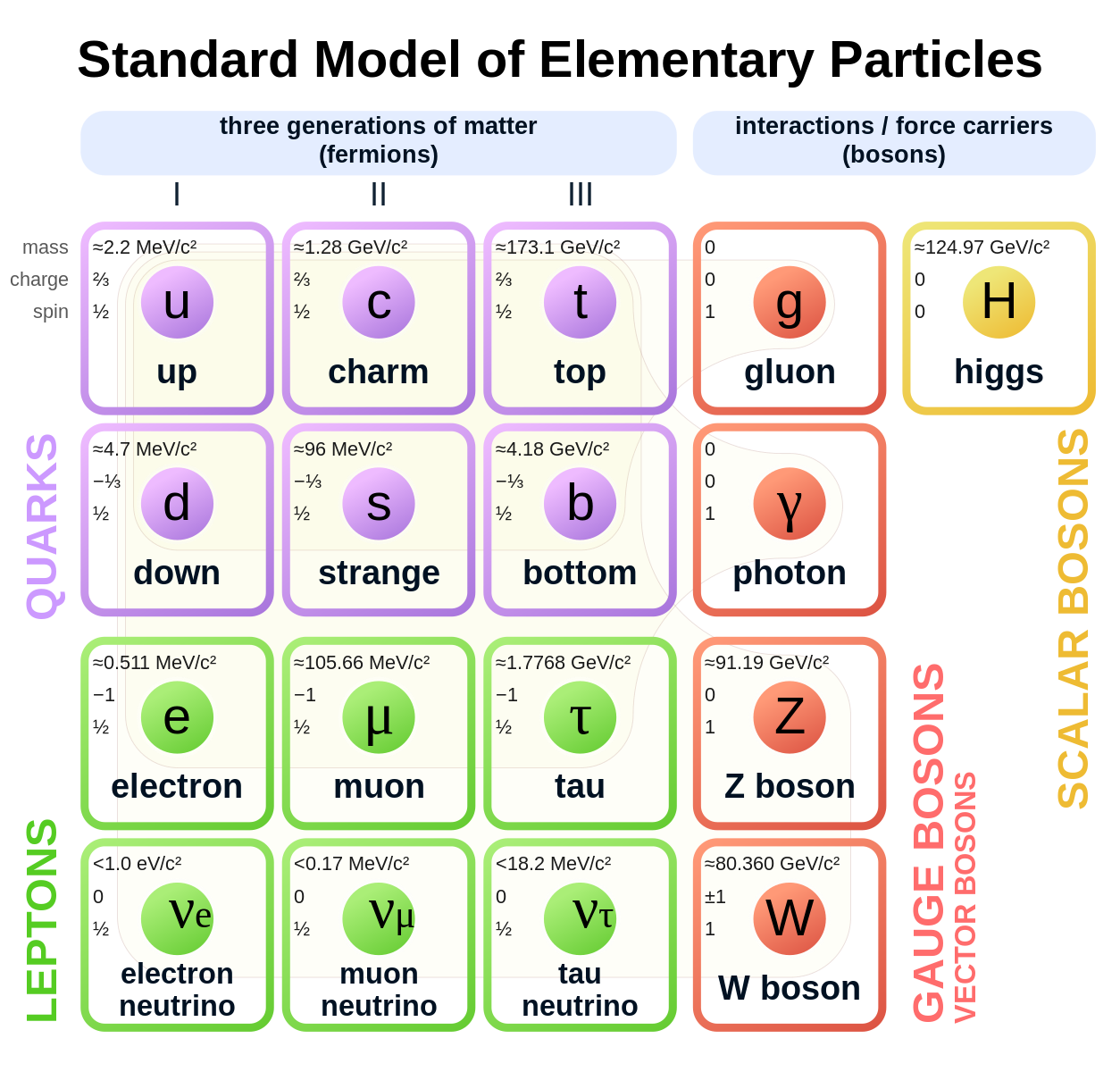} \caption{\centering Elementary Particles in the Standard Model.}
 {This figure is taken from Wikipedia\cite{standardmodelwiki}.}
 \label{fig:particles}
\end{figure} 

While unlike the gauge bosons the fermions carry color charge, we will only be concerned here with their classifications under the electroweak group. In the SM, only the singlet, doublet, and vector representations of $SU(2)_L$ are relevant.

\section{The Matter Field}

The Standard Model contains 12 kinds of elementary fermions. They are classified into 3 generations. The first 3 columns in Figure~\ref{fig:particles} represents the 3 generations. Particles in each generation appear to differ by their masses and are assigned with different flavor quantum numbers. In general, the fermions are represented by $(4\times 1)$ Dirac spinors which under Lorentz transformations transform as
\begin{equation}
\Psi_D(x)=\left(\begin{array}{c}
\psi_L(x) \\
\psi_R(x)
\end{array}\right) \rightarrow\left(\begin{array}{cc}
\Lambda_L & 0 \\
0 & \Lambda_R
\end{array}\right)\left(\begin{array}{c}
\psi_L(x) \\
\psi_R(x)
\end{array}\right),
\end{equation}
where,
\begin{equation}
\Lambda_{L(R)}= e^{\frac{1}{2}\left(i\vec{\theta}\cdot\vec{\tau}\mp\vec{\beta}\cdot \vec{\tau}\right)}, 
\end{equation}
here $\tau_i$, $i=1,2,3$, represent the three Pauli matrices, the $\theta$'s are rotation angles, and $\vec{\beta}$ is the rapidity vector. The fields $\psi_L(x)$ and $\psi_R(x)$ form the $(j^+,0)$ and $(0,j^-)$  representations of the Lorentz group respectively, with $j^+=j^-=1/2$.  They are called left-handed and right-handed Weyl spinors. The Dirac fermions thus form a reducible representation of the Lorentz transformations. The left-handedness and right-handedness refer to the $(1/2,0)$ and $(0,1/2)$ representations of the Lorentz group. The handedness is also called chirality. Since in the SM only the left-handed components participate in the weak interactions, it is called a chiral theory. The top component of the Dirac spinor $\Psi_D(x)$, $\psi_L(x)$, is a doublet under the gauge group $SU(2)_L$ while the bottom component, $\psi_R(x)$, is a singlet.

In summary, the transformations under $SU(2)_L\times U(1)_Y$ gauge symmetry for Weyl spinors are \begin{equation}\label{tlh}
\psi_L \rightarrow e^{-i \frac{1}{2}\vec{\theta}(x) \cdot \vec{\tau}-i \frac{1}{2}\omega(x) Y} \psi_L, \quad \psi_R \rightarrow e^{-i \frac{1}{2} \omega(x) Y} \psi_R,
\end{equation}
where $\omega(x)$ and $\vec{\theta}(x)$ are local gauge group transformation parameters depending on spacetime coordinates. 

We have seen that the elementary particles form irreducible representations of the electroweak group $SU(2)_L\times U(1)_Y$. We summarize this in Table~\ref{table:fermion}, where $i$ runs over the 3 generations. 

\begin{table}[H]
\centering
\begin{tabular}{|c|c|c|}
\hline
Fields  & $SU(2)_L$ & $U(1)_Y$ \\
\hline $Q_i$ & 2 & $1/6$ \\
\hline $L_i$ & 2 & $-1/2$ \\
\hline $U_i=u_{R,i}$ & 1 & $2/3$ \\
\hline $D_i=d_{R,i}$ & 1 & $-1/3$\\
\hline $E_i=e_{R,i}$ & 1 & -1\\
\hline
\end{tabular}
\caption[Elementary fermions grouped by transformation property and hyperchage.]{\centering
  Elementary fermions grouped by transformation property.}
  \label{table:fermion}
\end{table}

\begin{equation}
Q_i =\left(\begin{array}{c}u_{L, i} \\ d_{L, i}\end{array}\right), \quad L_i=\left(\begin{array}{c}\nu_{ L, i} \\ e_{L, i}\end{array}\right)
\end{equation}
The left-handed fields form the doublet representation of $SU(2)_L$, while the right-handed fields form the singlet representation of $SU(2)_L$. $Q_i$'s are left-handed quark doublets, where $u_i$ is the up-type quark and $d_i$ is the down-type quark for each generation. $L_i$'s are left-handed lepton doublets, where $e_i$ represents electron, muon and tauon in each generation and $\nu_i$ represents the associated neutrinos. $U_i$, $D_i$ and $E_i$ represent the right-handed singlets. 

The Lagrangian invariant under the gauge transformation \eqref{tlh} can be written as, \begin{equation}
\mathcal{L}_f=i \bar{\psi}_L \not D \psi_L+i \bar{\psi}_R \not D \psi_R,
\end{equation}
where we use the Feynman slash notation $\not D=\gamma^\mu D_\mu$. The covariant derivatives are, \begin{equation}
D_\mu \psi_L=\left(\partial_\mu-\frac{i}{2} g  Y B_\mu-\frac{i}{2} g' W_\mu^a \tau_a \right) \psi_L, \quad D_\mu \psi_R=\left(\partial_\mu-\frac{i}{2} g' Y B_\mu\right) \psi_R.
\end{equation}
\newpage
\section{The Higgs Mechanism}

The Lagrangian for the electroweak gauge bosons is 
\begin{equation}
\mathcal{L}=-\frac{1}{4} W_{\mu \nu}^a W^{a \mu \nu}-\frac{1}{4} B_{\mu \nu} B^{\mu \nu},
\end{equation}
where \begin{equation}
W_{\mu \nu}^a=\partial_\mu A_\nu^a-\partial_\nu A_\mu^a+g \epsilon^{a b c} A_\mu^b A_\nu^c, \quad B_{\mu \nu}=\partial_\mu B_\mu-\partial_\nu B_\mu.
\end{equation}
Terms quadratic in the fields are prohibited by the gauge symmetry. Therefore, the gauge bosons at this stage represent massless states. While this is true for the photon, it is not true of the other gauge bosons. The finite range of the weak interactions imply that the associated force carriers must be massive. The resolution of this impasse lies in the elegant framework of the Higgs mechanism. The Higgs mechanism is an example of what is called Spontaneous Symmetry Breaking (SSB). SSB occurs when an underlying symmetry of the Lagrangian is not manifest in the state of lowest energy of the system. A simple example is that of a ferromagnet. Above the Curie temperature, the magnetic moments of the constituent atoms are randomly aligned, as in Figure~\ref{fig:ferro1}. The system exhibits complete spherical symmetry. The interaction Hamiltonian is invariant under spatial rotations. Below the Curie temperature, the magnetic moments tend to align in a particular but random direction, as in Figure~\ref{fig:ferro2}. Below the Curie temperature, the system settles into this state of lowest energy and no longer exhibits the rotational symmetry. The Hamiltonian remains the same and the underlying symmetry is still there in the sense that any direction of alignment is as valid as any other.  Although it is common to refer to this as Spontaneous Symmetry Breaking, a more proper term might be spontaneously hidden symmetry.

\begin{figure}[hbt!]
    \centering
    \begin{minipage}{.5\textwidth}
        \centering
   \includegraphics[scale=0.4]{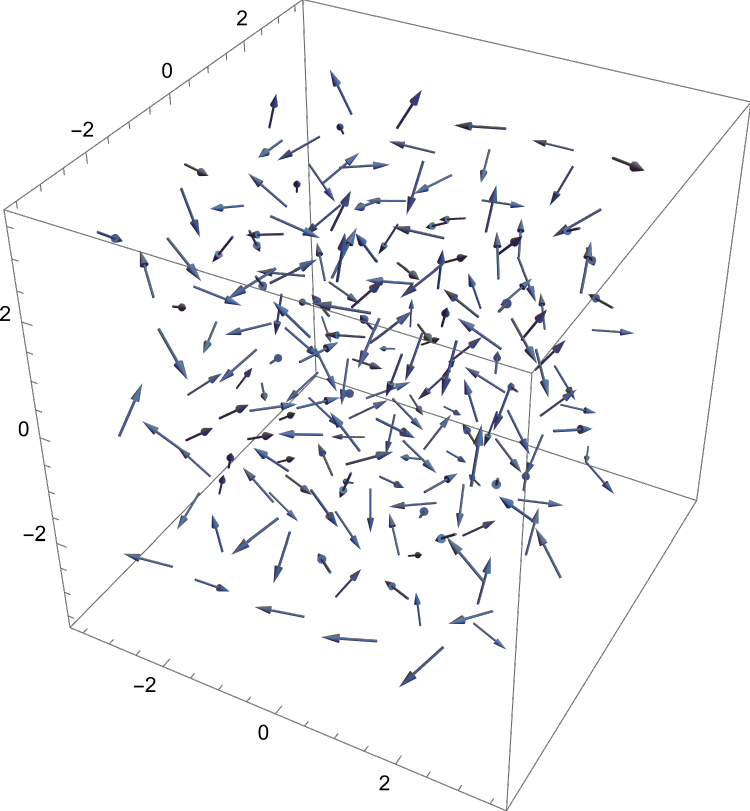} 
        \caption{Magnetic moments of the constituent atoms in a ferromagnet when $T>T_c$.}
        \label{fig:ferro1}
    \end{minipage}%
    \begin{minipage}{0.5\textwidth}
        \centering
   \includegraphics[scale=0.42]{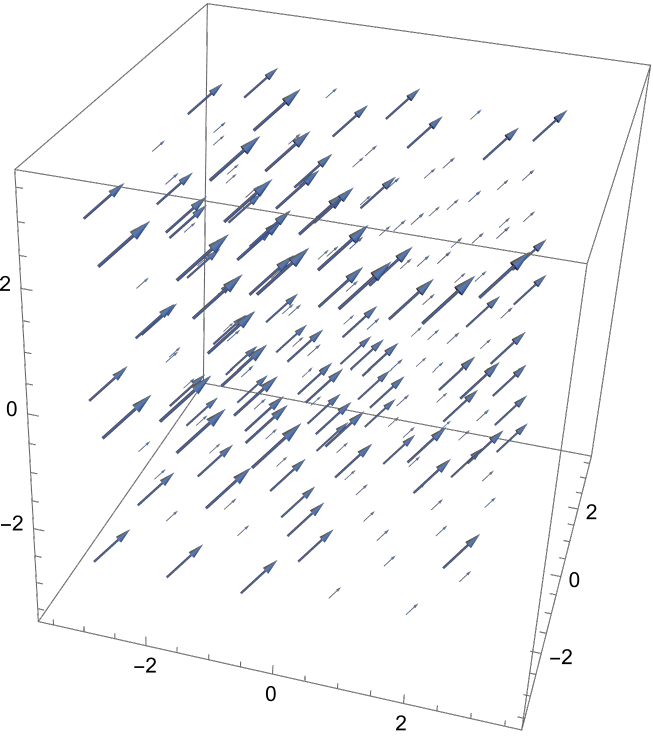} 
        \caption{Magnetic moments of the constituent atoms in a ferromagnet when $T<T_c$.}
        \label{fig:ferro2}
    \end{minipage}
\end{figure}

In the SM, the state of lowest energy, referred to as the vacuum state, is not invariant under part of the gauge symmetry. This breaking will manifest itself in the particle spectrum because we build the one particle states by action of creation operators on the vacuum state. If the vacuum state violates the gauge symmetry it will be reflected in the one-particle states. This is how the mass of the gauge bosons and the matter particles comes about.

As a simple example of spontaneous symmetry breaking in QFT, we consider a toy model with one complex scalar field $\phi(x)$ coupled to a gauge field field $A_\mu$. This model is known as scalar QED. The Lagrangian for scalar QED is given by, \begin{equation}\label{lagex1}
\mathcal{L}=-\frac{1}{4} F_{\mu \nu} F^{\mu \nu}+(D_\mu\phi)^\dagger D^\mu\phi-V(\phi),
\end{equation}
where the field strength is \begin{equation}
F_{\mu \nu}=\partial_\mu A_\mu-\partial_\nu A_\mu.
\end{equation}
The covariant derivative is \begin{equation}
D_\mu=\partial_\mu-i g A_\mu.
\end{equation}
The covariant derivative has been defined precisely so that
the Lagrangian is invariant under a $U(1)$ gauge transformations \begin{equation}\label{gt}
\phi(x) \rightarrow e^{i \alpha(x)} \phi(x), \quad A_\mu(x) \rightarrow A_\mu(x)+\frac{1}{g} \partial_\mu \alpha(x).
\end{equation}
The coupling to the gauge field arises from cross terms in the kinetic energy term which contains the covariant derivative. It is in this sense that we say that the gauge symmetry determines the dynamics. In non-relativistic Quantum Mechanics this is called minimal coupling.

The most general gauge-invariant Hermitian scalar potential with terms of dimension four or less can be written as
\begin{equation}
V(\phi)=-\mu^2 \phi^* \phi+\lambda\left(\phi^* \phi\right)^2.
\end{equation}
Unitarity requires both $\mu^2$ and $\lambda$ to be real while stability of the ground state requires $\lambda$ to be positive. We assume that $\mu^2>0$ so that the vacuum expectation value of the scalar field is different from zero and breaks the gauge symmetry.

\begin{figure}[hbt!]
 \centering
 \includegraphics[scale=0.4]{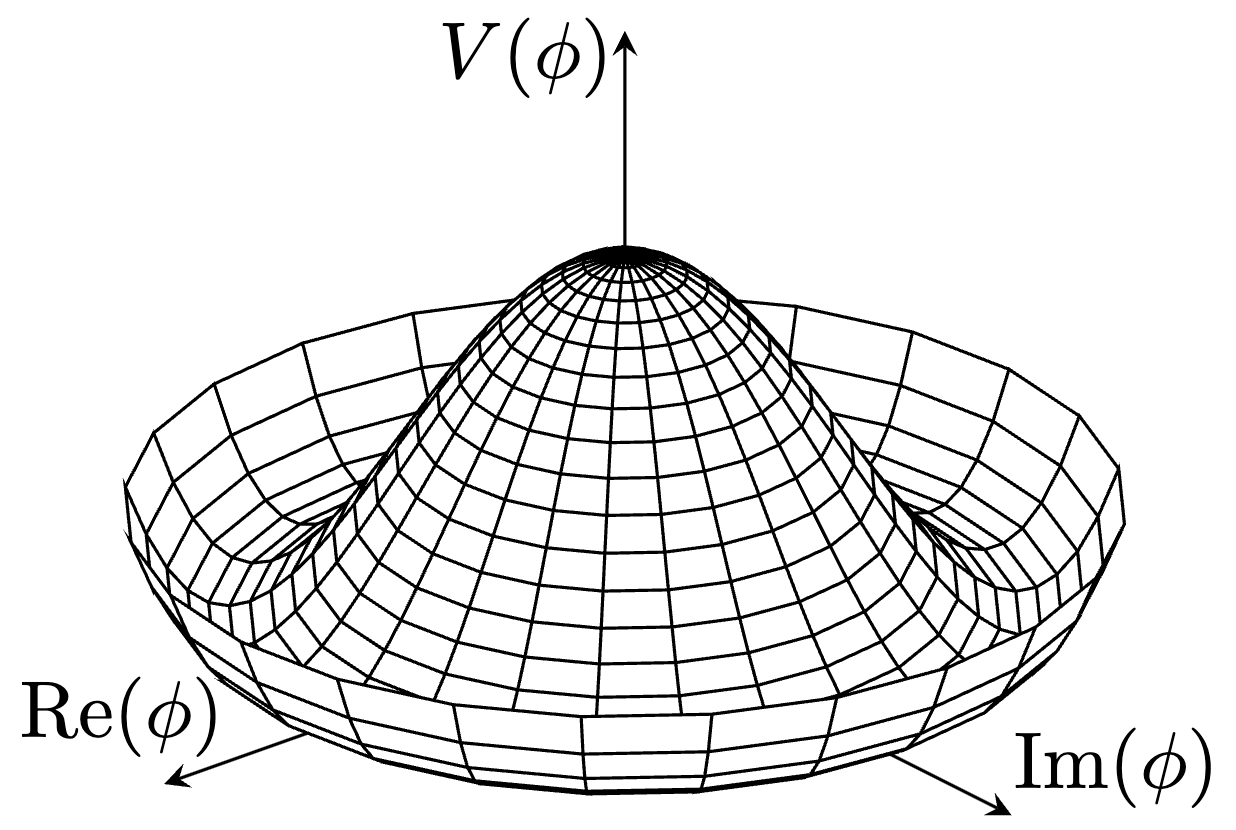} 
\caption{\centering The Mexican hat potential of the SM Higgs field.}
 {This figure is a courtesy of \cite{mexhat}.}
\label{fig:mexpot}
\end{figure} 

Figure~\ref{fig:mexpot} illustrates the so-called Mexican-hat potential for this toy model.
To find the value of $\phi(x)$ which minimizes the potential, it is simplest to use the parameterization
\begin{equation}\label{ex1}
\phi(x)=\frac{1}{\sqrt{2}}e^{i\xi(x)}h(x),
\end{equation}
where $h(x)$ and $\xi(x)$ are real fields. It is then easy to see that the minimum of the potential ocurrs when 
\begin{equation}\label{vev}
\Bigl. h(x)\Bigr\lvert_{min}=\sqrt{\frac{\mu^2}{ \lambda}}\equiv v.
\end{equation}
When the minimum the value of $h(x)$ is fixed, however, the $\xi(x)$ field can take any value. The field $\xi(x)$ just distinguishes between an infinite number of equivalent ground states. In Figure~\ref{fig:mexpot} we can see that as we vary the $\xi(x)$ field, we move along the circle of radius  $|\phi(x)|^2=\mu^2/2\lambda$. It costs no energy to move from one particular field configuration of $\xi(x)$ to another, so this field must represent a massless degree of freedom. This is the essence of Goldstone's theorem\cite{Goldstone:1962es}. One can verify these statements by substituting (\ref{ex1}) back into the Lagrangian. Of course, to study perturbations about the minimum, we also have to shift the scalar field, $h(x)\rightarrow h(x)+v$. Having done this, we would find a kinetic term for the $\xi(x)$ field with no corresponding quadratic term, verifying our expectation that the $\xi(x)$ fields represent massless states.  Upon these substitutions, we would also find that the $\xi(x)$ field mixes with the gauge boson $A_\mu(x)$; that is, there will be a quadratic term of the form $\xi(x) A_\mu(x)$. It is precisely this mixing which leads to massive gauge bosons. It is simplest to take advantage of the gauge invariance and set $\alpha(x)=-\xi(x)$ in transformation \eqref{gt},
\begin{equation}
\phi(x)\rightarrow e^{-i\xi(x)}\phi(x).
\end{equation}
This is called the unitary gauge, it has the effect equivalent to setting $\xi(x)=0$ at the minimum of the potential. Furthermore, in the unitary gauge, the $\xi(x)$ field completely disappears from the Lagrangian. It would appear as though picking the unitary gauge has the effect of reducing the number of degrees of freedom in the theory.  However, the number of degrees of freedom is preserved because in the process a mass term for the $A_\mu(x)$ gauge boson is generated. A massless gauge boson has only two degrees of freedom: its two transverse polarizations. A massive gauge boson has three degrees of freedom: two transverse and one longitudinal polarizations. The degree of freedom represented by the $\xi(x)$ is absorbed into the longitudinal component of the field $A_\mu(x)$. To make this explicit, we focus on the scalar kinetic energy term in (\ref{lagex1}),
\[
\Bigl.(D_\mu\phi)^\dagger D^\mu\phi\Bigr\rvert_{h(x)\rightarrow h(x)+v}=\partial_\mu h(x)\partial^\mu h(x) +\frac{1}{2}g^2v^2A_\mu(x)A^\mu(x) + \text{interaction terms}.
\]
The mass of the gauge boson is $m_A=gv$.

In the unitary gauge, the scalar potential takes the form,
\begin{equation}
V(\phi)=-\frac{\mu^4}{4\lambda}+\frac{1}{2}\cdot 2\mu^2h^2+\mathcal{O}(h^3).
\end{equation}
We note that the field $h(x)$ also acquires a mass of $\sqrt{2}\mu$. 

In the SM model the gauge bosons develop their mass in pretty much the same way. Rather than the simple $U(1)$, the SM exhibits a non-abelian symmetry represented by the product group, $SU(2)_L\times U(1)_Y$, i.e. the matter fields transform like the left-handed field in \eqref{tlh}. Since this product group has four generators, there will be four corresponding gauge bosons\footnote{For $SU(2)_L$ the gauge bosons transform under the adjoint representation.}. In the scalar QED example, there was only one generator, $Y$, and one gauge boson $A_\mu$. When the $U(1)$ gauge symmetry is broken, a single boson inherits a mass. In the SM three generators will be broken in the sense that they do not leave the vacuum invariant. Therefore, three of the four gauge bosons will then develop a mass and one will remain massless. The massless gauge boson is identified as the photon.

The scalar field is chosen to be in a doublet representation of $SU(2)_L$ with a vacuum expectation value such that the $SU(2)_L\times U(1)_Y$ symmetry is reduced to a $U(1)$ symmetry. In other words, the state of lowest energy is only invariant under $U(1)$. Since charge is absolutely conserved, this group must correspond to the electromagnetic group $U(1)_{Q}$. This is consistent with identifying of the massless gauge boson with the photon. In general, massless gauge bosons are associated with the generators that are unbroken. This process is referred as the Higgs mechanism. 

To break the $SU(2)_L\times U(1)_Y$ gauge symmetry, the simplest option is to utilize a complex scalar that transforms as a doublet under $SU(2)_L$ (as in (2.9)) and has hypercharge $Y$. A complex doublet has four degrees of freedom or four independent real fields. Using a parametrization similar to what we used in scalar QED, we write this complex scalar doublet as,
\begin{equation}\label{phix}
\Phi(x)=e^{i\frac{1}{2}\vec{\xi}(x)\cdot\vec{\tau}}
  \begin{pmatrix}
0 \\ H(x)
  \end{pmatrix}
\end{equation}
where $\tau_i$ are Pauli matrices, $\xi_1$, $\xi_2$, $\xi_3$, and $H(x)$ are four real scalar fields. As in scalar QED, the $\xi$-fields represent the Goldstone bosons.  

The Lagrangian of the SM scalar sector has the same form as in scalar QED,
\begin{equation}\label{smlag}
\mathcal{L}=\left(D_\mu \Phi\right)^{\dagger}\left(D_\mu \Phi\right)-V(\Phi),
\end{equation}
where $D_\mu$ is the covariant derivative. Here the covariant derivative for a $SU(2)\times U(1)$ theory, has the form, 
\begin{equation}\label{cov}
D_\mu \Phi=\partial_\mu \Phi-\frac{i}{2} g W_\mu^a \tau_a \Phi-\frac{i}{2} g' B_\mu \Phi.
\end{equation}
The Higgs potential again has the form,
 \begin{equation}
V(\Phi)=-\mu^2 \Phi^{\dagger} \Phi+\lambda\left(\Phi^{\dagger} \Phi\right)^2.
\end{equation}
The domain of the potential is a four dimensional space so it is difficult to visualize. However, the minimum is easily found by using the parametrization of \eqref{phix}. At the minimum the field $\Phi(x)$ satisfies,
\begin{equation}
  \label{eq:1}
  |\Phi_0(x)|=\frac{v}{\sqrt{2}},
\end{equation}
where $v$ is defined in (\ref{vev}). This defines the surface of a solid sphere in four dimensional space. This surface corresponds to the three dimensional space defined by the three fields $\vec{\xi}(x)$. As in scalar QED, the field $H(x)$ is equal to $v$ at the minimum, but the $\xi$-fields can take any value. Likewise, taking advantage of the $SU(2)_L$ gauge invariance, the three massless Goldstone bosons $\xi_i(x)$ can be gauged away by working in the unitary gauge and setting $\theta_i(x)=-\xi_i(x)$ in \eqref{tlh}, 
\begin{equation}
\Phi(x) \rightarrow e^{-i\frac{1}{2}\vec{\xi}(x)\cdot \vec{\tau}} \Phi(x).
\end{equation}
After this transformation, the Goldstone bosons do not appear in the Lagrangian explicitly. We then shift the $\Phi(x)$ field by the vacuum expectation value to get the perturbative mass spectrum. The $\Phi(x)$ field now takes the form
\begin{equation}
\Phi'(x)=\frac{1}{\sqrt{2}}\left(\begin{array}{c}
0 \\
H(x)+v
\end{array}\right).
\end{equation}
There remains only one physical scalar, referred to as the Higgs particle. By working in the unitary gauge, we have essentially selected a special vacuum point with $\xi_i(x)=0$ at that point. The symmetry of the theory is no longer apparent as the ground state $\Phi'_0$ is not invariant under original $SU(2)_L\times U(1)_Y$ transformation:
\begin{equation}
\tau_i \langle 0|\Phi'| 0\rangle\neq0, \quad Y \langle 0|\Phi'| 0\rangle\neq0,
\end{equation}
However, for $Y=1$, the combination $\tau_3+Y$ has the form 
\begin{equation}
\tau_3+Y=\left(\begin{array}{ll}
1 & 0 \\
0 & 0
\end{array}\right),
\end{equation}
which leaves $\Phi_0$ invariant:
\begin{equation}
\left(\tau_3+Y\right)\langle\Phi\rangle_0=\left(\begin{array}{ll}
1 & 0 \\
0 & 0
\end{array}\right) \frac{1}{\sqrt{2}}\left(\begin{array}{l}
0 \\
v
\end{array}\right)=0.
\end{equation}
Since we know that a $U(1)_Q$ symmetry must remain, the unbroken generator must correspond to the charge, 
\begin{equation}
Q=\tau_3+Y.
\end{equation}
This confirms that the gauge group $SU(2)_L\times U(1)_Y$ breaks down to $U(1)_Q$. 

We can now proceed as in the scalar QED case and rewrite the Lagrangian in (\ref{smlag}) in the unitary gauge with the scalar field shifted.
Ignoring overall constants the Lagrangian takes the form,
\begin{equation}\label{lag2}
\begin{aligned}
\mathcal{L}_{\text{Higgs}}&=\left(D_\mu \Phi'\right)^{\dagger}\left(D^\mu \Phi'\right)-V(\Phi')\\&=  \frac{1}{2} \partial_\mu H \partial^\mu H-\frac{1}{8}(v+H)^2 g^2\left(W_\mu^1-i W_\mu^2\right)\left(W^{1 \mu}+i W^{2 \mu}\right) \\
& -\frac{1}{8}(v+H)^2\left(-g W^{3 \mu}+g' B^\mu\right)\left(-g W_{\mu}^3+g' B_\mu\right)\\ &-\frac{1}{2}2\mu^2 H^2-\lambda v H^3-\frac{\lambda}{4} H^4.
\end{aligned}
\end{equation}
The Higgs mass can immediately be read off as $\sqrt{2}\mu$.
The gauge boson masses can be obtained by expanding the terms in the second and third lines,
\begin{equation}\label{massterms}
\frac{1}{8} g^2 v^2\left|W_\mu^1-i W_\mu^2\right|^2+\frac{1}{8} v^2\left(g W_\mu^3-g^{\prime}B_\mu\right)^2.
\end{equation}
The first term in \eqref{massterms} can be rewritten by defining,

\begin{equation}
  \label{eq:2}
 W_\mu^{ \pm}=\frac{1}{\sqrt{2}}\left(W_\mu^{1} \mp i W_\mu^{2}\right).   
\end{equation}
These fields represent vector bosons with mass,
\begin{equation}
  \label{eq:3}
  m_W^2=\frac{1}{4} g^2 v^2.
\end{equation}
We will see shortly that the fields $W_\mu^\pm(x)$ correspond to charged gauge bosons. The second term in \eqref{massterms} mixes the $W_\mu^3$ and $B_\mu$ fields. This term can be rewritten in matrix form,
\begin{equation}\label{mc}
\frac{v^2}{8}
\begin{pmatrix}
    W_\mu^3 & B_\mu
\end{pmatrix}
\begin{pmatrix}
  g^2 & -g g' \\
 -g g' & g^{\prime2}
\end{pmatrix}
\begin{pmatrix}
  W_\mu^3 \\ B_\mu
\end{pmatrix}
\end{equation}
We note that the determinant of this $2\times 2$ mass matrix is zero. This indicates that one of the mass eigenvalues is zero. The eigenvector corresponding to this zero eigenvalue is the photon field. Upon diagonalizing this matrix, we obtain the two physical gauge bosons and their corresponding mass eigenvalue,
\begin{align}
Z_\mu = W_\mu^3 \cos \theta_{\mathrm{W}}-B_\mu \sin \theta_{\mathrm{W}}, & \qquad m_Z^2=\frac{1}{4}\left(g^{2}+g^{\prime2}\right) v^2 \\
A_\mu = W_\mu^3 \sin \theta_{\mathrm{W}}+B_\mu \cos \theta_{\mathrm{W}}, & 
\qquad m_A^2=0
\end{align}
where \begin{equation}
\sin \theta_{\mathrm{W}}=\frac{g^{\prime}}{\sqrt{g^2+g^{\prime 2}}}, \quad \cos \theta_{\mathrm{w}}=\frac{g}{\sqrt{g^2+g^{\prime2}}}.
\end{equation}
Upon replacing these definitions into \eqref{lag2}, we can verify that the $W_\mu^\pm(x)$ fields couple to the photon field $A_\mu(x)$ with strength $e=g\sin{\theta_W}$, they represent the charged gauge bosons.  The field $Z_\mu$ does not couple to the photon field and represents a neutral gauge boson.  Also note that the gauge boson masses are not independent,
\begin{equation}
  \label{eq:4}
  m_W^2=m_Z\cos{\theta_W}.
\end{equation}
As we discuss in the next section, this relation is a consequence of an accidental global symmetry in the SM. It is accidental in the sense that we do not a priori impose the symmetry, but is a consequence of the structure imposed by the requirements of gauge invariance.

\section{Custodial $SU(2)$}

To see the accidental custodial symmetry in the SM, we first write $\Phi$ as a bi-doublet 
\begin{equation}
\tilde{\Phi}=\left(\begin{array}{cc}
\phi^{0 *} & \phi^{+} \\
-\phi^{+*} & \phi^0
\end{array}\right).
\end{equation}
The second column is the original doublet $\Phi$, and the first column is the conjugate doublet \begin{equation}
\tilde{\Phi}=i\tau_2\Phi^*.
\end{equation}
In this representation, the scalar potential is \begin{equation}
V=-\frac{\mu^2}{2} \operatorname{Tr}\left(\tilde{\Phi}^{\dagger} \tilde{\Phi}\right)+\frac{\lambda}{4}\left[\operatorname{Tr}\left(\tilde{\Phi}^{\dagger} \tilde{\Phi}\right)\right]^2.
\end{equation}
The scalar potential is invariant under $SU(2)_L\times SU(2)_R$ transformations
\begin{equation}
\tilde{\Phi} \rightarrow e^{i\frac{1}{2}\vec{\theta}_L\times\vec{\tau}_a} \tilde{\Phi} e^{-i\frac{1}{2}\vec{\theta}_R\times\vec{\tau}_b}.
\end{equation}
Here, the "L" and "R" labels in this context have nothing to do with chirality and the parameters $\vec{\theta}_L$ and $\vec{\theta}_R$ are constants independent of spacetime coordinates. This symmetry $SU(2)_L\times SU(2)_R$ is a global symmetry. The Higgs mechanism also breaks this global $SU(2)_L\times SU(2)_R$ symmetry to a $SU(2)$ subgroup. We call this leftover symmetry custodial symmetry. The v.e.v of the Higgs bi-doublet takes the form \begin{equation}
\langle\tilde{\Phi}\rangle=\frac{v}{\sqrt{2}}\left(\begin{array}{ll}
1 & 0 \\
0 & 1
\end{array}\right).
\end{equation}
In the SM electroweak sector, the mass matrix squared in \eqref{mc} tells us that \begin{equation}
m_{W_\mu^{\pm}}^2=m_W^2=\frac{1}{4}g^2v^2
\end{equation}
and the $W^\pm$ bosons do not mix with anything else. $W_\mu^3$ and $B_\mu$ mix by an angle \begin{equation}
\theta_W=\arctan\left(\frac{g'}{g}\right)
\end{equation}
to produce massive a $Z$ boson and a photon. The custodial symmetry is hence not exact. It is violated by the hypercharge term which leads to a difference between $M_W$ and $M_Z$. In the limit $g'\rightarrow0$, the Lagrangian of the SM is invariant under a global $SU(2)_L\times SU(2)_R$ transformation. For this symmetry to be manifest, terms like $D_\mu\Phi(x)$ must transform like the field $\Phi(x)$. Therefore, the term $\tau^a W_\mu^a\Phi(x)$ contained in $D_\mu\Phi(x)$ must also transform like $\Phi(x)$. This imposes a transformation rule on $W^a_\mu$. After SSB, we have seen that the symmetry reduces to just the custodial symmetry, $SU(2)_C$. Since the generators $\tau^a$ transform as a triplet under $SU(2)$, this implies that $W^a_\mu$ must transform as a triplet. When $g’\neq 0$, \eqref{mc} shows that the field $B_\mu$, which does not have a well defined transformation under $SU(2)_C$, mixes with the field $W^3_\mu$. This mixing breaks the custodial symmetry, but in a particularly well-defined way. Once we diagonalize the $2\times2$ submatrix and change from the $B_\mu-W^3_\mu$ basis to the $A_\mu-Z_\mu$ basis, a remnant of the $SU(2)_C $ symmetry is still manifest in the relationship between $M_W$ and $M_Z$
\begin{equation}
\rho=\frac{m_W^2}{m_Z^2\cos^2\theta_W}=1.
\end{equation}

\section{Fermion Masses}

The fermion mass terms $m\bar{\psi}\psi$ are also absent in the Lagrangian of the SM. These terms are not invariant under $SU(2)_L$ transformations as $\bar{\psi}{\psi}=\bar{\psi}_L\psi_R\neq\bar{\psi}_R\psi_L$. It is said they violate chirality. The terms quadratic in fermion fields come from the Yukawa couplings after shifting the scalar fields by $v$. Therefore, fermion fields acquire masses proportional to the Higgs v.e.v. Let $Y_{mn}$ be the Yukawa couplings matrix; the relevant terms are \begin{equation}
\mathcal{L}_Y=Y_m^e \bar{E}_{i} \Phi^{\dagger} L_i+Y_{mn}^d \bar{D}_{i,m} \Phi^{\dagger} Q_{i,n}+Y_{mn}^u \bar{U}_{i,m} \tilde{\Phi}^{\dagger} Q_{i,n}+\text { h.c., } \quad\left(\text { where } \tilde{\Phi}=i \tau_2 \Phi^*\right),
\end{equation}
where we use the same notation as in Table 2.1. In the unitary gauge, we have \begin{equation}
\Phi=\frac{1}{\sqrt{2}}\left(\begin{array}{c}
0 \\
v+H(x)
\end{array}\right), \quad \tilde{\Phi}=\frac{1}{\sqrt{2}}\left(\begin{array}{c}
v+H(x) \\
0
\end{array}\right)
\end{equation}
Therefore, fermion masses are given by \begin{equation}
M_m^e=\frac{1}{\sqrt{2}} Y_m^e v, \quad M_{mn}^d=\frac{1}{\sqrt{2}} Y_{m n}^d v, \quad M_{mn}^u=\frac{1}{\sqrt{2}} Y_{mn}^u v
\end{equation}
We have reviewed the Spontaneous Symmetry Breaking and Higgs Mechanism in the SM, and how the Higgs mechanism generates masses for gauge bosons and fermions. The Higgs Mechanism in SM, is the simplest case since it only contains a complex doublet representation. We may add higher representations with caution, i.e. $\rho\approx1$ must be maintained and the custodial $SU(2)$ symmetry must be preserved. A natural extension to the SM Higgs sector is to add triplet representations, leading to the Georgi-Machacek Model. 
 \begin{singlespace}
\chapter{The Georgi-Machacek Model}\label{chapter:GM}
\end{singlespace}

The Georgi-Machacek model is a BSM model that adds electroweak triplets in its Higgs sector in a way that preserves the custodial symmetry. It provides natural settings for Type-II Seesaw Mechanisms that generate Majorana masses for neutrinos\cite{Gunion:1989ci}. The GM model also has two interesting features phenomenologically. Firstly, the coupling strengths of the SM-like Higgs $h$ to $W^\pm$ or $Z$ bosons can be larger than in the SM\cite{Hartling:2014aga}. Such enhancements would be impossible in a BSM model that only contains electroweak doublets. Secondly, a doubly charged Higgs boson is present in this model. 

\section{The Higgs Sector}

The Higgs sector of the GM model consists of the SM doublet Higgs field with hypercharge $Y=1$ and two new triplet Higgs fields: a complex field $\chi$ with $Y=1$ and a real field $\xi$ with $Y=0$. In matrix form, these fields can be expressed as \cite{Georgi:1985nv} \begin{equation}
\Phi=\left(\begin{array}{cc}
\phi^{0 *} & \phi^{+} \\
-\phi^{+*} & \phi^0
\end{array}\right),\ \quad
X=\left(\begin{array}{ccc}
\chi^{0 *} & \xi^{+} & \chi^{++} \\
-\chi^{+*} & \xi^0 & \chi^{+} \\
\chi^{++*} & -\xi^{+*} & \chi^0
\end{array}\right).
\end{equation}
$\Phi$ and $X$ transform under the global $SU(2)_L\times SU(2)_R$ as \begin{equation}
\Phi \rightarrow \mathcal{U}_L \Phi \mathcal{U}_R^{\dagger}, \quad X\rightarrow \mathcal{U}_L \Delta \mathcal{U}_R^{\dagger}.
\end{equation}
The kinetic term of the Lagrangian in this model is \begin{equation}
\mathcal{L}_{\mathrm{kin}}=\frac{1}{2} \operatorname{Tr}\left[\left(D_\mu \Phi\right)^{\dagger}\left(D_\mu \Phi\right)\right]+\frac{1}{2} \operatorname{Tr}\left[\left(D_\mu X\right)^{\dagger}\left(D_\mu X\right)\right],
\end{equation}
where \begin{equation}
D_\mu \Phi \equiv \partial_\mu \Phi+i \frac{1}{2} gW^a_\mu\tau_a \Phi-i \frac{1}{2}g^{\prime} B\Phi ,
\end{equation}
\begin{equation}
D_\mu X\equiv \partial_\mu X+i\frac{1}{2} g W^a_\mu\tau_aX-i \frac{1}{2} g^{\prime} B t_3 X.
\end{equation}
Here, $\tau_a$ are the Pauli matrices, the generators for the $(\mathbf{2},\bar{\mathbf{2}})$ representation of $SU(2)$. The generators for the $(\mathbf{3},\bar{\mathbf{3}})$ representation are \begin{equation}
t^1=\frac{1}{\sqrt{2}}\left(\begin{array}{lll}
0 & 1 & 0 \\
1 & 0 & 1 \\
0 & 1 & 0
\end{array}\right), \quad t^2=\frac{1}{\sqrt{2}}\left(\begin{array}{ccc}
0 & -i & 0 \\
i & 0 & -i \\
0 & i & 0
\end{array}\right), \quad t^3=\left(\begin{array}{ccc}
1 & 0 & 0 \\
0 & 0 & 0 \\
0 & 0 & -1
\end{array}\right),
\end{equation}
The most general gauge-invariant Higgs scalar potential that preserves custodial $SU(2)$ symmetry is given by \cite{Hartling:2014zca} 
\begin{equation}\label{gm}
\begin{aligned}
V(\Phi, X)= & \frac{\mu_2^2}{2} \operatorname{Tr}\left(\Phi^{\dagger} \Phi\right)+\frac{\mu_3^2}{2} \operatorname{Tr}\left(X^{\dagger} X\right)+\lambda_1\left[\operatorname{Tr}\left(\Phi^{\dagger} \Phi\right)\right]^2+\lambda_2 \operatorname{Tr}\left(\Phi^{\dagger} \Phi\right) \operatorname{Tr}\left(X^{\dagger} X\right) \\
& +\lambda_3 \operatorname{Tr}\left(X^{\dagger} X X^{\dagger} X\right)+\lambda_4\left[\operatorname{Tr}\left(X^{\dagger} X\right)\right]^2-\lambda_5 \operatorname{Tr}\left(\Phi^{\dagger} \tau^a \Phi \tau^b\right) \operatorname{Tr}\left(X^{\dagger} t^a X t^b\right) \\
& -M_1 \operatorname{Tr}\left(\Phi^{\dagger} \tau^a \Phi \tau^b\right)\left(U X U^{\dagger}\right)_{a b}-M_2 \operatorname{Tr}\left(X^{\dagger} t^a X t^b\right)\left(U X U^{\dagger}\right)_{a b}.
\end{aligned}
\end{equation}
The matrix $U$, which rotates $X$ into the Cartesian basis, is given by
\begin{equation}
U=\left(\begin{array}{ccc}
-\frac{1}{\sqrt{2}} & 0 & \frac{1}{\sqrt{2}} \\
-\frac{i}{\sqrt{2}} & 0 & -\frac{i}{\sqrt{2}} \\
0 & 1 & 0
\end{array}\right).
\end{equation}

\section{The Mass Spectrum}

The potential in \eqref{gm} is such that it allows for minima for the potential at non-zero values of the fields. As we have seen, non-zero values for the fields would lead to spontaneous symmetry breaking of the electroweak gauge symmetry. As in the SM, the global $SU(2)_L\times SU(2)_R$ symmetry will also break to a $SU(2)_C$ subgroup in the GM model. We define the vacuum expectation values of the neutral fields as 
\begin{equation}
\langle\phi_0\rangle=v_\phi/\sqrt{2}, \quad \langle\xi_0\rangle=v_\xi, \quad \langle \chi_0\rangle=v_\chi.
\end{equation}
Charged fields can only have zero vacuum expectation value, otherwise the $U(1)_Q$ symmetry would be broken and charge would not be conserved. As the v.e.v's minimize the scalar potential, we obtain the tadpole equations 
\begin{equation}
0=\frac{\partial V}{\partial v_\phi}, \quad 0=\frac{\partial V}{\partial v_\xi}, \quad 0=\frac{\partial V}{\partial v_\chi}.
\end{equation}
Under the custodial symmetry, v.e.v's must take the identity form 
\begin{equation}
\langle\Phi\rangle=\frac{1}{\sqrt{2}}\left(\begin{array}{cc}
v_\phi & 0 \\
0 & v_\phi
\end{array}\right), \quad\langle\Delta\rangle=\left(\begin{array}{ccc}
v_{\Delta} & 0 & 0 \\
0 & v_{\Delta} & 0 \\
0 & 0 & v_{\Delta}
\end{array}\right).
\end{equation}
Or simply $v_\chi=v_\xi=v_{\Delta}$. Both the doublet and triplet v.e.v's contribute to the EWSB, 
\begin{equation}
v_\phi^2+8 v_\chi^2 \equiv v^2=\frac{1}{\sqrt{2}G_F}=\frac{4 M_W^2}{g^2} \approx(246 \mathrm{GeV})^2.
\end{equation}
Custodial symmetry requires $v_\xi=v_\chi$ so only 2 tadpole equations are left. We can use the 2 tadpole equations to substitute 2 free parameters in the scalar potential. To see this, we rewrite the scalar potential in terms of $v_\phi$ and $v_\chi$,
\begin{equation}
V\left(v_\phi, v_\chi\right)=\frac{\mu_2^2}{2} v_\phi^2+3 \frac{\mu_3^2}{2} v_\chi^2+\lambda_1 v_\phi^4+\frac{3}{2}\left(2 \lambda_2-\lambda_5\right) v_\phi^2 v_\chi^2+3\left(\lambda_3+3 \lambda_4\right) v_\chi^4-\frac{3}{4} M_1 v_\phi^2 v_\chi-6 M_2 v_\chi^3.
\end{equation}
The tadpole equations are
\begin{equation}
\begin{aligned}
\frac{\partial V}{\partial v_\phi}&=v_\phi\left[\mu_2^2+4 \lambda_1 v_\phi^2+3\left(2 \lambda_2-\lambda_5\right) v_\chi^2-\frac{3}{2} M_1 v_\chi\right]=0, \\
\frac{\partial V}{\partial v_\chi}&=3 \mu_3^2 v_\chi+3\left(2 \lambda_2-\lambda_5\right) v_\phi^2 v_\chi+12\left(\lambda_3+3 \lambda_4\right) v_\chi^3-\frac{3}{4} M_1 v_\phi^2-18 M_2 v_\chi^2=0
\end{aligned}
\end{equation}
We can then shift the scalar fields in terms of their v.e.v's and decompose the neutral fields into real and imaginary parts,
\begin{equation}
\phi^0 \longrightarrow \frac{v_\phi}{\sqrt{2}}+\frac{\phi^{0, r}+i \phi^{0, i}}{\sqrt{2}}, \quad \chi^0 \longrightarrow v_\chi+\frac{\chi^{0, r}+i \chi^{0, i}}{\sqrt{2}}, \quad \xi^0 \longrightarrow v_\chi+\xi^0.
\end{equation}
The physical field can be organized by their transformation properties under the custodial $SU(2)$ symmetry. Explicitly, we have $\mathbf{2} \otimes \mathbf{2}=\mathbf{1} \oplus \mathbf{3}$ and $\mathbf{3} \otimes \mathbf{3}=\mathbf{1} \oplus \mathbf{3} \oplus \mathbf{5}$. Thus, we have 2 singlets, 2 triplets and 1 quintuplet. These can be organized to 10 physical fields and 3 Goldstone bosons. The Goldstone bosons are given by \begin{equation}
\begin{aligned}
G^{\pm} & =c_H \phi^{\pm}+s_H \frac{\left(\chi^{\pm}+\xi^{\pm}\right)}{\sqrt{2}}, \\
G^0 & =c_H \phi^{0, i}+s_H \chi^{0, i},
\end{aligned}
\end{equation}
where \begin{equation}
c_H \equiv \cos \theta_H=\frac{v_\phi}{v}, \quad s_H \equiv \sin \theta_H=\frac{2 \sqrt{2} v_\chi}{v}.
\end{equation}
The W and Z bosons are given mass by absorbing these Goldstone bosons so that \begin{equation}
m_W^2=m_Z^2\cos^2\theta_W=\frac{1}{4}g^2v^2
\end{equation}
and $\rho_{tree}=1$ is preserved. The quintuplet and triplet physical states are \begin{equation}
\begin{aligned}
H_5^{\pm\pm} & =\chi^{\pm\pm}, \\
H_5^{\pm} & =\frac{\left(\chi^{\pm}-\xi^{\pm}\right)}{\sqrt{2}}, \\
H_5^0 & =\sqrt{\frac{2}{3}} \xi^0-\sqrt{\frac{1}{3}} \chi^{0, r}, \\
H_3^{\pm} & =-s_H \phi^{\pm}+c_H \frac{\left(\chi^{\pm}+\xi^{\pm}\right)}{\sqrt{2}}, \\
H_3^0 & =-s_H \phi^{0, i}+c_H \chi^{0, i},
\end{aligned}
\end{equation}  
Here, $H_5^0$ is CP-even, whereas $H_3^0$ is CP-odd. The quintuplet and triplet masses are \begin{equation}\label{m5}
m_5^2=\frac{M_1}{4 v_\chi} v_\phi^2+12 M_2 v_\chi+\frac{3}{2} \lambda_5 v_\phi^2+8 \lambda_3 v_\chi^2,
\end{equation}
\begin{equation}\label{m3}
m_3^2=\left(\frac{M_1}{4 v_\chi}+\frac{\lambda_5}{2}\right) v^2.
\end{equation}
The two singlet states in the gauge basis are given by \begin{equation}
\begin{aligned}
H_1^0 & =\phi^{0, r}, \\
H_1^{0 \prime} & =\sqrt{\frac{1}{3}} \xi^0+\sqrt{\frac{2}{3}} \chi^{0, r} .
\end{aligned}
\end{equation}
The physical states are \begin{equation}
\left(\phi^{0,r}, H_1^{0,\prime}\right)\left(\begin{array}{cc}
\mathcal{M}_{11}^2 & \mathcal{M}_{12}^2 \\
\mathcal{M}_{21}^2 & \mathcal{M}_{22}^2
\end{array}\right)\left(\begin{array}{c}
\phi^{0,r} \\
H_1^{0,\prime}
\end{array}\right)=\left(h, H\right)\left(\begin{array}{cc}
m_h^2 & 0 \\
0 & m_{H}^2
\end{array}\right)\left(\begin{array}{c}
h \\
H
\end{array}\right),
\end{equation}
The matrix elements are 
\begin{equation}
\begin{aligned}
& \mathcal{M}_{11}^2=8 \lambda_1 v_\phi^2, \\
& \mathcal{M}_{12}^2=\frac{\sqrt{3}}{2} v_\phi\left[-M_1+4\left(2 \lambda_2-\lambda_5\right) v_\chi\right], \\
& \mathcal{M}_{22}^2=\frac{M_1 v_\phi^2}{4 v_\chi}-6 M_2 v_\chi+8\left(\lambda_3+3 \lambda_4\right) v_\chi^2,
\end{aligned}
\end{equation}
The mixings are \begin{equation}
\begin{aligned}
h & =\cos \alpha H_1^0-\sin \alpha H_1^{0 \prime} \\
H & =\sin \alpha H_1^0+\cos \alpha H_1^{0 \prime}.
\end{aligned}
\end{equation}
The mixing angle is defined as \begin{equation}
\sin 2 \alpha  =\frac{2 \mathcal{M}_{12}^2}{m_H^2-m_h^2}, \quad
\cos 2 \alpha  =\frac{\mathcal{M}_{22}^2-\mathcal{M}_{11}^2}{m_H^2-m_h^2}
\end{equation} 
The masses of the two singlets are given by the eigenvalues of the mass-squared matrix
\begin{equation}
m_{h, H}^2=\frac{1}{2}\left[\mathcal{M}_{11}^2+\mathcal{M}_{22}^2 \mp \sqrt{\left(\mathcal{M}_{11}^2-\mathcal{M}_{22}^2\right)^2+4\left(\mathcal{M}_{12}^2\right)^2}\right].
\end{equation}
\begin{table}[H]
\centering
\begin{tabular}{|c|c|c|}
\hline
  & scalar & pseudoscalar \\
\hline singlet & $h, H$ & \\
\hline triplet & & $G_3, H_3$ \\
\hline quintuplet & $H_5$ & \\
\hline
\end{tabular}
 \caption[Particle Content in the GM model Higgs Sector.]{\centering
  The two singlet Higgs bosons, the triplet and the quintuplet are shown along with the Goldstone bosons.}
\end{table}
The particle content of the GM Higgs sector is summarized in Table 3.1. Compared with the SM, the GM model introduces 9 new parameters: the quadratic couplings $\mu_{2,3}^2$, the cubic couplings $M_{1,2}$ and the quartic couplings $\lambda_{1,2,3,4,5}$. We use tadpole conditions to replace $\mu_{2,3}^2$ with $v_\phi$ and $v_\Delta$. The EWSB and the Higgs mass will fix 2 of the 9 parameters as \begin{equation}
v_\phi^2+8 v_{\Delta}^2=v^2=(246 \mathrm{GeV})^2, \quad m_{h, H}=125 \mathrm{GeV}.
\end{equation}
As a result, we have 7 free parameters in the GM model. We pick them to be $\lambda_{1,2,3,4,5}$ and $M_{1,2}$. Like other non-supersymmetric models, the GM model suffers from the hierarchy problem and the naturalness problem. In addition, the model needs to be fined-tuned since the radiative corrections to $\rho\approx 1$ is infinite in this model\cite{Gunion:1990dt}. Supersymmetry, a symmetry that relates integer-spin bosons to half-integer spin fermions can alleviate the naturalness and hierarchy problem. The supersymmetric extension of the GM model, namely the Supersymmetric Custodial Triplet model, will be reviewed in next chapter.

 \begin{singlespace}
\chapter{The Supersymmetric Custodial Higgs Triplet Model}\label{chapter:SCTM}
\end{singlespace}

Supersymmetry is a rather radical extension of the Standard Model in that it postulates a symmetry that relates fermions and bosons. This extension essentially doubles the particle spectrum since each particle in the standard model inherits a superpartner. In the minimal case, it also doubles the number of fields in the scalar sector. For technical reasons, we need two distinct Higgs doublets to give mass to both the up-type and down-type quarks respectively. Of course, these extra scalar fields are also accompanied by their superpartners. The motivations for this extension are widely discussed in the literature and are well known. Although the theory starts from a relatively simple premise, it quickly becomes very complicated with the number of free parameters in the hundreds.

The supersymmetric extension of the GM model (Supersymmetric Georgi-Machacek Model, or simply SGM model) arises as the decoupling limit of the Supersymmetric Custodial Higgs Triplet model\cite{Vega:2017gkk}. It has a much more complex mass spectrum and richer particle content. In the SCTM model, the number of fields in the scalar sector is doubled. Whereas the GM model is populated by a complex scalar doublet, a real triplet with 0 hypercharge and a quintuplet with hypercharge 2, for a total of 9 physical fields, the SCTM model doubles that number. In addition, there are also the corresponding fermionic superpartners to these scalar fields. In this model, the problem of naturalness in the SM, and the problem of the $\rho$-parameter in the GM model are resolved. 

For this dissertation, we will not dwell on the technical details related to supersymmetry, but rather focus on the rich phenomenology of just the scalar sector of the model. In this section, we will review the scalar sector of the SCTM model\cite{Cort:2013foa}\cite{Garcia-Pepin:2014yfa} and discuss how we can use global fits to constrain or rule out this model.

\section{The Higgs Sector}

In the SCTM model, the Higgs sector is constructed to be invariant under the $SU(2)_R\times SU(2)_L$ symmetry. The MSSM Higgs sector\footnote{When supersymmetrizing the SM Higgs sector, we need to double the number of scalar fields, one for down-type $H_1$ and one for up-type $H_2$, and couple them separately to up-type and down-type superfields in the superpotential. The is a result of the Holomorphic Principle, which states that superpotential can only depend on chiral superfields, not their complex conjugates\cite{Seiberg:1993vc}.} is complemented with 3 $SU(2)_L$ triplets $\Sigma_+$, $\Sigma_0$ and $\Sigma_-$ with $Y=(+1,0,1)$, respectively. The $SU(2)_L$ triplets can be expressed as \begin{equation}\label{sigma}
\Sigma_{-}=\left(\begin{array}{cc}
\frac{\chi^{-}}{\sqrt{2}} & \chi^0 \\
\chi^{--} & -\frac{\chi^{-}}{\sqrt{2}}
\end{array}\right), \quad \Sigma_0=\left(\begin{array}{cc}
\frac{\phi^0}{\sqrt{2}} & \phi^{+} \\
\phi^{-} & -\frac{\phi^0}{\sqrt{2}}
\end{array}\right), \quad \Sigma_+=\left(\begin{array}{cc}
\frac{\psi^{+}}{\sqrt{2}} & \psi^{++} \\
\psi^0 & -\frac{\psi^{+}}{\sqrt{2}}
\end{array}\right)
\end{equation}
These fields can be organized into bi-doublet and bi-triplet representation as \begin{equation}\label{fields}
\bar{H}=\left(\begin{array}{c}
H_1 \\
H_2
\end{array}\right), \quad \bar{\Delta}=\left(\begin{array}{cc}
-\frac{\Sigma_0}{\sqrt{2}} & -\Sigma_{-1} \\
-\Sigma_1 & \frac{\Sigma_0}{\sqrt{2}}
\end{array}\right).
\end{equation}
The $SU(2)_R\times SU(2)_L$ invariant superpotential can be written in terms of $\bar{H}$ and $\bar{\Delta}$ as \begin{equation}
W_0=\lambda \bar{H} \cdot \bar{\Delta} \bar{H}+\frac{\lambda_\Delta}{3} \operatorname{Tr} (\bar{\Delta}\bar{\Delta}\bar{\Delta})+\frac{\mu}{2} \bar{H} \cdot \bar{H}+\frac{\mu_{\Delta}}{2} \operatorname{Tr} (\bar{\Delta}\bar{\Delta})
\end{equation}
Here, the anti-symmetric dot product is \begin{equation}
X \cdot Y=\epsilon^{a b} \epsilon_{i j} X_a^i X_b^j, \quad \epsilon^{12}=-\epsilon_{12}=1.
\end{equation}
The total potential is \begin{equation}
V=V_F+V_D+V_{\text{soft}}.
\end{equation}
The $F$-term potential is 
\begin{equation}
\begin{aligned}
V_F = \left| \frac{\partial W}{\partial \Phi_a} \right|^2 &= \mu^2 \bar{H}^{\dagger} \bar{H} + \mu_{\Delta}^2 \operatorname{Tr} \left( \bar{\Delta}^{\dagger} \bar{\Delta} \right) + 2 \lambda \mu \left( \bar{H}^{\dagger} \bar{\Delta} \bar{H} + \text{ c.c. } \right) \\
&\quad + \lambda^2 \left\{ 4 \operatorname{Tr} \left[ \left( \bar{\Delta} \bar{H} \right)^{\dagger} \bar{\Delta} \bar{H} \right] + \left( \bar{H}^{\dagger} \bar{H} \right)^2 - \frac{1}{4} | \bar{H} \cdot \bar{H} |^2 \right\} \\
&\quad + \lambda_{\Delta}^2 \left[ \operatorname{Tr} \left( \bar{\Delta}^{\dagger} \bar{\Delta}^{\dagger} \bar{\Delta} \bar{\Delta} \right) - \frac{1}{4} \operatorname{Tr} \left( \bar{\Delta}^{\dagger} \bar{\Delta}^{\dagger} \right) \operatorname{Tr} \left( \bar{\Delta} \bar{\Delta} \right) \right] \\
&\quad + \lambda \lambda_{\Delta} \left[ \bar{H} \cdot \bar{\Delta}^{\dagger} \bar{\Delta}^{\dagger} \bar{H} - \frac{1}{4} \bar{H} \cdot \bar{H} \operatorname{Tr} \left( \bar{\Delta}^{\dagger} \bar{\Delta}^{\dagger} \right) + \text{ h.c. } \right] \\
&\quad + \lambda \mu_{\Delta} \left( \bar{H} \cdot \bar{\Delta}^{\dagger} \bar{H} + \text{ c.c. } \right) + \lambda_{\Delta} \mu_{\Delta} \left[ \operatorname{Tr} \left( \bar{\Delta}^{\dagger} \bar{\Delta}^{\dagger} \bar{\Delta} \right) + \text{ h.c. } \right].
\end{aligned}
\end{equation}
The $D$-term potential is 
\begin{equation}
V_D=-\frac{g_1}{2}\left[\bar{H}^{\dagger} Y_H \bar{H}+\operatorname{Tr}\left(\bar{\Delta} Y_{\Delta} \bar{\Delta}\right)\right]-\frac{g_2}{2}\left[\bar{H}^{\dagger} Y_H \bar{H}+\operatorname{Tr}\left(\bar{\Delta} Y_{\Delta} \bar{\Delta}\right)\right].
\end{equation}
The soft supersymmetry breaking terms are 
\begin{equation}
\begin{aligned}
V_{\text {soft }} & =m_H^2\bar{H}^\dagger\bar{H}+m_{\Delta}^2 \operatorname{Tr}(\bar{\Delta}^\dagger\bar{\Delta})+\frac{1}{2} B \bar{H} \cdot \bar{H} \\
& +\left\{\frac{1}{2} B_{\Delta} \operatorname{Tr} (\bar{\Delta}\bar{\Delta})+A \bar{H} \cdot \bar{\Delta} \bar{H}+\frac{1}{3} A_{\Delta} \operatorname{Tr}(\bar{\Delta}\bar{\Delta}\bar{\Delta})+\text { h.c. }\right\}
\end{aligned}
\end{equation}

\section{The Mass Spectrum}

The neutral component of these fields develop v.e.v.s as \begin{equation}
\begin{aligned}
\left\langle H_1^0\right\rangle&=v_1 / \sqrt{2},\quad\left\langle H_2^0\right\rangle=v_2 / \sqrt{2}, \\
\left\langle\chi^0\right\rangle&=v_\chi / \sqrt{2}, \quad\left\langle\psi^0\right\rangle=v_\psi / \sqrt{2}, \quad\left\langle\phi^0\right\rangle=v_\phi / \sqrt{2}.
\end{aligned}
\end{equation}
Again, custodial symmetry requires \begin{equation}
\begin{aligned}
v_1&=v_2\equiv v_H,\\
v_\xi&=v_\psi=v_\phi\equiv v_\Delta,
\end{aligned}
\end{equation}
where \begin{equation}
v^2=2v_H^2+8v_\Delta^2=\frac{1}{\sqrt{2} G_F}=\frac{4 M_W^2}{g^2} \approx(246 \mathrm{GeV})^2, \quad m_W=\frac{g_2 v}{2}, \quad m_Z=\frac{\sqrt{g_1^2+g_2^2} v}{2}.
\end{equation}
Compared with the GM model, we note that the v.e.v.'s are related as \begin{equation}
v_\phi= \sqrt{2}{v_H}, \quad v_\xi=v_\Delta.
\end{equation}
The tadpole conditions are  \begin{equation}
\frac{\partial V}{\partial v_H}=\frac{\partial V}{\partial v_\Delta}=0.
\end{equation}
We can use these tadpole conditions to eliminate $m_H^2$ and $m_\Delta^2$ \begin{equation}
\begin{aligned}
& m_H^2=B+3 v_{\Delta}\left[\lambda\left(\Delta v_{\Delta}-\mu_{\Delta}\right)-A\right]+6 \lambda \mu v_{\Delta}-3 \lambda^2\left(v_H^2+3 v_{\Delta}^2\right)-\mu^2, \\
& m_{\Delta}^2=\frac{v_H^2\left(2 \lambda \mu-6 \lambda^2 v_{\Delta}-A\right)-v_{\Delta} B_{\Delta}+\left(2 \Delta v_{\Delta}-\mu_{\Delta}\right)\left[\lambda v_H^2-\left(\Delta v_{\Delta}-\mu_{\Delta}\right) v_{\Delta}\right]-A_{\Delta} v_{\Delta}^2}{v_{\Delta}}.
\end{aligned}
\end{equation}
We decompose the Higgs fields representations as $\mathbf{2} \otimes \mathbf{2}=\mathbf{1} \oplus \mathbf{3}$ and $\mathbf{3} \otimes \mathbf{3}=\mathbf{1} \oplus \mathbf{3} \oplus \mathbf{5}$, namely \begin{equation}
\bar{H}=h_1 \oplus h_3, \quad \bar{\Delta}=\delta_1 \oplus \delta_3 \oplus \delta_5,
\end{equation}
where \begin{equation}
\begin{aligned}
& h_1^0=\frac{1}{\sqrt{2}}\left(H_1^0+H_2^0\right) \\
& h_3^{+}=H_2^{+}, \quad h_3^0=\frac{1}{\sqrt{2}}\left(H_1^0-H_2^0\right), \quad h_3^{-}=H_1^{-},
\end{aligned}
\end{equation}
and 
\begin{equation}
\begin{aligned}
\delta_1^0 & =\frac{\phi^0+\chi^0+\psi^0}{\sqrt{3}} \\
\delta_3^{+} & =\frac{\psi^{+}-\phi^{+}}{\sqrt{2}}, \delta_3^0=\frac{\chi^0-\psi^0}{\sqrt{2}}, \delta_3^{-}=\frac{\phi^{-}-\chi^{-}}{\sqrt{2}} \\
\delta_5^{++} & =\psi^{++}, \delta_5^{+}=\frac{\phi^{+}+\psi^{+}}{\sqrt{2}}, \delta_5^0=\frac{-2 \phi^0+\psi^0+\chi^0}{\sqrt{6}}, \delta_5^{-}=\frac{\phi^{-}+\chi^{-}}{\sqrt{2}}, \delta_5^{--}=\chi^{--}.
\end{aligned}
\end{equation}
Just as in the discussion of the GM model, these fields can be shifted around their v.e.v's and decomposed as 
\begin{equation}
\begin{aligned}
& h_1^0=v_H+\frac{h_{1}^{0,r}+i h_{1}^{0,i}}{\sqrt{2}}, \quad \delta_1^0=\sqrt{\frac{3}{2}} v_\Delta+\frac{\delta_{1}^{0,r}+i \delta_{1}^{0,i}}{\sqrt{2}}, \\
& h_3^{a}=\frac{h_{3}^{a,r}+i h_{3}^{a,i}}{\sqrt{2}}, \quad \delta_3^a=\frac{\delta_{3}^{a,r}+i \delta_{3}^{i}}{\sqrt{2}}, \quad(a=+, 0,-), \\
& \delta_5^a=\frac{\delta_{5}^{a,r}+i \delta_{5}^i}{\sqrt{2}}, \quad(a=++,+, 0,-,--) .
\end{aligned}
\end{equation}

\subsection{$SU(2)_C$ Triplet}

There are two pseudoscalars in the triplet sector; one is the massless Goldstone boson \begin{equation}
\begin{aligned}
G^0 & =c_T h_{3}^{0,i}+s_T \delta_{3}^{0,i}, \\
G^{\mp} & =c_T\frac{h_3^{ \pm *}-h_3^{\mp}}{\sqrt{2}}+s_T \frac{\delta_3^{ \pm *}-\delta_3^{\mp}}{\sqrt{2}},
\end{aligned}
\end{equation}
and another is a massive triplet \begin{equation}
\begin{aligned}
A^0 & =-s_T h_{3}^{0,i}+c_T \delta_{3}^{0,i}, \\
A^{\mp} & =-s_T \frac{h_3^{ \pm *}-h_3^{\mp}}{\sqrt{2}}+c_T \frac{\delta_3^{ \pm *}-\delta_3^{\mp}}{\sqrt{2}},
\end{aligned}
\end{equation}
where we define \begin{equation}
s_T=\sin \alpha_T=\frac{2 \sqrt{2} v_{\Delta}}{v}, \quad c_T=\cos \alpha_T=\frac{\sqrt{2} v_H}{v} .
\end{equation}
The mass of the pseudoscalar is \begin{equation}
m_A^2=\frac{v_H^2+4 v_{\Delta}^2}{v_{\Delta}}\left(\lambda\left[2 \mu-\mu_{\Delta}-\left(2 \lambda-\lambda_\Delta\right) v_{\Delta}\right]-A\right).
\end{equation}
We expect $v_H>v_\Delta$. The mass can be expanded around $v_\Delta\approx0$. The power series is \begin{equation}
m_A^2=\frac{v_H^2}{v_{\Delta}}\left[\lambda\left(2 \mu-\mu_{\Delta}\right)-A\right]-\lambda\left(2 \lambda-\lambda_\Delta\right) v_H^2+\mathcal{O}\left(v_{\Delta}\right).
\end{equation}
There are also two scalars in the triplet sector: 
\begin{equation}
T_H=\left(\begin{array}{c}
\frac{1}{\sqrt{2}}\left(h_3^{+}+h_3^{-*}\right) \\
h_{3}^{0,r} \\
\frac{1}{\sqrt{2}}\left(h_3^{-}+h_3^{+*}\right)
\end{array}\right), \quad T_{\Delta}=\left(\begin{array}{c}
\frac{1}{\sqrt{2}}\left(\delta_3^{+}+\delta_3^{-*}\right) \\
\delta_{3}^{0,r} \\
\frac{1}{\sqrt{2}}\left(\delta_3^{-}+\delta_3^{+*}\right)
\end{array}\right).
\end{equation}
They are mixed by a squared mass matrix $\mathcal{M}^2$,
\begin{equation}
\left(h_{3 R}^a, \delta_{3 R}^a\right)\left(\begin{array}{cc}
\mathcal{M}_{11}^2 & \mathcal{M}_{12}^2 \\
\mathcal{M}_{21}^2 & \mathcal{M}_{22}^2
\end{array}\right)\left(\begin{array}{c}
h_{3 R}^a \\
\delta_{3 R}^a
\end{array}\right)=\left(T_H^a, T_\Delta^a\right)\left(\begin{array}{cc}
M_3^2 & 0 \\
0 & M_{3^{\prime}}^2
\end{array}\right)\left(\begin{array}{c}
T_H^a \\
T_\Delta^a
\end{array}\right),
\end{equation}
where $a$ represents $\pm$ and $0$ charge. The matrix elements are \begin{equation}
\begin{aligned}
\mathcal{M}_{11}^2 & =G v_H^2+2 \lambda\left(-4 \lambda v_{\Delta}^2+\lambda v_H^2+4 \mu v_\Delta\right)+2 B-2 v_\Delta\left[A+\lambda\left(\mu_{\Delta}-\lambda_\Delta v_{\Delta}\right)\right], \\
\mathcal{M}_{22}^2 & =4 G^2 v_\Delta-\left[2 B_{\Delta}-3 \lambda \lambda_\Delta v_H^2+2 v_\Delta\left(\lambda_\Delta^3 v_{\Delta}-A_\Delta\right)\right] \\
& -\frac{1}{v_\Delta}\left[2 \lambda v_H^2\left(\lambda v_\Delta-\mu\right)-\left(\lambda v_H^2-2 \lambda_\Delta v_\Delta^2\right) \mu_{\Delta}-v_H^2 A\right] \\
\mathcal{M}_{12}^2 & =\mathcal{M}_{21}^2=2 v_H\left[-A+v_\Delta\left(G^2-4 \lambda^2-\lambda \lambda_\Delta\right)+\lambda \mu_{\Delta}\right] .
\end{aligned}
\end{equation}
where $G=g_2^2$ for charged fields and $G=g_1^2+g_2^2$ for neutral fields. The rotation angle is given by \begin{equation}
\sin 2 \alpha_T=\frac{2 \mathcal{M}_{12}^2}{\sqrt{\operatorname{Tr}^2 \mathcal{M}^2-4 \operatorname{det} \mathcal{M}^2}}, \quad \cos 2 \alpha_T=\frac{\mathcal{M}_{22}^2-\mathcal{M}_{11}^2}{\sqrt{\operatorname{Tr}^2 \mathcal{M}^2-4 \operatorname{det} \mathcal{M}^2}}
\end{equation}
The $v_\Delta$ series expansions of the mass eigenstates and rotation angles are \begin{equation}
\begin{aligned}
m_{T_H}^2 & =G^2 v_H^2+2 B+2 \lambda^2 v_H^2+\mathcal{O}\left(v_{\Delta}\right), \\
m_{T_\Delta}^2 & =\frac{v_H^2}{v_{\Delta}}\left[\lambda\left(2 \mu-\mu_{\Delta}\right)-A\right]-2 B_{\Delta}-2 \lambda^2 v_H^2+3 \lambda \lambda_\Delta v_H^2+\mathcal{O}\left(v_{\Delta}\right), \\
\sin \alpha_T & =\frac{2\left(\lambda \mu_{\Delta}-A\right)}{\lambda\left(2 \mu-\mu_{\Delta}\right)-A} \frac{v_{\Delta}}{v_H}+\mathcal{O}\left(v_{\Delta}^2\right).
\end{aligned}
\end{equation}

\subsection{$SU(2)_C$ Quintuplet}

The quintuplets are given by \begin{equation}
F_{\text{scalar}}=\left(\begin{array}{c}
\frac{1}{\sqrt{2}}\left(\delta_5^{++}+\delta_5^{--*}\right) \\
\frac{1}{\sqrt{2}}\left(\delta_5^{+}+\delta_5^{-*}\right) \\
\delta_{5 R}^0 \\
\frac{1}{\sqrt{2}}\left(\delta_5^{-}+\delta_5^{+*}\right) \\
\frac{1}{\sqrt{2}}\left(\delta_5^{--}+\delta_5^{++*}\right)
\end{array}\right), \quad F_{\text{pseudoscalar}}=\left(\begin{array}{c}
\frac{1}{\sqrt{2}}\left(\delta_5^{++}-\delta_5^{--*}\right) \\
\frac{1}{\sqrt{2}}\left(\delta_5^{+}-\delta_5^{-*}\right) \\
\delta_{5 R}^0 \\
\frac{1}{\sqrt{2}}\left(\delta_5^{-}-\delta_5^{+*}\right) \\
\frac{1}{\sqrt{2}}\left(\delta_5^{--}-\delta_5^{++*}\right)
\end{array}\right)
\end{equation}
The masses are \begin{equation}
\begin{aligned}
& m_F^2=\frac{v_H^2\left[\lambda\left(2 \mu-\mu_{\Delta}\right)-A\right]}{\sqrt{2} v_{\Delta}}+\frac{3}{2} \lambda v_H^2\left(\lambda_{\Delta}-2 \lambda\right) \sqrt{2} v_{\Delta}\left(3 \lambda_{\Delta} \mu_{\Delta}+A_{\Delta}\right)-v_{\Delta}^2 \lambda_{\Delta}^2 \\
& M_F^2=\frac{v_H^2\left[\lambda\left(2 \mu+\mu_{\Delta}\right)+A\right]}{\sqrt{2} v_0}-\frac{1}{2} v_H^2\left(6 \lambda+\lambda_\Delta\right)+2 \sqrt{2} \lambda_\Delta v_\Delta \mu_{\Delta}-2 B_{\Delta} .
\end{aligned}
\end{equation}
Here, the scalar quintuplet is the quintuplet emerges in the GM model with squared mass $m_F^2$, whereas the pseudoscalars are mirror particles with squared mass $M_F^2$. The power series expansion reads
\begin{equation}
\begin{aligned}
& m_{F}^2=\frac{v_H^2}{v_{\Delta}}\left[\lambda\left(2 \mu-\mu_{\Delta}\right)-A\right]-3 \lambda\left(2 \lambda-\lambda_\Delta\right) v_H^2+\mathcal{O}\left(v_{\Delta}\right) \\
& M_{F}^2=\frac{v_H^2}{v_{\Delta}}\left[\lambda\left(2 \mu-\mu_{\Delta}\right)-A\right]-2 B_{\Delta}-\lambda\left(6 \lambda-\lambda_\Delta\right) v_H^2+\mathcal{O}\left(v_{\Delta}\right)
\end{aligned}
\end{equation}

\subsection{$SU(2)_C$ Singlelet}

There are two neutral real scalar singlets mixed by $\mathcal{M}_S$
\begin{equation}
\left(h_{1}^{0,r}, \delta_{1}^{0,r}\right) \mathcal{M}_S^2\left(\begin{array}{c}
h_{1}^{0,r} \\
\delta_{1}^{0,r}
\end{array}\right)
\end{equation}
The matrix elements are \begin{equation}
\begin{aligned}
& (\mathcal{M}_S)^2_{11}=6 \lambda^2 v_H^2, \\
& (\mathcal{M}_S)^2_{22}=\frac{v_H^2\left[\lambda\left(2 \mu-\mu_{\Delta}\right)-A\right]+v_{\Delta}^2\left[-A_\Delta+\lambda_\Delta\left(4 \lambda_\Delta v_{\Delta}-3 \mu_{\Delta}\right)\right]}{v_{\Delta}}, \\
& (\mathcal{M}_S)^2_{12}=\left(\mathcal{M}_S^2\right)_{21}=\sqrt{6} v_H\left[A+\lambda\left(6 \lambda v_{\Delta}-2 \lambda_\Delta v_{\Delta}-2 \mu+\mu_{\Delta}\right)\right] .
\end{aligned}
\end{equation}
The expansions of the mass eigenstates and mixing angle around $v_\Delta\approx0$ are \begin{equation}
\begin{aligned}
m_{S_1}^2 & =6 \lambda^2 v_H^2+\mathcal{O}\left(v_{\Delta}\right), \\
m_{S_2}^2 & =\frac{\lambda\left(2 \mu-\mu_{\Delta}\right)-A}{v_{\Delta}}+\mathcal{O}\left(v_{\Delta}\right), \\
\sin \alpha_S & =-\sqrt{6} \frac{v_{\Delta}}{v_H}+\mathcal{O}\left(v_{\Delta}^2\right).
\end{aligned}
\end{equation}
There are also two neutral pseudoscalar singlets mixed by $\mathcal{M}_P$
\begin{equation}
\left(h_{1}^{0,i} \delta_{1}^{0,i}\right) \mathcal{M}_P^2\left(\begin{array}{c}
h_{1}^{0,i} \\
\delta_{1}^{0,i}
\end{array}\right).
\end{equation}
The matrix elements are \begin{equation}
\begin{aligned}
& (\mathcal{M}_P)^2_{11}=2\left(B-3 v_{\Delta}\left[A+\lambda\left(-\lambda_\Delta v_{\Delta}+\mu_{\Delta}\right)\right]\right), \\
& (\mathcal{M}_P^2)_{22}=-\frac{v_H^2\left(A-2 \lambda \mu\right)+v_{\Delta}\left(-3 A_\Delta v_{\Delta}+2 B_{\Delta}-4 \lambda \lambda_\Delta v_H^2\right)+\left(\lambda v_H^2-\lambda_\Delta v_{\Delta}^2\right) \mu_{\Delta}}{v_{\Delta}}, \\
& (\mathcal{M}_P)^2_{12}=\left(\mathcal{M}_P^2\right)_{21}=\sqrt{6} v_H\left[\lambda\left(-2 \lambda_\Delta v_{\Delta}+\mu_{\Delta}\right)-A\right].
\end{aligned} 
\end{equation}
The expansions of the mass eigenstates and mixing angle around $v_\Delta\approx0$ are \begin{equation}\begin{aligned}
m_{P_1}^2 & =2 m_3^2+\mathcal{O}\left(v_{\Delta}\right), \\
m_{P_2}^2 & =\frac{v_H^2\left[\lambda\left(2 \mu-\mu_{\Delta}\right)-A\right]}{v_{\Delta}}-2 B_{\Delta}+4 \lambda \lambda_\Delta v_H^2+\mathcal{O}\left(v_{\Delta}\right), \\
\sin \alpha_P & =\frac{2\left(\lambda \mu_{\Delta}-A\right)}{\lambda\left(2 \mu-\mu_{\Delta}\right)-A} \frac{v_{\Delta}}{v_H}+\mathcal{O}\left(v_{\Delta}^2\right).
\end{aligned}
\end{equation}

\section{The Supersymmetric GM Model and the GM Model}

\begin{table}[H]
\centering
\begin{tabular}{|c|c|c|c|c|}
\hline & \multicolumn{2}{|c|}{ GM } & \multicolumn{2}{|c|}{ SCTM } \\
\hline & scalar & pseudoscalar & scalar & pseudoscalar \\
\hline singlet & $h, H$ & & $S_{1,2}$ & $P_{1,2}$ \\
\hline triplet & & $G, H_3$ & $T_{H,\Delta}$ & $G, A$ \\
\hline quintuplet & $H_5$ & & $F_s$ & $F_p$ \\
\hline
\end{tabular}
\caption[GM and SCTM Particle Content in Higgs Sector.]{\centering
 The particle content in the GM model compared to SCTM model.}
\end{table}
In the GM model, we have a complex doublet, a real triplet and a complex triplet in the Higgs sector. In the SCTM model, we need two complex doublets and three complex triplets in the Higgs sector. In the SCTM model, we decompose these fields in terms of $\mathbf{2} \otimes \mathbf{2}=\mathbf{1} \oplus \mathbf{3}$ and $\mathbf{3} \otimes \mathbf{3}=\mathbf{1} \oplus \mathbf{3} \oplus \mathbf{5}$. For the quintuplets, we have one CP-even state $F_S$ matching the quintuplet $H_5$ in the GM model, and the imaginary part $F_P$ is treated as mirror particle of $F_S$. For the triplets, we have two CP-odd states, one being the Goldstone boson and the other matching the CP-odd triplet $H_3$ in the GM model. The CP-even states $T_{H,\Delta}$ are also treated as mirror particles of $G$ and $A$. Similarly, the two scalar singlets in the SCTM match the GM model singlets $h$ and $H$, and the two pseudoscalars being mirror particles. Now, we also note that the squared masses of only these mirror particles contain the supersymmetry soft breaking parameter $B$ and $B_\Delta$. We then take the limit $|B| \sim\left|B_{\Delta}\right| \rightarrow \infty$. In this limit, \begin{equation}
m^2_{S_{1,2}, A, F_s}\ll m^2_{P_{1,2}, T_{H,\Delta}, F_p}\longrightarrow\infty.
\end{equation}
The mirror particles become so massive that they decouple from the original scalar spectrum. The SCTM model behaves exactly like the GM model. We call this limit the decoupling limit and refer to this henceforth as the supersymmetric Georgi-Machacek model\cite{Vega:2017gkk}. 

To mimic the effect of taking the decoupling limit in the SCTM model, we can rewrite the GM model potential in terms of the bi-doublet and bi-triplet fields in \eqref{sigma} and \eqref{fields}, along with the conditions in \eqref{co}
\begin{equation}\label{co}
\bar{\Delta}^{\dagger}=\bar{\Delta}, \quad H_2=-i \sigma_2 H_1^* .
\end{equation}
The Higgs potential of the GM model then takes the form, \begin{equation}
\begin{aligned}
V_{\mathrm{GM}} & =\frac{1}{2} \mu_2^2 \bar{H}^{\dagger} \bar{H}+\frac{1}{2} \mu_3^2 \operatorname{Tr}[\bar{\Delta} \bar{\Delta}]+\lambda_1\left(\bar{H}^{\dagger} \bar{H}\right)^2+\left(\lambda_4-\frac{1}{4} \lambda_5\right)\left(\bar{H}^{\dagger} \bar{H}\right) \operatorname{Tr}[\bar{\Delta} \bar{\Delta}] \\
& -2 \lambda_3 \operatorname{Tr}\left[(\bar{\Delta} \bar{\Delta})^2\right]+\left(\lambda_2+\frac{3}{2} \lambda_3\right) \operatorname{Tr}[\bar{\Delta} \bar{\Delta}]^2+\lambda_5 \bar{H}^{\dagger} \bar{\Delta} \bar{\Delta} \bar{H} \\
& -\frac{M_1}{2} \bar{H}^{\dagger} \bar{\Delta} \bar{H}-2 M_2 \operatorname{Tr}[\bar{\Delta} \bar{\Delta} \bar{\Delta}] .
\end{aligned}
\end{equation}
Comparing the coefficients of each term in $V_{H,\text{GM}}$ and $V_{H,\text{SCTM}}$, we have the following mappings between the two models:
\begin{equation}\label{llr}
\begin{aligned}
\lambda_1 & =\frac{3}{4} \lambda^2, \\
\lambda_2 &=\frac{1}{2} \lambda_{\Delta}^2, \\
\lambda_3 &=-\frac{1}{2} \lambda_{\Delta}^2, \\
\lambda_4 &=\lambda^2, \\
\lambda_5 &=-2 \lambda\left(\lambda_{\Delta}-2 \lambda\right) \\
M_1 & =-4\left[\lambda\left(2 \mu-\mu_{\Delta}\right)-A_\lambda\right],\\ M_2 & =-\frac{1}{3}\left(3 \lambda_{\Delta} \mu_{\Delta}+A_{\Delta}\right) \\
\mu_2^2 & =2\left(\mu^2+m_H^2\right)+B, \\
\mu_3^2 &=2\left(\mu_{\Delta}^2+m_{\Delta}^2\right)+B_{\Delta} .
\end{aligned}
\end{equation}
From \eqref{llr}, we can read off the constraints on $\lambda$ coefficients imposed by supersymmetry
\begin{equation}
\lambda_1=\frac{3}{4}\lambda_4,\quad \lambda_2=-\lambda_3,\quad\lambda_5=4\lambda_4-2\sqrt{2\lambda_2\lambda_4}.
\end{equation}
In the SGM model, supersymmetry imposes new constraints between the quartic couplings. We now only have 4 free parameters in the scalar potential. And we pick them to be $\lambda_{1,2}$ and $M_{1,2}$. By far, we have reviewed the GM and SGM model, and identified the free parameters in both models. These parameters are subject both to the theoretical constraints and experimental constraints. We use to open-source packages, \texttt{GMCalc} and \texttt{HiggsTools} to test these parameters against theoretical constraints and most recent available experimental data. The scan setup is discussed in next chapter.

 \begin{singlespace}
\chapter{Scan Setup}\label{chapter:package}
\end{singlespace}

In the supersymmetric version of the GM model, supersymmetry imposes new constraints between $\lambda_i$'s. These constraints reduce the number of free parameters in the supersymmetric Georgi-Machacek model to 4. Before confronting the model with experiment, we first must ensure that the free parameters satisfy theoretical constraints. These constraints arise from the requirement that the model satisfy unitarity and that the vacuum state, the state of lowest energy, is stable. In the global fits of the SGM model, we use $\texttt{GMCalc}$\cite{Hartling:2014xma} to test sets of input parameters for theoretical constraints. If a set of parameters passes the theoretical constraints, we use $\texttt{GMCalc}$ to calculate other necessary model parameters and physical observables, including scalar masses, branching ratios and cross sections, etc. We then feed these data to $\texttt{HiggsTools}$\cite{Bahl:2022igd} and test this set of parameter against LHC experimental data. In this way, we produce allowed regions at the $95\%$ confidence level for physical observables, such as exotic Higgs masses, vacuum expectation values, and other model parameters for the SGM model. For comparison purposes, we also present global fits results for the GM model. 

\section{Theoretical Constraints}

\subsection{GMCalc}

$\texttt{GMCalc}$ is a \texttt{Fortran} program specific to the GM model that determines the particle spectrum and couplings for a specific set of values for parameters of the model. Given a set of parameters, $\texttt{GMcalc}$ tests their values against theoretical constraints from unitarity and vacuum stability requirements. If the set of data passes the theoretical constraints, the program continues to compute the decay widths, branching ratios, and production cross sections for a variety of processes. We modify \texttt{GMCalc} in order to incorporate analysis of the SGM model.

\subsection{Unitarity Bounds}

There are two theoretical constraints considered in the \texttt{GMcalc} program: pertubative unitarity and vacuum stablility. Perturbative unitarity is imposed by requiring the S-matrix of $2\rightarrow2$ scalar scattering processes to be unitary. This imposes constraints on the quartic coupings in the GM model. The dimensional couplings, $\mu_{2,3}^2$ and $M_{1,2}$, are not constrained by perturbative unitarity, since only tree-level diagrams involving four-point scalar couplings contribute to the the $2\rightarrow2$ scalar scattering process. Additionally, scaterring processes involving transversely polarized gauge bosons or fermions are ignored. Under these conditions, only $a_0$, the zeroth partial wave amplitude, contributes to $\mathcal{M}$. Unitarity bounds requires \begin{equation}
\left|\operatorname{Re} a_0\right| \leq \frac{1}{2}\quad\Longleftrightarrow\quad |\mathcal{M}|<8\pi.
\end{equation}
The partial wave amplitude $a_0$ is related to the matrix element of $\mathcal{M}$ of the process by \begin{equation}
\mathcal{M}=16 \pi \sum_J(2 J+1) a_J P_J(\cos \theta),
\end{equation}
where $J$ is the orbital angular momentum and $P_J$ are the Legendre polynomials. We take the constraints from the unitarity bounds directly from Ref \cite{Hartling:2014zca}. 
\begin{equation}\label{lamp}
\begin{aligned}
\lambda_1 &\in\left(-\frac{1}{3} \pi, \frac{1}{3} \pi\right) \approx(-1.05,1.05),\\
\lambda_2 &\in\left(-\frac{2}{3} \pi, \frac{2}{3} \pi\right) \approx(-2.09,2.09),\\
\lambda_3 &\in\left(-\frac{4}{5} \pi, \frac{4}{5} \pi\right) \approx(-2.51,2.51),\\
\lambda_4 &\in\left(-\frac{16}{25} \pi, \frac{16}{25} \pi\right) \approx(-2.01,2.01),\\
\lambda_5 &\in\left(-\frac{8}{3} \pi, \frac{8}{3} \pi\right) \approx(-8.38,8.38).
\end{aligned}
\end{equation}
Combining with additional constraints we derived in Eq.(5.39), in the SGM model, we have \begin{equation}
\Lambda \in\left[0, \frac{4}{3} \sqrt{\pi}\right], \quad \Lambda_\Delta \in\left[0, \frac{2}{\sqrt{\pi}}\right].
\end{equation}
Here we use capital Greek letters to distinguish $\lambda$ couplings from the SGM model to the GM model.

\subsection{Vacuum Stability}

The GM model has a rich number of scalar fields in its Higgs sector. To determine the v.e.v's in the GM model, we need to find the minimum of the scalar potential. The existence of such a global minimum, is called the "bounded-from-below" requirement on the scalar potential. This is another theoretical bound implemented in \texttt{GMCalc}. To check whether a set of input parameters has well-defined global minimum, the scalar potential is parametrized as, \begin{equation}
\begin{aligned}
r & \equiv \sqrt{\operatorname{Tr}\left(\Phi^{\dagger} \Phi\right)+\operatorname{Tr}\left(X^{\dagger} X\right)}, \\
r^2 \cos ^2 \gamma & \equiv \operatorname{Tr}\left(\Phi^{\dagger} \Phi\right) \\
r^2 \sin ^2 \gamma & \equiv \operatorname{Tr}\left(X^{\dagger} X\right) \\
\zeta & \equiv \frac{\operatorname{Tr}\left(X^{\dagger} X X^{\dagger} X\right)}{\left[\operatorname{Tr}\left(X^{\dagger} X\right)\right]^2}, \\
\omega & \equiv \frac{\operatorname{Tr}\left(\Phi^{\dagger} \tau^a \Phi \tau^b\right) \operatorname{Tr}\left(X^{\dagger} t^a X t^b\right)}{\operatorname{Tr}\left(\Phi^{\dagger} \Phi\right) \operatorname{Tr}\left(X^{\dagger} X\right)} .
\end{aligned}
\end{equation}
The parameter ranges are given in Ref. \cite{Hartling:2014zca}
\begin{equation}
r \in[0, \infty), \quad \gamma \in\left[0, \frac{\pi}{2}\right], \quad \zeta \in\left[\frac{1}{3}, 1\right] \quad \text { and } \quad \omega \in\left[-\frac{1}{4}, \frac{1}{2}\right].
\end{equation}
The constraints on the quartic coefficients $\lambda_{1,2,3,4,5}$ are translated as
\begin{equation}
\begin{aligned}
& \lambda_1>0 \\
& \lambda_4> \begin{cases}-\frac{1}{3} \lambda_3 & \text { for } \lambda_3 \geq 0, \\
-\lambda_3 & \text { for } \lambda_3<0,\end{cases} \\
& \lambda_2> \begin{cases}\frac{1}{2} \lambda_5-2 \sqrt{\lambda_1\left(\frac{1}{3} \lambda_3+\lambda_4\right)} & \text { for } \lambda_5 \geq 0 \text { and } \lambda_3 \geq 0, \\
\omega_{+}(\zeta) \lambda_5-2 \sqrt{\lambda_1\left(\zeta \lambda_3+\lambda_4\right)} & \text { for } \lambda_5 \geq 0 \text { and } \lambda_3<0, \\
\omega_{-}(\zeta) \lambda_5-2 \sqrt{\lambda_1\left(\zeta \lambda_3+\lambda_4\right)} & \text { for } \lambda_5<0,\end{cases} \\
&
\end{aligned}
\end{equation}
where \begin{equation}
\omega_{ \pm}(\zeta)=\frac{1}{6}(1-B) \pm \frac{\sqrt{2}}{3}\left[(1-B)\left(\frac{1}{2}+B\right)\right]^{1 / 2},
\end{equation}
with 
\begin{equation}
B \equiv \sqrt{\frac{3}{2}\left(\zeta-\frac{1}{3}\right)} \in[0,1].
\end{equation}
The bounded-from-below requirements modify the numerical constraints in \eqref{lamp} to
\begin{equation}\label{lambda}
\begin{aligned}
& \lambda_1 \in\left(0, \frac{1}{3} \pi\right) \approx(0,1.05),\\
& \lambda_2 \in\left(-\frac{2}{3} \pi, \frac{2}{3} \pi\right) \approx(-2.09,2.09),\\
& \lambda_3 \in\left(-\frac{1}{2} \pi, \frac{3}{5} \pi\right) \approx(-1.57,1.88), \\
& \lambda_4 \in\left(-\frac{1}{5} \pi, \frac{1}{2} \pi\right) \approx(-0.628,1.57),\\
& \lambda_5 \in\left(-\frac{8}{3} \pi, \frac{8}{3} \pi\right) \approx(-8.38,8.38).
\end{aligned}
\end{equation}
The existence of a minimum hence imposes a few new numerical constraints on the quartic coefficients in the Lagrangian. To verify whether it is a true global minimum, \texttt{GMCalc}, checks the depth of alternative minima numerically by scanning over a parametrization parameter $\theta$ and minimizing the scalar potential at each $\theta$, where $\theta$ is defined as \begin{equation}
\operatorname{Re} \chi^0=\frac{1}{\sqrt{2}} \sin \theta, \quad \xi^0=\cos \theta.
\end{equation}

\subsection{Scan Range}

In our fits, we require the lighter singlet mass to be $125.09\pm0.24$GeV and $v=\sqrt{v_\phi^2+8v_\xi^2}=246$GeV. After imposing the supersymmetric constraints, only 4 parameters in the scalar potential of the SGM model remain free. The scan ranges consistent with \eqref{lambda} are then, \begin{equation}
\begin{aligned}
0\leq & \lambda_1\leq 1.05,\\
0\leq & \lambda_2\leq 2.09,\\
-1500\text{GeV}\leq & M_1 \leq 1500\text{GeV},\\
-1500\text{GeV}\leq & M_2 \leq 1500\text{GeV}.
\end{aligned}
\end{equation}
Here, unitarity requires $\lambda_{2}>0$.

For comparison, we also conduct a global fit on the GM model. In the GM model, there are 3 additional free parameters; we choose these to be $\lambda_3$, $\lambda_4$, and $\lambda_5$. The scan range for $\lambda_{1,2,3,4,5}$ is fixed by \eqref{lambda}. In both the GM model and SGM model, there are no theoretical limits on $M_{1,2}$. We choose to scan over the range of $-1500\text{GeV}\leq M_{1,2} \leq 1500\text{GeV}$, which is of the same order of the heavy scalars masses of interest.

\section{Experimental Constraints}

\subsection{HiggsTools}

\texttt{HiggsTools} is an open-source multi-purpose toolbox for comparing a wide class of BSM models to all available experimental results from searches for new scalar particles and measurements of the 125GeV SM-like Higgs boson at colliders. We use this package to check whether the input set of GM or SGM parameters is excluded by current experimental results. It is composed of three sublibraries, \texttt{HiggsPredictions}, \texttt{HiggsBounds} and \texttt{HiggsSignals}. After the particle spectrum, tree-level couplings, branching ratios and decay width of scalars are calculated by \texttt{GMCalc}, we input these results through \texttt{HiggsPredictions}. \texttt{HiggsPredictions} calculates all the necessary model predictions, such as cross sections of heavy scalars that have topologies analogous to those in the SM-like Higgs boson. The results are then passed to \texttt{HiggsBounds} and \texttt{HiggsSignals} where a global fit to current experimental data is performed. For those topologies that have no analog in the SM, the production and decay numbers must be added separately. 

\texttt{HiggsBounds} takes the model predictions from \texttt{HiggsPredictions} and tests them against its experimental database. Its workflow goes as follows\cite{Bechtle:2008jh}: let $X_i=X(H_i)$ be the application of the limits from a particular Higgs search experiment to each neutral scalar of the model under study, or the ``analysis application\footnote{In our model, there are four neutral Higgs scalars, $h$, $H$, $H_3$ and $H_5$. Suppose there are two neutral Higgs searches, $A_{1}$ and $A_{2}$, then there are eight possible analysis applications $X\in\{A_1(h), A_1(H), A_1(H_3), A_1{H_5}, A_2(h), A_2(H), A_2(H_3), A_2(H_5)\}$ for our model.}." Here, $i$ denotes a specific decay channel. Depending on whether the exclusion limit for an experiment is relative or absolute, the program then evaluates \begin{equation}
Q_{\text{model}}(X_i)=\frac{[\sigma_i \times \mathrm{BR}_i]_{\text{model}}}{[\sigma \times \mathrm{BR}]_{\text{ref}}} \quad \text{or} \quad [\sigma_i \times \mathrm{BR}_i]_{\text {model }}.
\end{equation}
When a decay channel analogous to the SM exists, we take $[\sigma\times\operatorname{BR}]_{\text{ref}}=[\sigma\times\operatorname{BR}]_{\text{SM}}$. When such a decay channel does not exists, we simply use $\left[\sigma_i \times \mathrm{BR}_i\right]_{\text {model }}$. $Q_{\text{exp}}(X_i)$ and $Q_{\text{obs}}(X_i)$ are the corresponding expected and observed limits associated with a specific decay channel, see Appendix~\ref{appendix:A} for details. \texttt{HiggsBounds} then selects the analysis application with the highest statistical sensitivity for the model point under consideration based on the ratio between the model predictions and the expected limit. 
\begin{equation}
X_{i}^*=\max\left\{X_i:\ R_{i,\text{exp}}(X_i)=\frac{Q_{\text{model}}(X_i)}{Q_{\text{exp}}(X_i)}\right\},
\end{equation}
where $R_{i,\text{exp}}$ is called the expected ratio. The analysis application that maximizes the expected ratio is selected as the exclusion criteria. \texttt{Higgsbounds} then performs an exclusion test by comparing the ratio of $Q_{\text{model}}$ to the observed limit of the most sensitive channel,
\begin{equation}
R_{\text{obs}}=\frac{Q_{\text {model}}\left(X_i^*\right)}{Q_{\text{obs}}\left(X_i^*\right)}.
\end{equation}
$R_{\text{obs}}$ is called the observed ratio. If $R_{\text{obs}}>1$, \texttt{HiggsBounds} concludes the input data point is excluded at $95\%$ confidence level under the modified frequentist approach. A $\chi^2$ value of the SM-like Higgs boson masses, compared to experimental data at some fixed value of Higgs masses \cite{Stal:2013hwa}, is then produced to quantify the agreement of model predictions with measurements by \texttt{HiggsSignals}. The statistical details for the modified frequentist method are presented in Appendix~\ref{appendix:A}. Due to the fact that a majority of experimental limits built in \texttt{HiggsTools} applies to a mass range below 1 TeV, we choose this mass as the cutoff mass of our global fits.

\subsection{Higgs Searches}

A large variety of direct heavy scalar searches performed in LHC are implemented in \text{HiggsTools}. We summarize the ATLAS and CMS experiments relevant to the GM and SGM scalars with a variety of final states in Appendix~\ref{appendix:exp}. The production channels of the triplet states and quintuplet states are listed below. 

\begin{itemize}
\item\textbf{Drell-Yan Process:} 
The triplet and quintuplet states can be produced via the well-known Drell-Yan process via gauge bosons. In this process, a quark and an anti-quark annihilate, creating a virtual gauge boson, which then decays into a pair of oppositely charged scalars.
\item\textbf{Gluon Fusion:} 
At tree level, gluon pairs can interact and produce a heavy scalar in association with a $b\bar{b}$ or $t\bar{t}$ pair. Gluons can also produce a single neutral heavy scalar via a one loop triangle diagram. We denote these processes as $bbH$, $ttH$ and $ggH$ production, respectively.
\item\textbf{Top Quark Decay:} 
The SM predicts a nearly $100\%$ branching ratio of top quark decaying into a $W$ boson and a $b$ quark. The branching ratio of a top quark decaying into a scalar and a quark can be enhanced to $\mathcal{O}(10^{-3})$ in models incorporating two Higgs doublets\cite{Aguilar-Saavedra:2004mfd}.
\item\textbf{Vector Boson Fusion:} 
Pairs of vector bosons radiated by quarks in incoming protons can interact and create a heavy scalar. We denote this by $\mathrm{vbf}H$. 
\item\textbf{Higgs Bremsstrahlung:} 
A single heavy scalar can also be created in association with a gauge boson, by the process $qq'\rightarrow HV$. 
\end{itemize}

\begin{figure}[hbt!]
 \centering
 \includegraphics[scale=0.35]{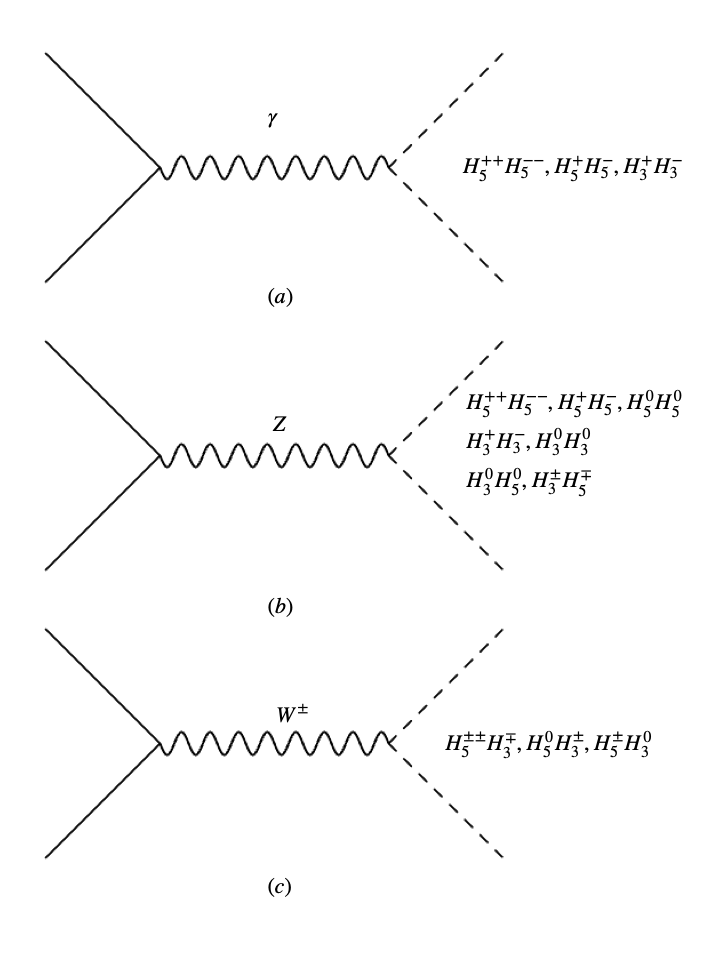} \caption[ Feynman Diagrams for Drell-Yan-like production processes for the GM and SGM scalars.]
  { \centering Feynman Diagrams for Drell-Yan-like production processes for the GM and SGM scalars via gauge bosons. In (a), we have a quark-anti-quark pair annihilates, creating a virtual photon, which decays into $H_5^{++} H_5^{--}$, $H_5^{+} H_5^{-}$, or $H_3^{+} H_3^{-}$. In (b), a quark-anti-quark pair annihilates, creating a virtual $Z$, which can into $H_5^{++} H_5^{--}$, $H_5^{+} H_5^{-}$, $H_5^0 H_5^0$, $H_3^{+} H_3^{-}$, $H_3^0 H_3^0$, $H_3^0 H_5^0$ or $H_3^{ \pm} H_5^{\mp}$. In (c), a quark-anti-quark pair annihilates, creating a virtual $W^\pm$, which can into $H_5^{ \pm \pm} H_3^{\mp}$, $H_5^0 H_3^{ \pm}$ or $H_5^{ \pm} H_3^0$.} 
 \label{fig:DY}
\end{figure} 

\begin{figure}[hbt!]
 \centering
 \includegraphics[scale=0.35]{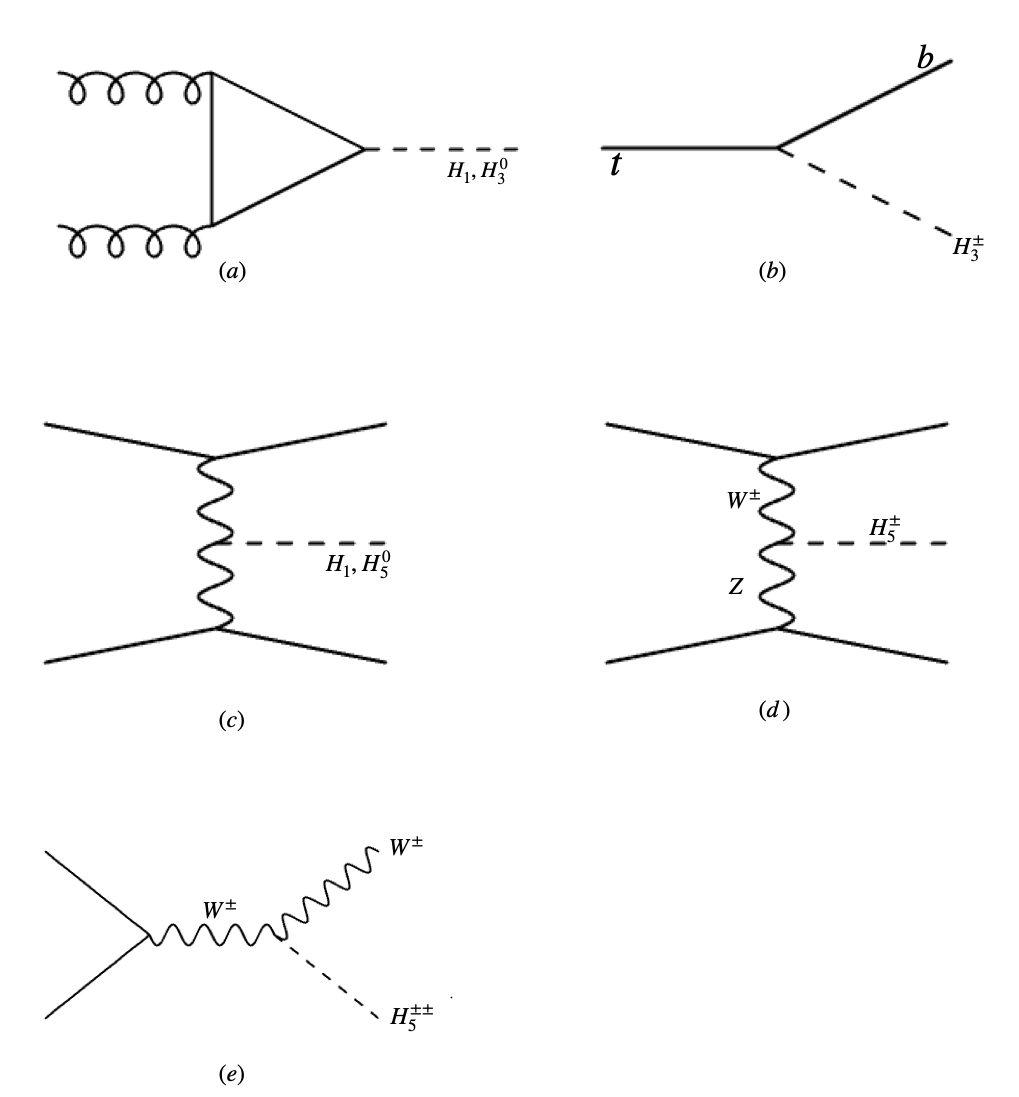} \caption[ Feynman Diagrams for other Higgs production processes for the GM and SGM scalars.]
  { \centering Feynman Diagrams for other production process for the GM and SGM scalars. In (a), we have gluon fusion. $H_1$ and $H_3^0$ can be produced in this way. In (b), when $m_{H_3}<m_t$, $H_3^\pm$ can be produced via top quark decay. In (c) and (d), we have the vector boson fusion process. In (e), we have the Higgs Bremsstrahlung for the production of the doubly charged $H_5^{\pm\pm}$.} 
 \label{fig:hp}
\end{figure} 

The scalar spectrum of the GM and SGM model exhibits a rich spectrum of scalar particles. The heavy singlet $H$ can be produced just as the SM Higgs. We consider the production channel $ggH$, $\mathrm{vbf}H$ and Higgs bremstrahlung for the heavy singlet. At the mass of $125$ GeV, the production cross section $\sigma_{H}$ is around 50 pb. $\sigma_H$ goes down to around 6 pb when $m_H=200$ GeV and only around $500$ fb when $m_H=500$ GeV. We cut off the scan when $m_H=1$ TeV. And the production at the cutoff mass is around 10 fb. The triplets and quintuplets, $H_3$ and $H_5$, can be produced via the Drell-Yan process, $pp\rightarrow H_i H_j$ for $i,j=3,5$. The Feynman diagrams for the Drell-Yan-like processes and final products are shown in Figure~\ref{fig:DY}. The production cross section is determined by the gauge couplings and the masses of these exotic scalars. Figure~\ref{fig:h5pp} shows the production cross section of the Drell-Yan like process for $H_5^{\pm\pm}$ pair production. For the Drell-Yan pair production, the residual renormalization and factorization scale dependence at NLO amounts
to about $5–10\%$ and serves as an estimate of the theoretical uncertainties. The total theoretical uncertainties, including the errors of the parton densities, can be estimated
to be $10–15\%$\cite{Muhlleitner:2003me}. $H_3$ and $H_5$ can also be produced together through this process, but are suppressed by large $v_\Delta$. 

\begin{figure}[hbt!]
 \centering
 \includegraphics[scale=0.4]{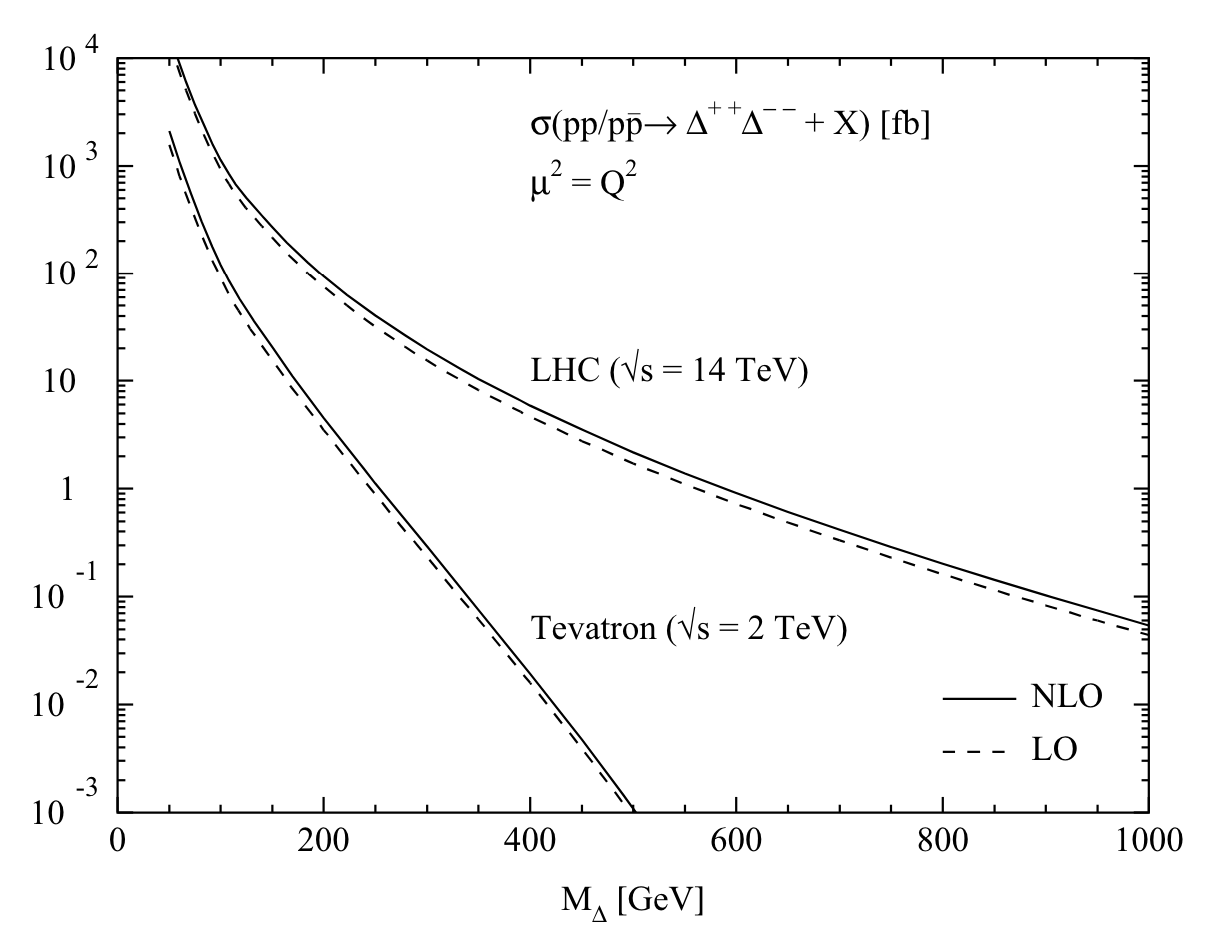} \caption[ The production cross section for the doubly charged quintuplet state pair production at the Tevatron and LHC.]
  { \centering The production cross section for the doubly charged quintuplet state pair production at the Tevatron and LHC\cite{Muhlleitner:2003me}.} 
 \label{fig:h5pp}
\end{figure} 

There is an interesting dichotomy in the model that influences the search strategies. The triplets scalars, $H_3$, couple to fermions, but not to gauge boson pairs. In contrast, the quintuplet scalars, $H_5$, only couple to gauge bosons. Since the $H_5$ does not couple to fermions it cannot be produced via top decay or gluon fusion.  When the mass of the singly charged $H_3^\pm$ is lighter than the top quark mass, $H_3^\pm$ can be produced via, $t\rightarrow b H_3^\pm$, with a production cross section 3.6 pb, at a mass of $200$ GeV\cite{Krab:2022lih}. The neutral $H_3^0$ can be produced via gluon fusion. At $m_{H_3}=200$ GeV, The production cross section for $ggH$ is around 4 pb. This goes down to 200 fb when $m_{H_3}$ is at 500 GeV. The quintuplet states can be produced via the vector boson fusion process $qq'\rightarrow H_5$ and Higgs Bremsstrahlung $qq'\rightarrow H_5 V$. The production channel involves vector bosons only applies to the $H_5$ quintuplet as the triplet $H_3$ does not couple to vector bosons. The production cross section for vector boson associated processes are proportional to $v_\Delta^2$ and can be significant in the large $v_\Delta$ case\cite{Chiang:2012cn}. The neutral $H_5^0$ is mostly produced by vector boson fusion and vector boson associated processes. Therefore, the production cross section are significantly lower. At $m_{H_5}=200$ GeV, the production cross section is only around 200 fb. $\sigma_{H_5}$ is around 50 fb when $m_{H_5}=500$ GeV, and around 5 fb at 1 TeV. And the singly charged $H_5^{\pm}$ have production cross sections that are slightly higher, but of the same order as that of $H_5^0$.

We have discussed how the theoretical constraints affect the quartic couplings in the scalar potential of the GM and SGM model. We also show how \texttt{HiggsTools} excludes a set of parameter against experimental data using the modified frequentist method. The production modes are discussed and the production cross sections for exotic scalars in the GM and SGM model are estimated in the mass range to our interest. Results of the global fits are presented in next chapter.

 \begin{singlespace}
\chapter{Scan Results}\label{chapter:Results}
\end{singlespace}

In this section, we present the impact of all the theoretical and experimental constraints aforementioned on the SGM model. In the scan of the SGM model, we randomly generate combinations of $\lambda_{1,2}$ and $M_{1,2}$ while fixing the SM-like Higgs mass $m_h=125.09$ GeV and $v^2=246^2$ $($GeV$)^2$. We also perform a complementary scan on the GM model by scanning over $\lambda_{1,2,3,4,5}$ and $M_{1,2}$. For the SGM model, we generate 2,500,000 parameter combinations and test them against both the theoretical and experimental constraints. The shown points in each plots pass the theoretical constraints and the experimental constraints, and are the same for all plots. We also construct the allowed regions, at the $95\%$ confidence level, for the set of the free parameters in model, based on these points. In all figures, allowed regions for the GM model are presented as yellow background, and allowed regions for the SGM model are mapped to color gradients based on the observed ratios $0<R_{\text{obs}}<1$. The higher the observed ratios, the closer a point is to the exclusion limit. 

In the global fits of the SGM model, we also incorporate the decays of heavy scalar particles into light Higgsino pairs. In the SGM model, the lightest stable Higgsino mass is predicted to be 45-50 GeV\cite{Vega:2017gkk}. This decay channel contributes to an invisible decay width less than 2 GeV in the mass range $[150,1000]$ GeV for heavy scalars. It lowers the branching ratios of other decay channels and relaxes the experimental constraints a bit. We still find that the experimental limits constrain the theoretically allowed regions significantly. Although theoretically, v.e.v.'s of up to 84 GeV are allowed, current experimental data places an upper limit of just 31 GeV on the GM model and a little lower, to 28 GeV, in the SGM model. In both the GM and SGM model, the masses can be as low as $150$ GeV. However, the masses of exotic scalars tend to be degenerate in the SGM model. Therefore, some of the decay channels allowed in the GM model, such as $H'\rightarrow HH$ and $H'\rightarrow HV$, where $H$ denotes a heavy scalar and $V$ a vector boson, are not allowed kinematically. The bounds of the $\lambda_i$ parameters from the Lagrangian are also greatly tightened in the SGM model as compared to the GM model. We find $\lambda_1$ is constrained to a maximum value of $0.31$ and $\lambda_2$ to $0.39$ in the SGM model. In the GM model, the allowed range for $\lambda_1$ is the same as in the SGM model, but the allowed range for $\lambda_2$ is $[-0.49,1.93]$. 

\begin{figure}[hbt!]
 \centering
 \includegraphics[scale=0.8]{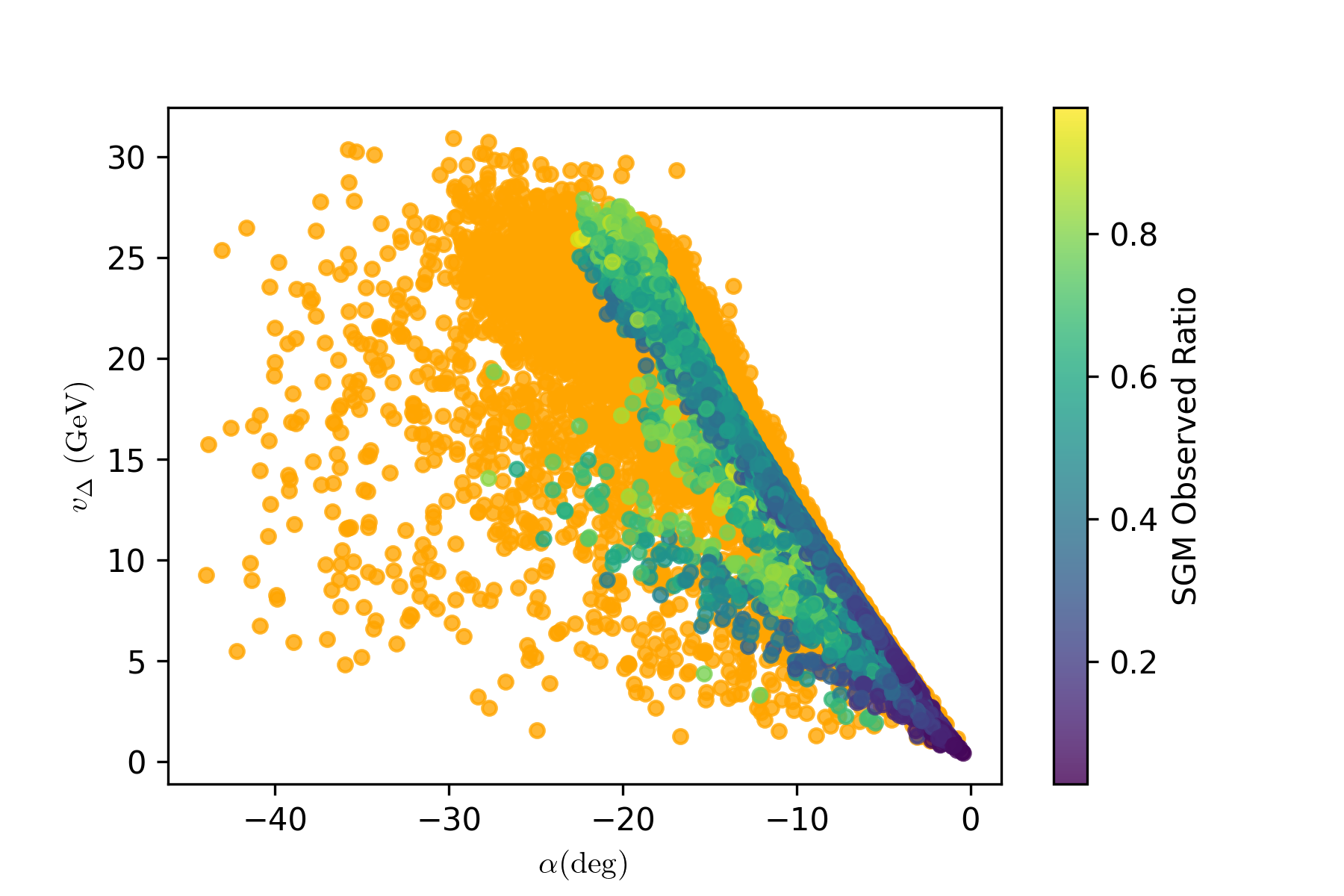} \caption[Allowed region at the $95\%$ confidence level in the $\alpha$-$v_\Delta$ plane for the GM model and the SGM model.]
{ \centering The allowed region at the $95\%$ confidence level in the $\alpha$-$v_\Delta$ plane. The color gradient ranks the observed ratios for SGM scalars, and the yellow background illustrates the allowed region for the GM model.}
\label{fig:alphavd}
\end{figure} 

In Figure~\ref{fig:alphavd}, we show the allowed region at the $95\%$ confidence level in the $\alpha$ versus $v_\Delta$ plane. Here, $\alpha\approx0$ corresponds to the decoupling limit of the GM model, in which the heavy scalar states are so massive that they decouple from the particle spectrum and the GM model behaves exactly like the SM. The allowed region on the GM model as well as in the SGM model constrain $\alpha$ to negative values. We can see from the figure that $\alpha$ is constrained to $\alpha>-44^\circ$ in the GM model, while $\alpha>-28^\circ$ in the SGM model. We also note that $v_\Delta$ cannot exceed $31$ GeV in the GM model. In the SGM model, the limit on $v_\Delta$ is slightly lower. The highest allowed value is $28$ GeV. Generally, as $v_\Delta$ approaches $28$ GeV, points in the allowed region are very close to exclusion limit. We also note that none of our points passes the decoupling limit in this global fit, which corresponds to $\alpha\approx0$, mainly because the mass range we scanned does not exceed $1$ TeV. 

\begin{figure}[hbt!]
 \centering
 \includegraphics[scale=0.4]{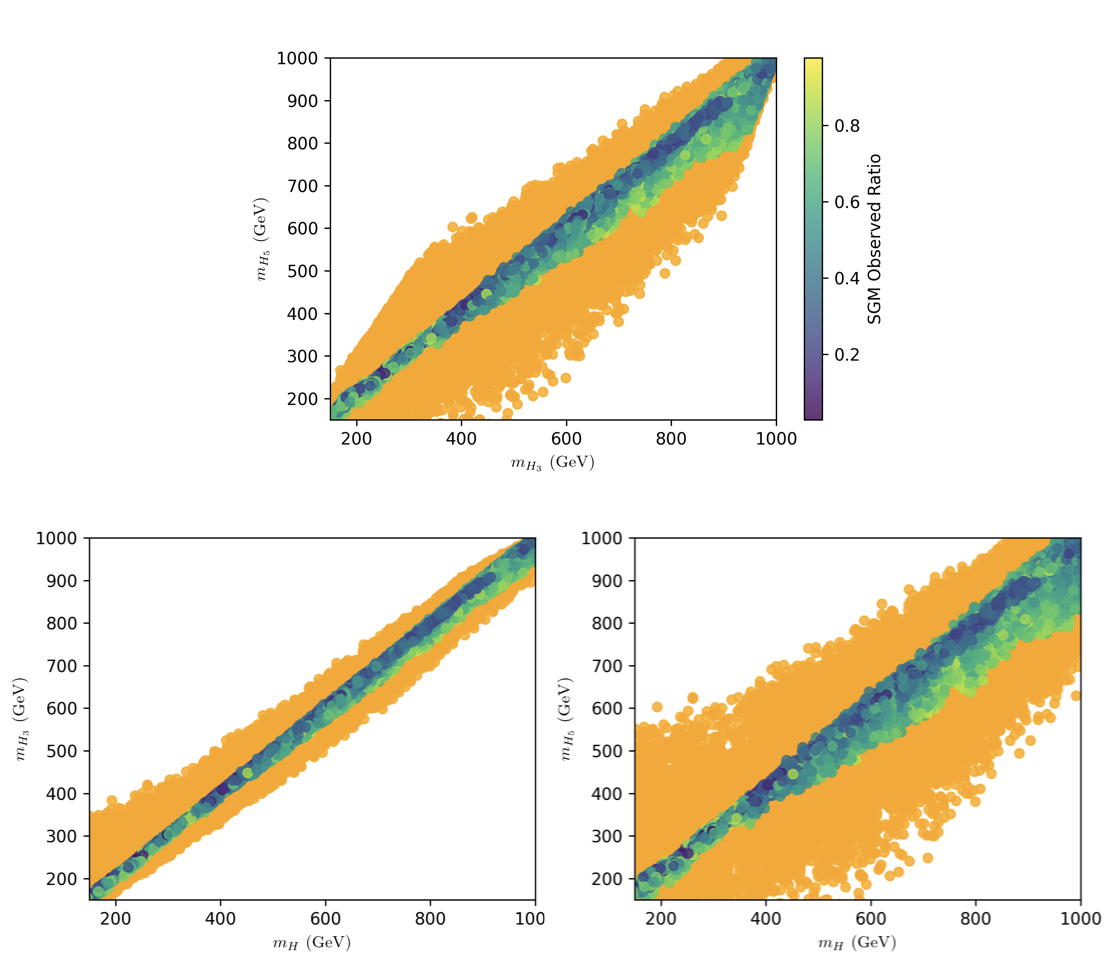} \caption[ Allowed regions at the $95\%$ confidence level for scalar masses. ]
 { \centering In the upper graph, we plot the allowed region at the $95\%$ confidence level in $m_H$-$m_{H3}$ plane. We present the allowed region at the $95\%$ confidence level in $m_{H}$-$m_{H5}$ plane on the lower left panel and the allowed region at the $95\%$ confidence level in $m_{H3}$-$m_{H5}$ plane on the lower right panel.} 
\label{fig:massr}
\end{figure} 

In Figure~\ref{fig:massr}, we present the allowed regions at the $95\%$ confidence level in $m_H$ versus $m_{H3}$, $m_H$ versus $m_{H5}$ and $m_{H3}$ versus $m_{H5}$ planes, respectively. We can see from the figure that, for both the GM and SGM model, the allowed mass spans the entire range over which we scan. The seemingly linear relation between the pairs of masses indicates that these masses are dominated by a single large mass dimensional parameter $M_1$, see \eqref{m5}\eqref{m3}. 

\begin{figure}[hbt!]
 \centering
 \includegraphics[scale=0.4]{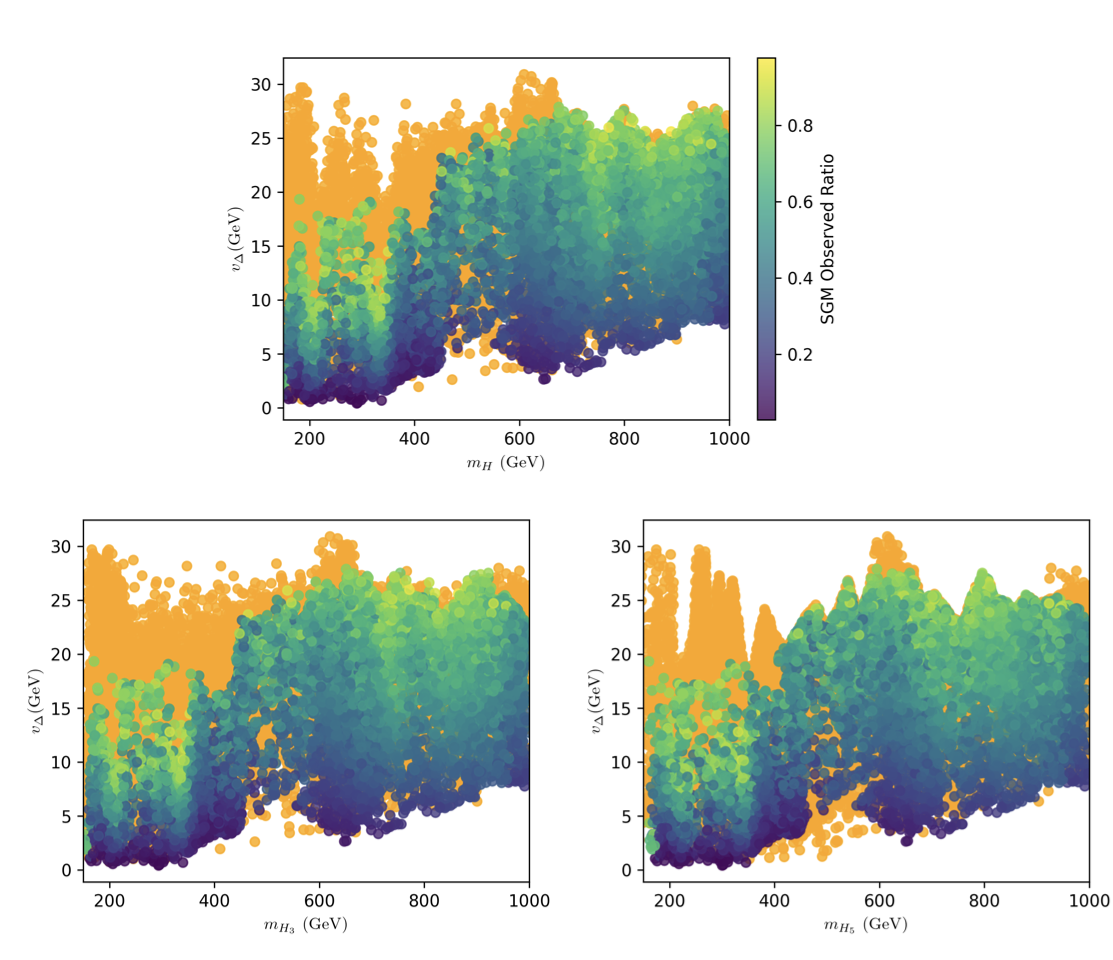} \caption[Allowed regions at the $95\%$ confidence level in the scalar masses versus $v_\Delta$ planes.]
 { \centering In the upper graph, we plot the allowed region at the $95\%$ confidence level in $m_H$ versus $v_{\Delta}$ plane. We present the allowed region at the $95\%$ confidence level in $m_{H3}$ versus $v_{\Delta}$ plane on the lower left panel and the allowed region at the $95\%$ confidence level in $m_{H5}$ versus $v_{\Delta}$ plane on the lower right panel.} 
 \label{fig:mvd}
\end{figure}

\begin{figure}[hbt!]
 \centering
 \includegraphics[scale=0.4]{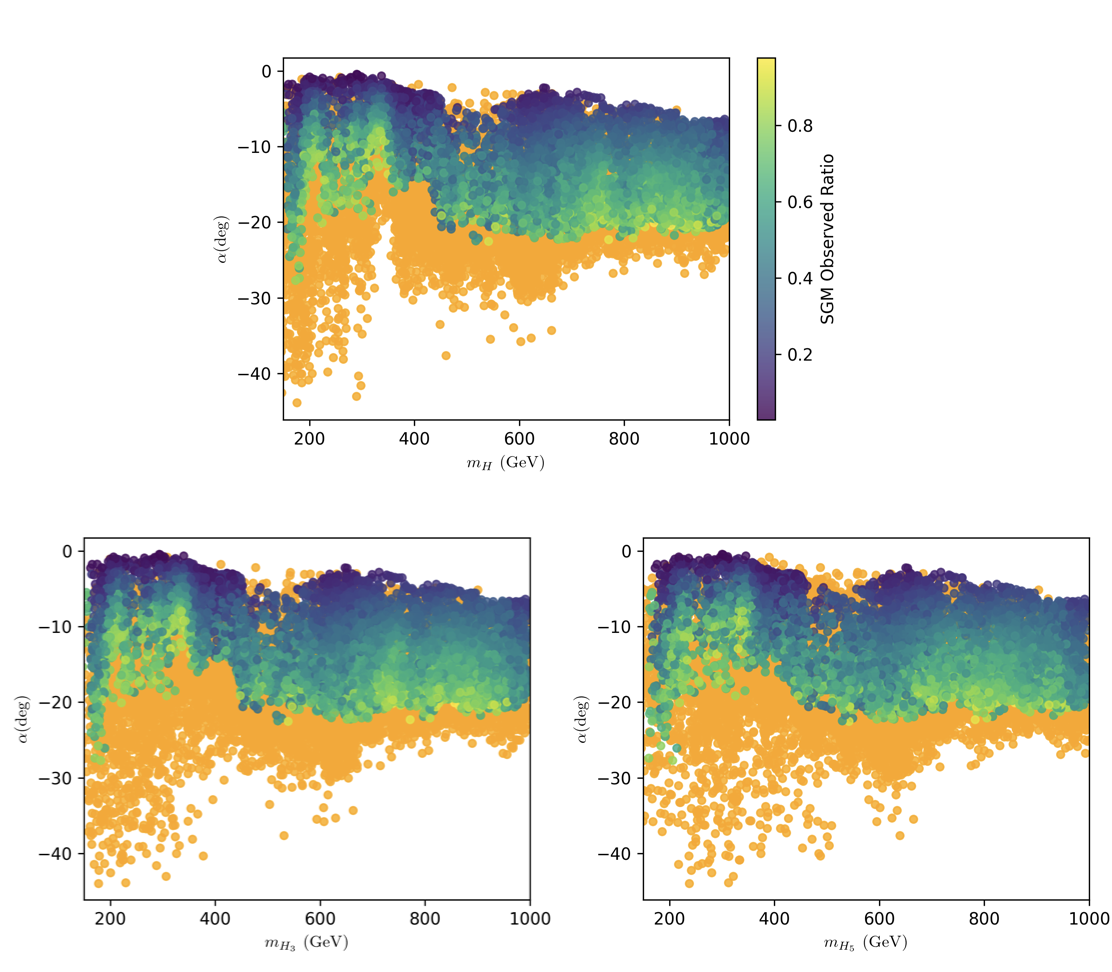} \caption{ Allowed regions at the $95\%$ confidence level in the scalar masses versus $\alpha$ planes.}
 \label{fig:mal}
\end{figure} 

Figure~\ref{fig:mvd} illustrates the allowed regions at the $95\%$ confidence level in the mass of the heavy singlet state $m_H$, the triplet state $m_{H3}$, and the quintuplet state $m_{H5}$ versus $v_\Delta$ plane, respectively. The allowed the $95\%$ regions in the plane of these masses v.s. the mixing angle $\alpha$ are presented in Figure~\ref{fig:mal}. We observe that in the SGM model, experimental constraints favor smaller $v_\Delta$ values and $\alpha$ value close to 0. When the exotic Higgs masses are below 450 GeV, $v_\Delta$ is constrained to be below $20$ GeV in the SGM model. In the GM model, the peak of $v_\Delta$ occurs when the exotic Higgs masses are 600 GeV, whereas in the SGM model, similar peaks occur when the Higgs masses are round 600 GeV and 800 GeV. In the SGM, $\alpha<22^\circ$ only occurs when the Higgs masses are below 200 GeV. When the Higgs masses are greater than 200 GeV, $\alpha$ does not go below $22^\circ$.

\begin{figure}[hbt!]
 \centering
 \includegraphics[scale=0.4]{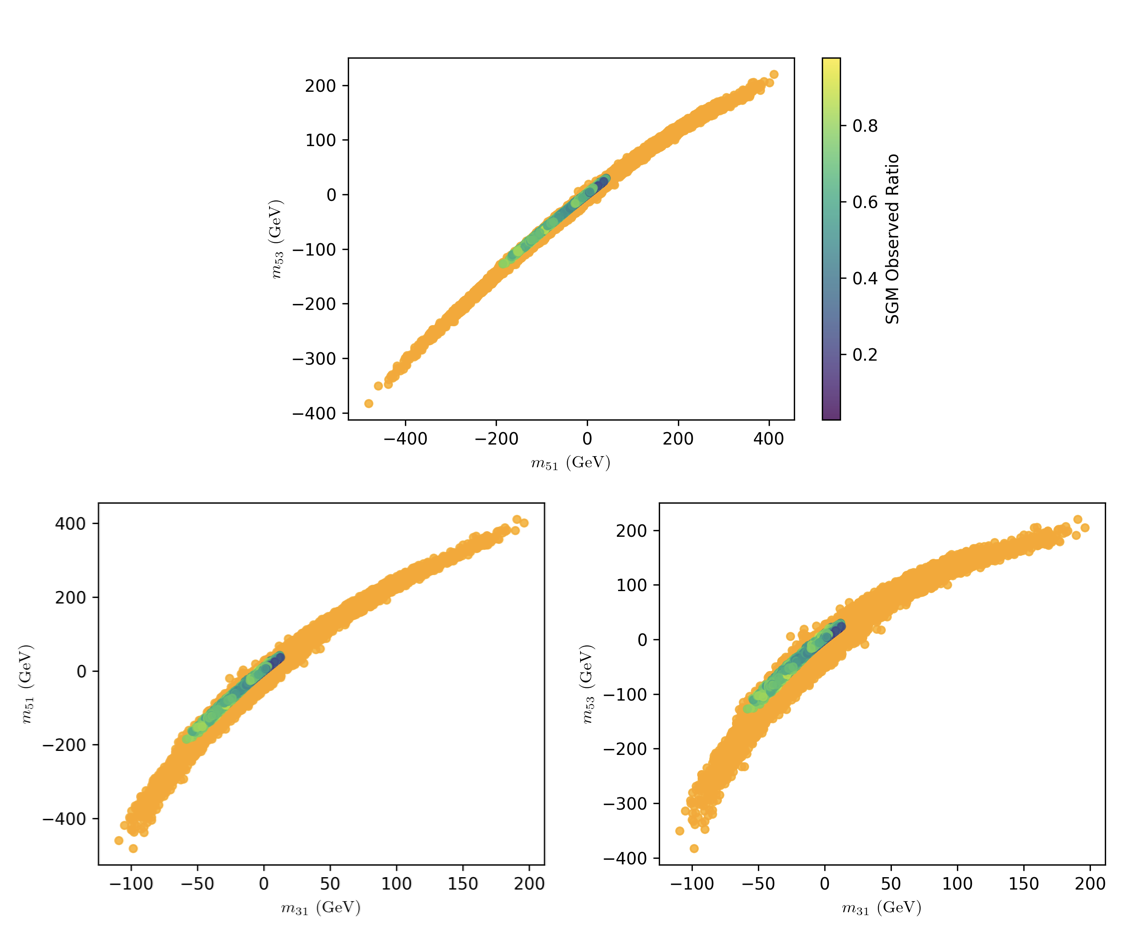} \caption[ Allowed regions at the $95\%$ confidence level for mass differences.]
 { \centering In the upper graph, we plot the allowed region at the $95\%$ confidence level in $m_{51}$-$m_{53}$ plane. We present the The allowed region at the $95\%$ confidence level in $m_{31}$-$m_{51}$ plane on the lower left panel and theThe allowed region at the $95\%$ confidence level in $m_{31}$-$m_{53}$ plane on the lower right panel.} 
 \label{fig:massdiff}
\end{figure} 

To get a better picture of the mass spectrum, we define the mass differences,
\begin{equation}
m_{31}=m_{H3}-m_{H},\quad m_{51}=m_{H5}-m_{H},\quad m_{53}=m_{H5}-m_{H3}.
\end{equation}
Figure~\ref{fig:massdiff} elucidates the impact of all the theoretical and experimental constraints on the mass differences of these exotic Higgs bosons in the GM and SGM model. Here, we can see that not only are the allowed mass range in the SGM model at the $95\%$ confidence level more restricted than that in the GM model, but the exotic Higgs boson masses tend to be degenerate in the SGM model. In the GM model, the mass difference between the triplet state and singlet state is less than $200$ GeV. The largest discrepancy occurs when $m_{H5}$ is around $550$GeV. The mass difference between the quintuplet state and the singlet state reaches a maximum at $460$ GeV when the triplet mass is around $600$ GeV. The mass difference between the triplet state and quintuplet state extends up to $360$ GeV, this occurs when $m_H$ is $730$ GeV. These regions shrink to thin strips after we impose the supersymmetric constraints on $\lambda_i$. The mass difference between the heavy singlet state and triplet state is now $60$ GeV at maximum. $|m_5-m_3|$ are below $180$ GeV and $|m_5-m_1|$ are bounded above by $130$ GeV. The upper limit of mass differences are summarized in Table~\ref{table:md}. After imposing the supersymmetric constraints, the data predicts the mass hierarchy $m_H>m_{H_3}>m_{H5}$.  This mass hierarchy prohibits some of the decay channels in the form $H\rightarrow H'V$ at the $95\%$ confidence level. The upper limit of $50$ GeV on $m_1-m_3$, in the SGM model, excludes heavy singlet decay channels $H_1\rightarrow H_3H_3$, $H_1\rightarrow H_3^+H_3^-$,  $H_1\rightarrow H_3 Z$ and $H_1\rightarrow H_3^+ W^-$ at the $95\%$ confidence level. For the quintuplet, decay channels $H_5^{++}\rightarrow H_3^+ W^+$, $H_5^+\rightarrow H_3^+ Z/H_3 W^+$, $H_5\rightarrow H_3^\pm/H_3Z$ are also excluded. For the triplet, $H_3^+\rightarrow H_1 W^+$ and $H_3\rightarrow H_1Z$ are excluded, but $H_3^+\rightarrow H_5^{++} W^-/H_5^+ Z/H_5W^+$ and $H_3\rightarrow H_5^\pm W^\mp/H_5Z$ are still allowed. 

\begin{table}[hbt!]
\centering
\begin{tabular}{|c|c|c|}
\hline
Mass Differences  & $95\%$ CL in the GM model & $95\%$ CL in the SGM model \\
\hline 
$m_1-m_3$ & 110GeV & 60GeV  \\ 
\hline 
$m_1-m_5$ & 480GeV & 180GeV  \\ 
\hline 
$m_3-m_1$ & 200GeV & 20GeV  \\ 
\hline 
$m_3-m_5$ & 380GeV & 130GeV  \\ 
\hline 
$m_5-m_1$ & 410GeV & 40GeV  \\ 
\hline 
$m_5-m_3$ & 220GeV & 30GeV \\ 
\hline 
\end{tabular}
 \caption[Allowed mass differences between scalars at the $95\%$ confidence level in the GM and SGM model.]{\centering
  Allowed  mass differences between scalars at the $95\%$ confidence level in the GM and SGM model.}
  \label{table:md}
\end{table} 

\begin{figure}[hbt!]
 \centering
 \includegraphics[scale=0.4]{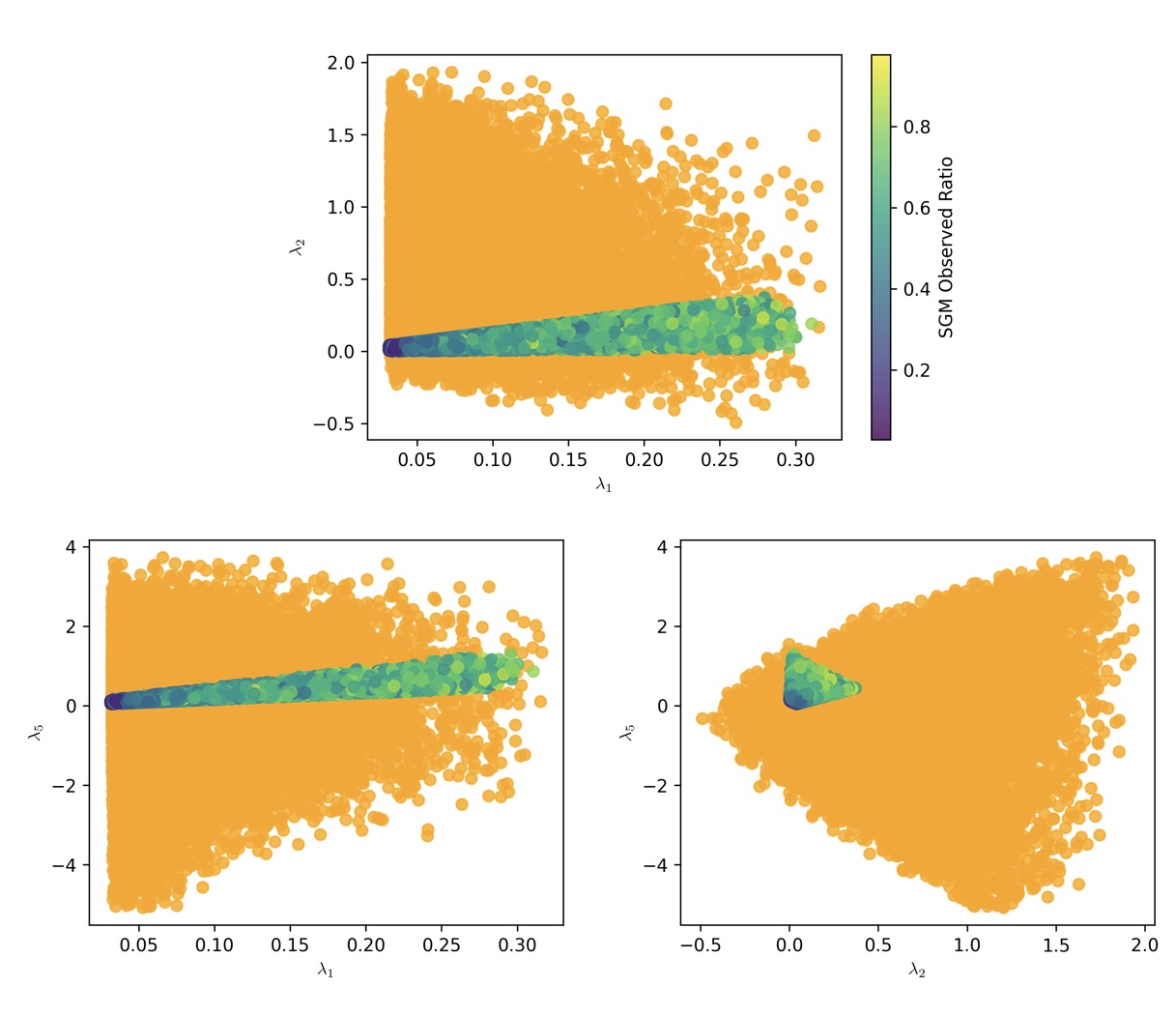} \caption[ Allowed regions at the $95\%$ confidence level for $\lambda_i$s.]
 {\centering
In the upper graph, we present the allowed regios at the $95\%$ confidence level constructed in the $\lambda_1$ versus $\lambda_2$ plane. The allowed regions in the $\lambda_{1}$ and $\lambda_{2}$ versus $\lambda_5$ planes are also presented on the lower left and right panels, respectively.}
\label{fig:ll}
\end{figure} 

In the GM model, we have five free $\lambda$ parameters. After imposing the supersymmetric constraints, only two $\lambda$ parameters are free, we select these to be $\lambda_1$ and $\lambda_2$. The allowed region at the $95\%$ confidence level in the $\lambda_1$ v.s. $\lambda_2$ plane is presented in Figure~\ref{fig:ll}. We also plot the  allowed region at the $95\%$ confidence level in the $\lambda_1$ v.s. $\lambda_5$ plane and $\lambda_2$ v.s. $\lambda_5$ plane, as $\lambda_5$ has a much larger allowed range theoretically. In both the GM and the SGM model, experimental constraints put more stringent limits on the allowed regions of $\lambda_i$ parameters. Theoretical constraints on $\lambda_1$ requires it to be less than $1.05$ in both the GM mode and the SGM model. Our scan shows that $\lambda_1<0.31$. This is a direct result of \begin{equation}
\lambda_1=\frac{1}{8 v_\phi^2}\left[m_h^2+\frac{\left(\mathcal{M}_{12}^2\right)^2}{\mathcal{M}_{22}^2-m_h^2}\right].
\end{equation}
$v_\Delta$ in the SGM model is restricted to be less than $28$GeV. Equivalently, $v_\phi$ can not be less than $233$GeV. That leads to a small $\lambda_1$ value. In the GM model, theoretical constraints require $\lambda_2\in[-2.09,2.09]$. Supersymmetry requires $\lambda_2$ to be positive in the SGM model. This constraint arises from the requirement that all parameters be real. The scan results show a much more constrained parameter space for $\lambda_2$ as well. In the GM model, $-0.49<\lambda_2<1.93$. In the SGM model, $\lambda_2>0$ cannot reach beyond $0.39$. $\lambda_5$ enjoys much more freedom than other $\lambda_i$'s. It can reach as low as $-5.07$ in the GM model, and it is capped at $3.72$ in the GM model. In the SGM model, $\lambda_5$ is strictly positive and bounded above by $1.30$.

\begin{figure}[hbt!]
 \centering
 \includegraphics[scale=0.8]{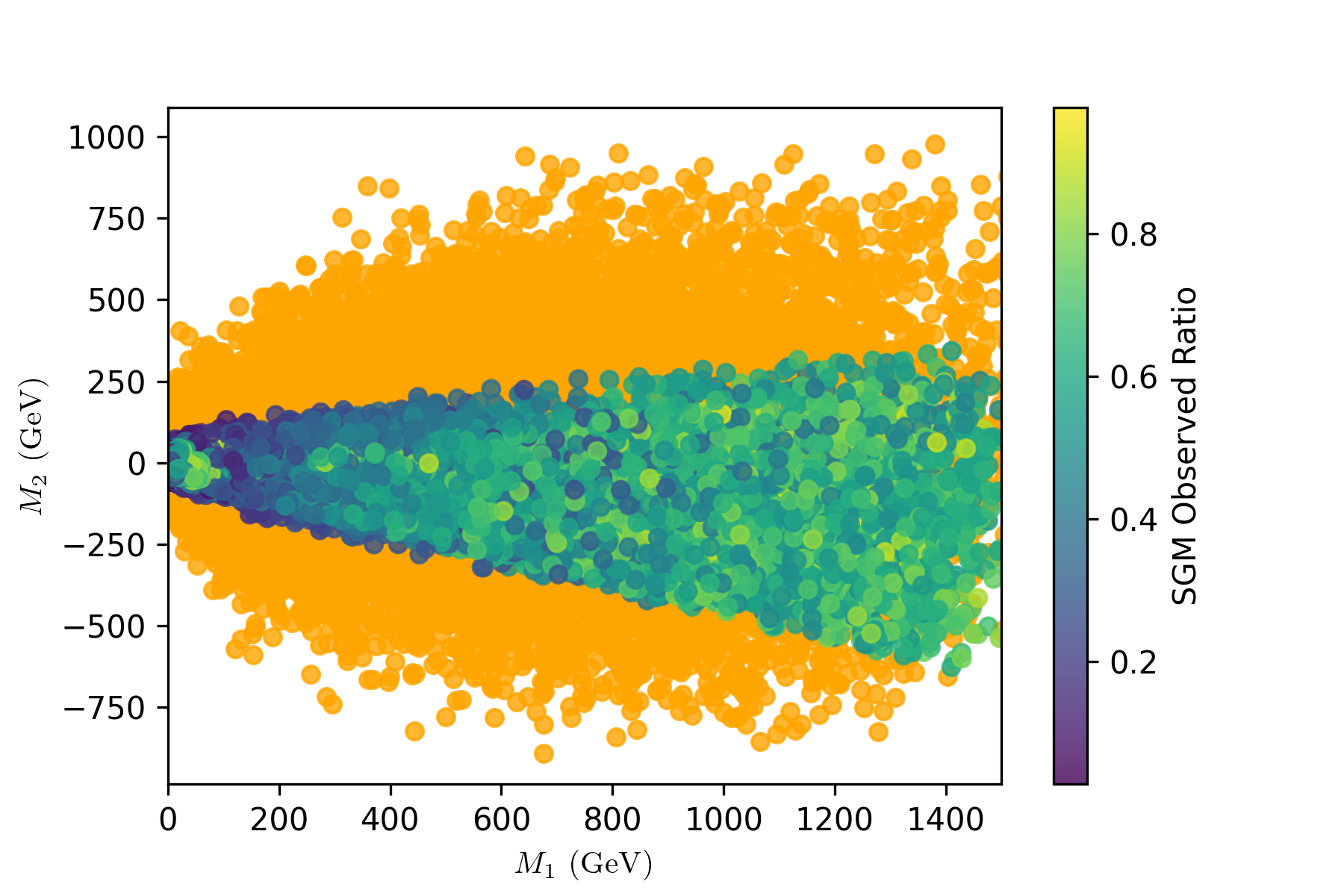} \caption{Allowed region at the $95\%$ confidence level in the $M_1$-$M_2$ plane.}
 \label{fig:mm}
\end{figure} 

\begin{figure}[hbt!]
 \centering
 \includegraphics[scale=0.4]{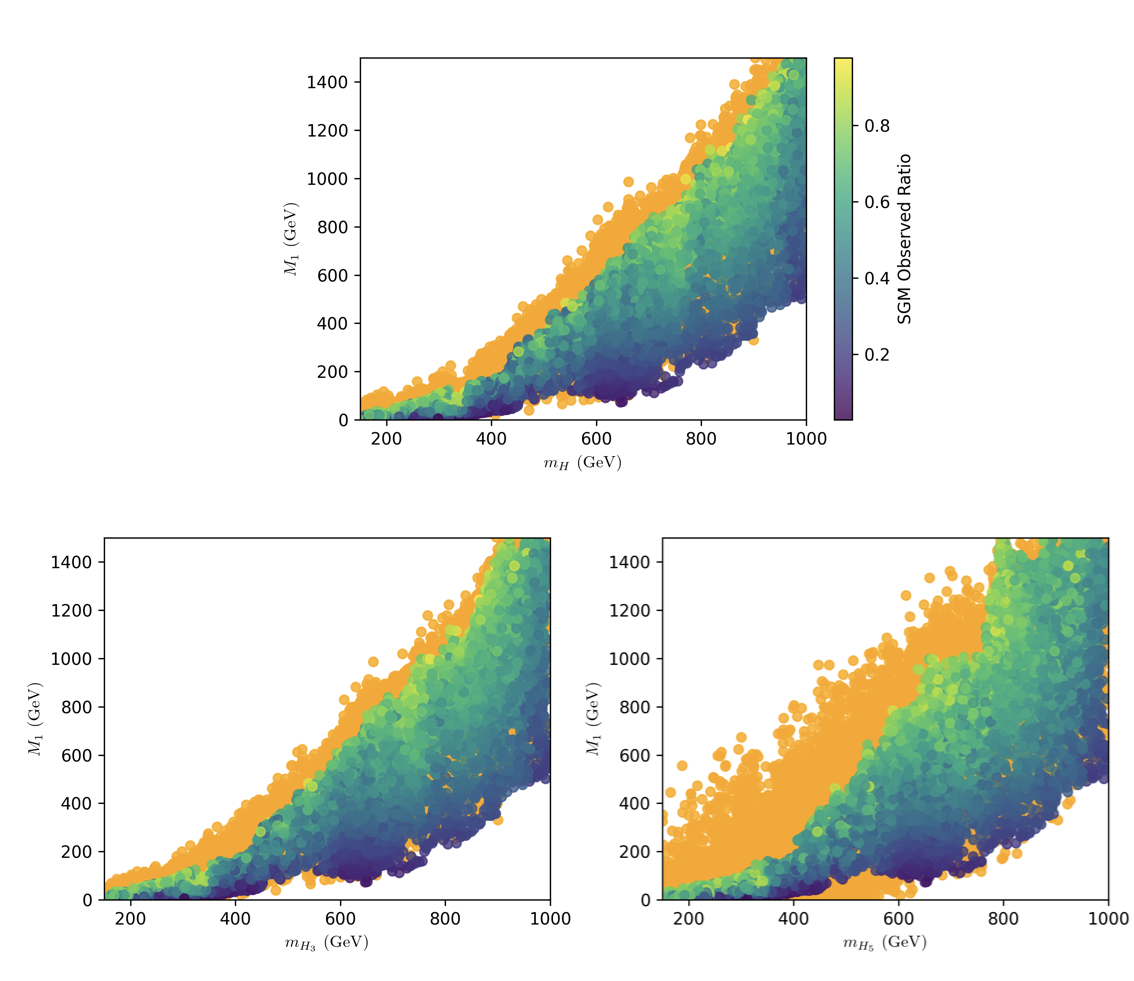} \caption[Allowed region at the $95\%$ confidence level in the mass of scalars v.s. $M_1$ planes.]
  { \centering In the upper graph, we plot the  allowed region at the $95\%$ confidence level in $m_H$-$M_1$ plane. We present the allowed region at the $95\%$ confidence level in $m_{H3}$-$M_1$ plane in the lower left panel and the  allowed region at the $95\%$ confidence level in $m_{H5}$-$M_1$ plane in the lower right panel.} 
  \label{fig:m1}
\end{figure} 

\begin{figure}[hbt!]
 \centering
 \includegraphics[scale=0.4]{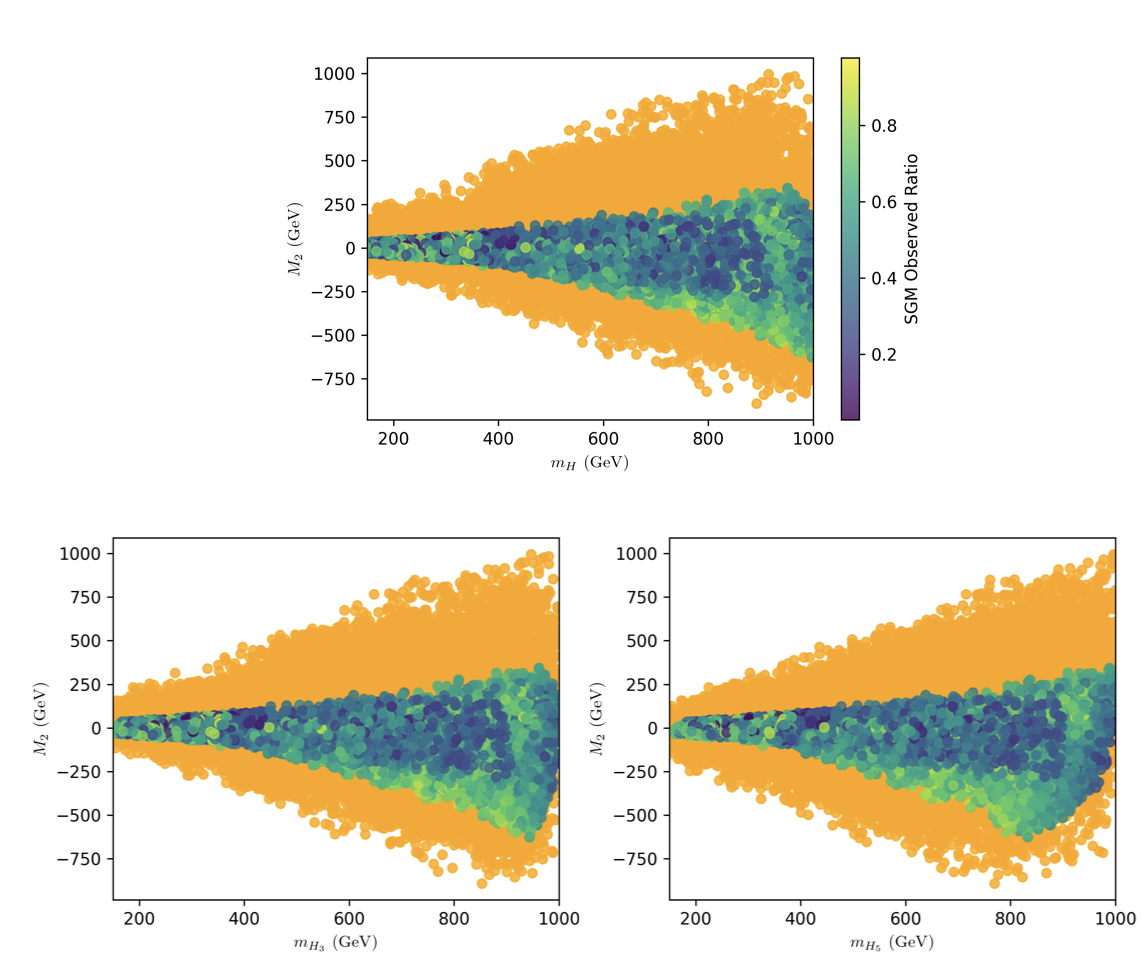} \caption[ Allowed region at the $95\%$ confidence level in the mass of scalars v.s. $M_1$ planes.]
  { \centering In the upper graph, we plot the allowed region at the $95\%$ confidence level in $m_H$-$M_2$ plane. We present the allowed region at the $95\%$ confidence level in $m_{H3}$-$M_2$ plane in the lower left panel and the  allowed region at the $95\%$ confidence level in $m_{H5}$-$M_2$ plane in the lower right panel.} 
  \label{fig:m2}
\end{figure} 

Figure~\ref{fig:mm} depicts the  allowed region at the $95\%$ confidence level in the $M_1$ v.s. $M_2$ plane. In Figure~\ref{fig:m1}, we plot the  allowed region at the $95\%$ confidence level for the masses of the 3 neutral exotics Higgs bosons with respect to $M_1$. In Figure~\ref{fig:m2}, we present the  allowed regions at the $95\%$ confidence level of the 3 scalar masses with respect to $M_2$. $M_1$ and $M_2$ are cubic couplings with dimension of mass in the scalar potential and there is no theoretical constraints for neither $M_1$ or $M_2$ in the GM model. We see that $M_1$ is purely positive and reach the maximum of our parameter input range in both the GM and the SGM model. This observation confirms that the exotic scalar masses are largely controlled by the cubic coupling, $M_1$. In the GM model, $M_2$ has an upper limit of $1000$ GeV. Unlike $M_1$, it can take negative values and go as low as $-900$ GeV. In the SGM model, the $95\%$ confidence interval shrinks to $[-600$ GeV$,350$ GeV$]$. 

All the limits on model parameters are summarized in Table~\ref{table:parameters}. In Table~\ref{table:H}-\ref{table:H5pp}, we present the limits on experiment related quantities, including total decay width and branching ratios for the heavy scalars $H_{1,3,5}$. We also present the allowed regions at the $95\%$ confidence level for branching ratios that exceed $5\%$ in Figure~\ref{fig:H}-\ref{fig:H5}. 

\begin{figure}[hbt!]
 \centering
 \includegraphics[scale=0.4]{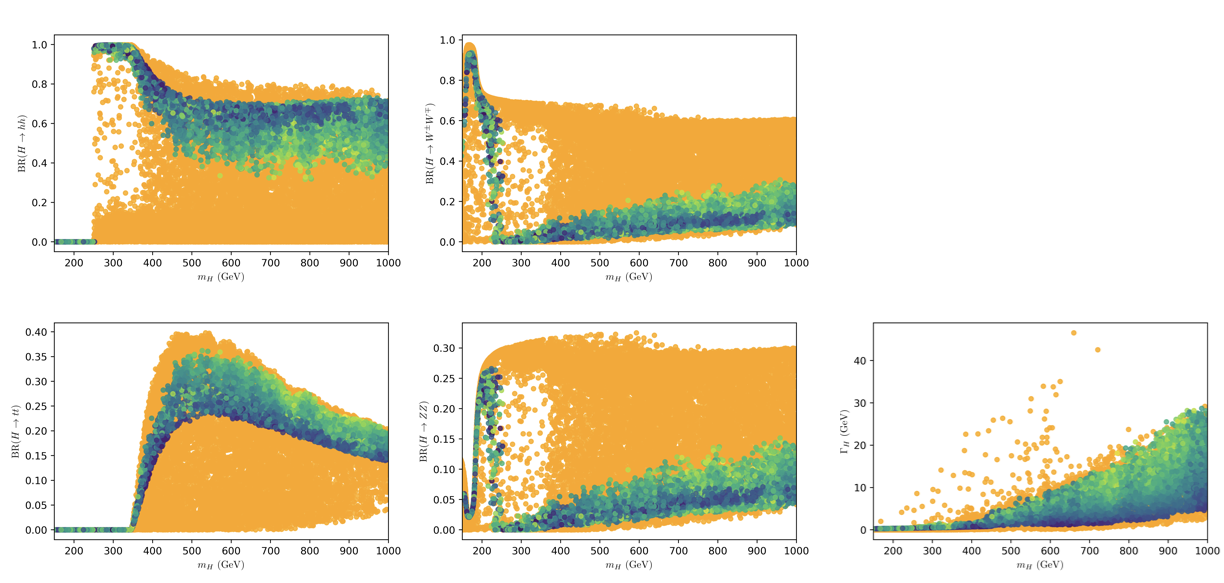} \caption[ Allowed regions at the $95\%$ confidence level for the total decay width and the largest branching ratios of the heavy singlet.]
{For the neutral triplet, we present the allowed regions at the $95\%$ confidence level for branching ratios of $H_1$ decaying to $hh$, $tt$, $WW$, $ZZ$ and the total decay width against its mass. }
\label{fig:H}
\end{figure} 

In Figure~\ref{fig:H}, we present the allowed regions at the $95\%$ confidence level for dominant branching ratios in the SGM models and the total decay width for the neutral heavy singlet $H_1$. When $m_{H1}<250$ GeV, the predominant decay channels of $H_1$ are $H_1\rightarrow WW$ and $ZZ$. The branching ratios of the decay mode $H_1\rightarrow WW$ can go up to $98\%$ in the GM model and $94\%$ in the SGM model. The branching ratio of $H_1\rightarrow ZZ$ can go up to $33\%$ in the GM model but is capped by $27\%$ in the SGM model. These two channels are heavily suppressed when $m_{H1}>250$ GeV. In the mass range 250 GeV$<m_{H1}<$350 GeV, the predominant decay mode is $H_1$ decaying to a pair of SM-like Higgs. After the $H_1\rightarrow t\bar{t}$ channel opens up around 350 GeV, the branching ratios of $H_1 \rightarrow hh$ channel quickly goes down. The branching ratio of $H_1 \rightarrow t\bar{t}$ is capped by $40\%$ in the GM model, and $37\%$ in the SGM model. Decay channels, such $H_1 \rightarrow H_{3}Z$ and $H_1 \rightarrow H_{3}^\pm W^\mp$, are allowed in the GM model but the branching ratios does not exceed $5\%$. As previously mentioned, these two channels are kinematically prohibited in the SGM model. 

\begin{figure}[hbt!]
 \centering
 \includegraphics[scale=0.4]{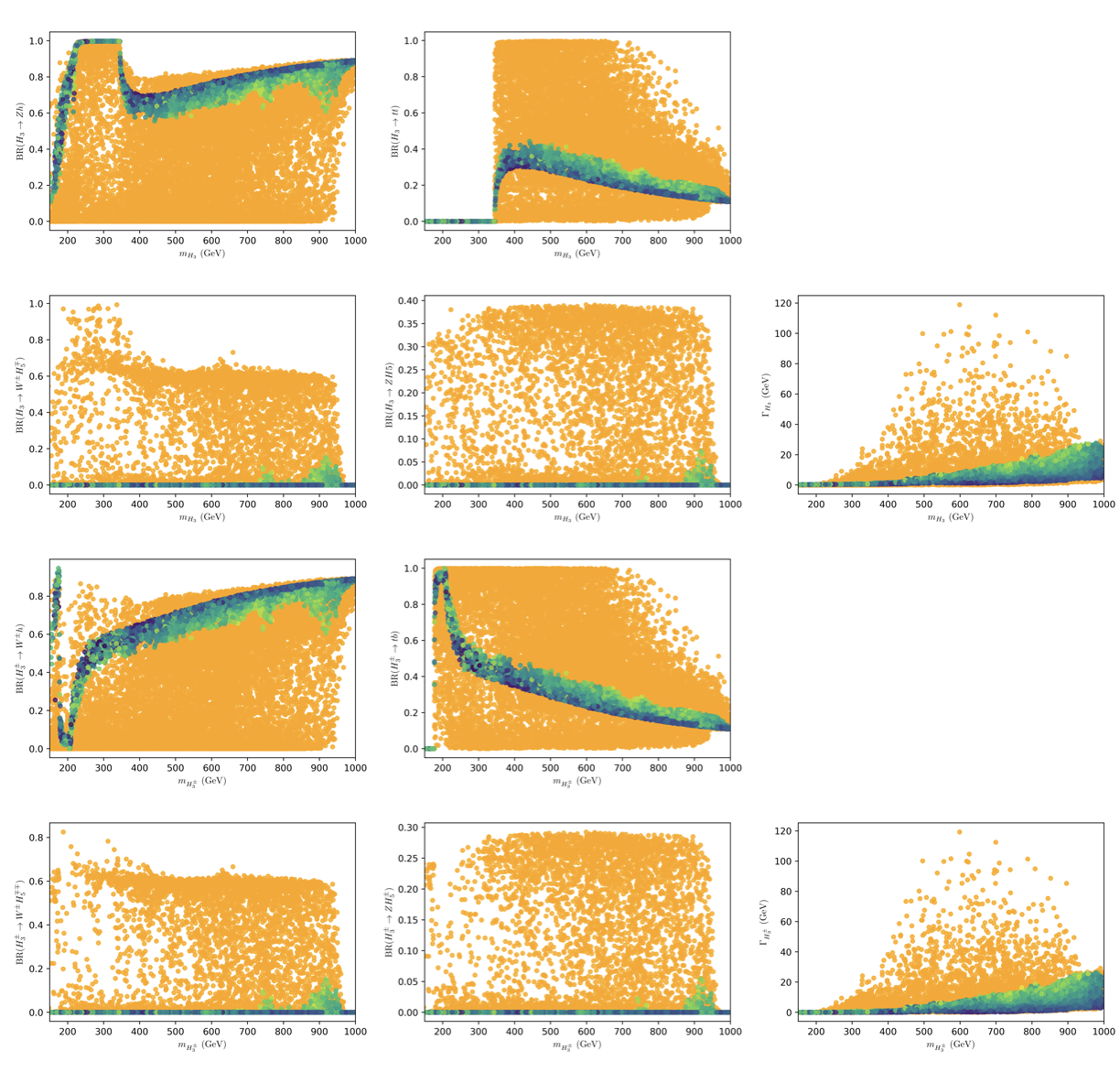} \caption[ Allowed regions at the $95\%$ confidence level for the total decay width and the largest branching ratios of the neutral and charged triplet.]
{For neutral heavy triplet, we present the allowed $95\%$ probability regions  the branching ratios of $H_3$ decaying to $Zh$, $tt$, $W^{\pm}H_5^{\mp}$, $ZH_5$. For the charged triplet, we present the branching ratios of $H_3^\pm$ decaying to $W^\pm h$, $tb$, $W^{\mp}H_5^{\pm\pm}$, $ZH_5^\pm$.}
\label{fig:H3}
\end{figure} 

In Figure~\ref{fig:H3}, we present the allowed region at the $95\%$ confidence level for dominant branching ratios in the SGM models and the total decay width for the neutral triplet $H_3$ and charged triplet $H_3^{\pm}$. The decay channels of $H_3\rightarrow H_1Z$, and $H_3^\pm \rightarrow H_1 W^{\pm}$ are kinematically prohibited in the SGM model. For the neutral triplet, the predominant decay modes are  $H_3\rightarrow Zh$ and $H_3\rightarrow t\bar{t}$ in the GM model. But the branching ratio of decay to $t\bar{t}$ pair is suppressed to $45\%$ in the SGM model. For the charged triplet state, the dominant decay modes are $H_3^{ \pm} \rightarrow tb$ and $W^{ \pm} h$. The branching ratio of the decay channel $H_3^{ \pm} \rightarrow W^{\mp} H_5^{ \pm \pm}$ can go up to $83\%$ in the GM model but only $15\%$ in the SGM model. The decay channel $H_3^\pm\rightarrow ZH_5^\pm$ is also suppressed in the SGM model, whose branching ratios barely exceeds $5\%$ in the SGM model but can be as large as $30\%$ in the GM model. The decay channel $H_3^\pm\rightarrow W^{\pm}H_5$ exceeds $5\%$ in the GM model as well, but are only around $3\%$ in the SGM model when the quintuplet mass is between $800$ GeV to $1$ TeV.

\begin{figure}[hbt!]
 \centering
 \includegraphics[scale=0.4]{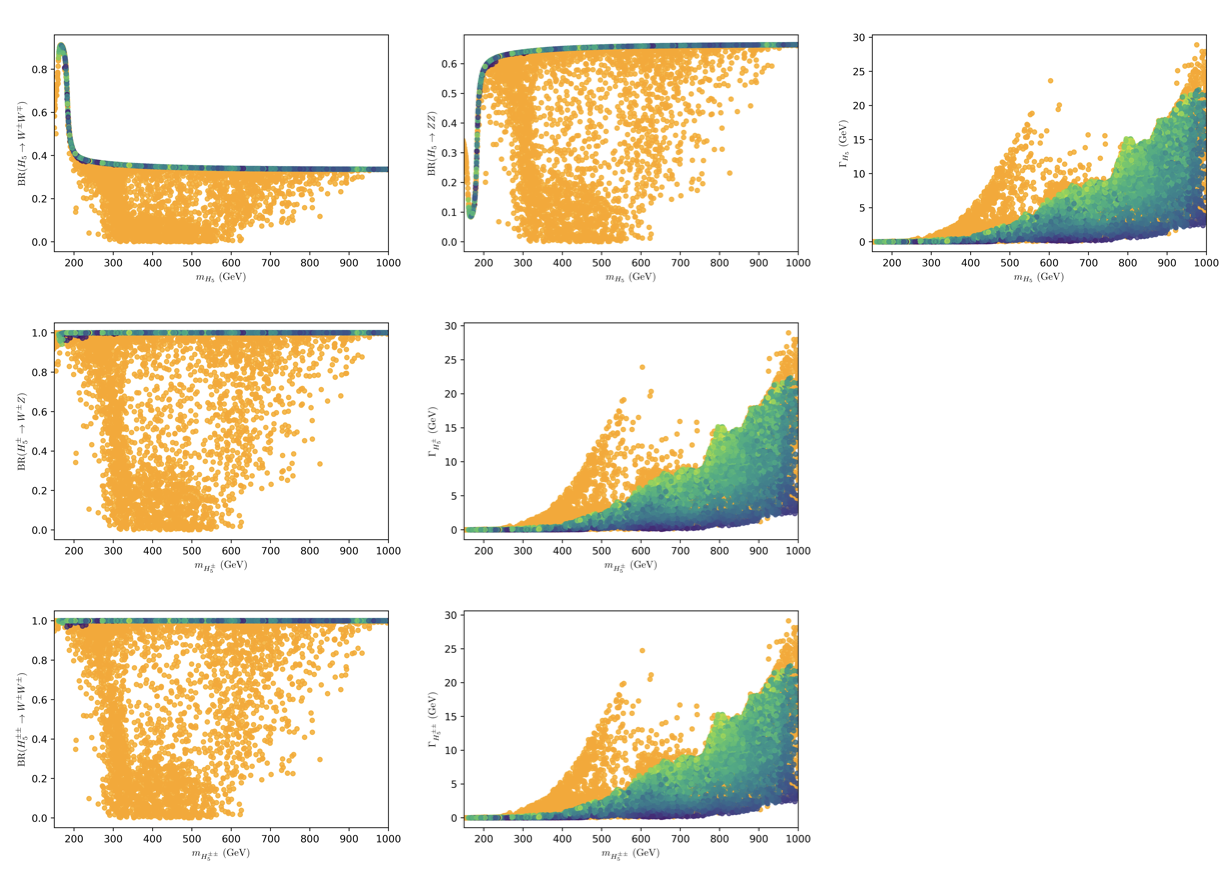} \caption[ Allowed regions at the $95\%$ confidence level for the total decay width and the largest branching ratios of the neutral, charged and doubly charged quintuplet.]
{For neutral heavy triplet, we present the allowed regions at the $95\%$ confidence level the branching ratios of $H_5$ decaying to $ZZ$, $W^\pm W^\pm$. For the charged triplet, we present the branching ratios of $H_5^\pm$ decaying to $W^\pm Z$. For the doubly charge quituplet, we present the branching ratios of $H_5^{\pm\pm}$ decaying to $W^{\pm}W^{\pm}$.}
\label{fig:H5}
\end{figure} 

In Figure~\ref{fig:H5}, we present the allowed region at the $95\%$ confidence level for dominant branching ratios in the SGM model and the total decay width for the neutral quintuplet $H_5$, singly charged quintuplet $H_5^{\pm}$ and doubly charged quintuplet $H_5^{\pm\pm}$. The quintuplet states only couple to vector bosons. In the SGM model, $H_5$ can only decay to $WW$ and $ZZ$ pairs. For the singly charged $H_5^{\pm}$, it can only decay to $W^{\pm}Z$. The doubly charged $H_5^{\pm\pm}$ only decays to same sign $W^{\pm}W^{\pm}$ pairs. The decay channels of $H_5 \rightarrow ZH_3$, $W^{ \pm} H_3^{\mp}$, $H_5^{ \pm} \rightarrow W^{ \pm} H_3$, $ZH_3^\pm$, and $H_5^{ \pm \pm} \rightarrow H_3^{ \pm} W^{ \pm}$, are not allowed in the SGM model because of the mass hierarchy. In the GM model, we have a few more decay channels with branching ratios greater than $5\%$. For the neutral quintuplet, the branching ratios of decay modes $H_5 \rightarrow ZH_3$ and $W^{ \pm} H_3^{\mp}$ can be as large as of $65\%$ and $92\%$, respectively. For the charged quintuplet, the branching ratios for the channel $H_5^\pm\rightarrow W^\pm H_3$ is capped by $96\%$ in the GM model, and the branching ratio of $H_5^\pm\rightarrow ZH_3^\pm$ can be as large as $48\%$. For the doubly charged state $H_5^{\pm\pm}$, the decay to the $W^\pm H_3^\pm$ can also dominate in the GM model, whose branching ratio can exceed $90\%$. However, this decay channel is kinematically prohibited in the SGM model, which is a notable feature that distinguishes the GM and SGM model.

The total decay width of the heavy singlet, triplet and quintuplet cannot exceed 47 GeV, 119 GeV and 29 GeV, respectively, in the GM model. In the SGM model, the allowed range shrinks heavily. The total decay width of the heavy singlet, triplet and quintuplet can only go up to 27 GeV, 26 GeV and 23 GeV, respectively. Furthermore, the total decay width for the SM-like Higgs now ranges between $3.7$ MeV to $4.6$ MeV, regardless of the supersymmetric constraints. The most sensitive channel for exclusion to the even-parity states $H_1$ and $H_5$ are the decays of heavy neutral scalars to weak vector bosons from ATLAS $139$fb$^{-1}$ experiments\cite{ATLAS:2020tlo,ATLAS:2020fry}. Both vector boson fusion and gluon fusion production channels contribute to the data set that included in the analysis. For the odd-parity state $H_3$, the most sensitive channel is $H_3$ decay to diphoton data from ATLAS $139$fb$^{-1}$ experiments\cite{ATLAS:2021uiz}. By incorporating the latest experimental data, we have been able to considerably shrink the allowed parameter space of the GM model from that obtained in previous work\cite{Chiang:2018cgb}.

\begin{table}[H]
\centering
\begin{tabular}{|c|c|c|c|}
\hline
Parameters  & $95\%$ CL in the GM model & $95\%$ CL in the SGM model & unit \\
\hline 
$\lambda_1$ & [0,0.31] & [0,0.31] & \\ 
\hline 
$\lambda_2$ & [-0.49,1.93] & [0,0.39] & \\ 
\hline 
$\lambda_3$ & [-1.52,1.78] & [-0.39,0] & \\ 
\hline 
$\lambda_4$ & [-0.57,1.54] & [0,0.42] & \\ 
\hline 
$\lambda_5$ & [-5.07,3.72] & [0,1.29] & \\ 
\hline 
$M_1$ & [0,1500] & [0,1500] &$\text{GeV}$  \\ 
\hline 
$M_2$ & [-900,1000] & [-630,350] & $\text{GeV}$ \\ 
\hline 
$\alpha$ & [-44,0] & [-28,0] & $\text{degree}$ \\ 
\hline
$v_\Delta$ & [0,31] & [0,28] &$\text{GeV}$ \\ 
\hline
\end{tabular}
 \caption[The $95\%$ confidence limits on model parameters in the GM and SGM model.]{\centering
  Allowed parameter range at the $95\%$ confidence level in the GM and SGM model.}
  \label{table:parameters}
\end{table}

\begin{table}[H]
\centering
\begin{tabular}{|c|c|c|}
\hline
$H_1$  & $95\%$ CL in the GM model & $95\%$ CL in the SGM model \\
\hline 
$\Gamma_1$ & $\leq 47$GeV & $\leq  27$GeV  \\ 
\hline \hline
$\operatorname{Br}(H_1\rightarrow hh)$ & $[0,100]\%$ & $[0,100]\%$\\

$\operatorname{Br}(H_1\rightarrow WW)$  & $[0,98]\%$ & $[0,94]\%$ \\

$\operatorname{Br}(H_1\rightarrow tt)$ & $[0,40]\%$ & $[0,37]\%$  \\

$\operatorname{Br}(H_1\rightarrow ZZ)$ & $[0,33]\%$ & $[0,27]\%$ \\
\hline
\end{tabular}
 \caption[ The $95\%$ confidence limits of the decay width and branching ratios in the GM and SGM models for the heavy singlet $H_1$.]{\centering
  The $95\%$ confidence limits of the decay width and branching ratios in the GM and SGM models, here we only consider branching ratios greater than $5\%$ in the SGM model.}
  \label{table:H}
\end{table}

\begin{table}[H]
\centering
\begin{tabular}{|c|c|c|}
\hline
$H_3$  & $95\%$ CL in the GM model & $95\%$ CL in the SGM model \\
\hline 
$\Gamma_3$ &  $\leq 119$GeV &  $\leq 26$GeV \\ 
\hline \hline
$\operatorname{Br}(H_3\rightarrow Zh)$ & $[0,100]\%$ & $[10,100]\%$ \\

$\operatorname{Br}(H_3\rightarrow tt)$ & $[0,100]\%$ & $[0,45]\%$ \\

$\operatorname{Br}(H_3\rightarrow W^{\pm}H_5^{\mp})$ & $[0,100]\%$ & $[0,16]\%$ \\

$\operatorname{Br}(H_3\rightarrow ZH_5)$ & $[0,40]\%$ & $[0,8]\%$ \\

\hline
\end{tabular}
 \caption[ The $95\%$ confidence limits of the decay width and branching ratios in the GM and SGM models for the neutral triplet $H_3$.]{\centering
 The $95\%$ confidence limits of the decay width and branching ratios in the GM and SGM models for the neutral triplet $H3$, here we only consider branching ratios greater than $5\%$ in the SGM model.}
  \label{table:H3}
\end{table}

\begin{table}[H]
\centering
\begin{tabular}{|c|c|c|}
\hline 
$H_3^\pm$  & $95\%$ CL in the GM model & $95\%$ CL in the SGM model \\
\hline 
$\Gamma_{3\pm}$ & $\leq 120$GeV & $\leq 26$GeV  \\ 
\hline \hline
$\operatorname{Br}(H_3^\pm\rightarrow tb)$ & $[0,100]\%$ & $[10,100]\%$ \\

$\operatorname{Br}(H_3^\pm\rightarrow W^\pm h)$ & $[0,95]\%$ & $[0,95]\%$ \\

$\operatorname{Br}(H_3^\pm\rightarrow W^\mp H_5^{\pm\pm})$ & $[0,83]\%$ & $[0,16]\%$ \\

$\operatorname{Br}(H_3^\pm\rightarrow Z H_5^\pm)$ & $[0,30]\%$ & $[0,6]\%$ \\

\hline 
\end{tabular}
 \caption[ The $95\%$ confidence limits of the decay width and branching ratios in the GM and SGM models for the singlet charged triplet state $H_3^\pm$.]{\centering
  The $95\%$ confidence limits of the decay width and branching ratios in the GM and SGM models, here we only consider branching ratios greater than $5\%$ in the SGM model.}
  \label{table:H3p}
\end{table}

\begin{table}[H]
\centering
\begin{tabular}{|c|c|c|}
\hline
$H_5$  & $95\%$ CL in the GM model & $95\%$ CL in the SGM model \\
\hline 
$\Gamma_5$ & $\leq 29$GeV & $\leq 23$GeV  \\ 
\hline \hline
$\operatorname{Br}(H_5\rightarrow ZZ)$ & $[0,67]\%$ & $[3,67]\%$ \\

$\operatorname{Br}(H_5\rightarrow W^\pm W^\mp)$ & $[0,92]\%$ & $[10,92]\%$ \\
\hline
\end{tabular}
 \caption[ The $95\%$ confidence limits of the decay width and branching ratios in the GM and SGM models for the neutral quintuplet $H_5$.]{\centering
  The $95\%$ confidence limits of the decay width and branching ratios in the GM and SGM models for the neutral quintuplet $H_5$, here we only consider branching ratios greater than $5\%$ in the SGM model.}
  \label{table:H5}
\end{table}

\begin{table}[H]
\centering
\begin{tabular}{|c|c|c|}
\hline 
$H_5^\pm$  & $95\%$ CL in the GM model & $95\%$ CL in the SGM model \\
\hline 
$\Gamma_{5\pm}$ & $\leq 29$GeV & $\leq 23$GeV \\ 
\hline \hline
$\operatorname{Br}(H_5^\pm\rightarrow W^\pm Z)$ & $[0,100]\%$ & $\sim 100\%$ \\
\hline
\end{tabular}
 \caption[ The $95\%$ confidence limits of the decay width and branching ratios in the GM and SGM models for the singly charged quintuplet $H_5^\pm$.]{\centering
  The $95\%$ confidence limits of the decay width and branching ratios in the GM and SGM models for the singlet charged quintuplet $H_5^\pm$, here we only consider branching ratios greater than $5\%$ in the SGM model.}
  \label{table:H5p}
\end{table}

\begin{table}[H]
\centering
\begin{tabular}{|c|c|c|}
\hline
$H_5^{\pm\pm}$  & $95\%$ CL in the GM model & $95\%$ CL in the SGM model \\
\hline 
$\Gamma_{5\pm\pm}$& $\leq 29$GeV & $\leq 23$GeV    \\ 
\hline \hline
$\operatorname{Br}(H_5^{\pm\pm}\rightarrow W^\pm W^\pm)$ &  $[0,100]\%$ &  $\sim 100\%$\\
\hline
\end{tabular}
 \caption[ The $95\%$ confidence limits of the decay width and branching ratios in the GM and SGM models for the doubly charged quintuplet $H_5^{\pm\pm}$.]{\centering
 The $95\%$ confidence limits of the decay width and branching ratios in the GM and SGM models for the doubly charged quintuplet $H_5^{\pm\pm}$, here we only consider branching ratios greater than $5\%$.}
  \label{table:H5pp}
\end{table}

 \begin{singlespace}
\chapter{A $95$ GeV Higgs Boson}\label{chapter:95GeV}
\end{singlespace}

CMS and ATLAS have reported small excesses in the search for low-mass Higgs bosons in the di-photon decay channel at exactly the same mass, 95.4 GeV, based on the full LHC Run 2 data\cite{CMS:2018cyk,ATLAS:2018xad}. Neglecting possible correlations, the combined signal strength for this di-photon excess was obtained \cite{Biekotter:2023oen},
\begin{equation}
\mu_{\gamma \gamma}^{\exp }=\mu_{\gamma \gamma}^{\mathrm{ATLAS}+\mathrm{CMS}}=0.24_{-0.08}^{+0.09},
\end{equation}
corresponding to an excess at $3.1\sigma$ at \begin{equation}
m_\phi \equiv m_\phi^{\mathrm{ATLAS}+\mathrm{CMS}}=95.4 \mathrm{GeV}.
\end{equation}
LEP also reported a local $2.3\sigma$ excess in the $e^{+} e^{-} \rightarrow Z(\phi \rightarrow b \bar{b})$ searches\cite{LEPWorkingGroupforHiggsbosonsearches:2003ing}, which is also consistent with a Higgs boson with a mass of 95.4 GeV. In addition to the di-photon excess, CMS also reported another excess compatible with a mass of 95.4 GeV in the search $p p \rightarrow \phi \rightarrow \tau^{+} \tau^{-}$\cite{CMS:2022goy}. Though the excess in the $\tau\tau$ channel peaked around 100 GeV, with a local significance at $3.1\sigma$, it is also compatible with the excess at 95.4 GeV, with a local significance of $2.6\sigma$. 

Given all the excesses reported above occurs at a similar mass around 95 GeV, we perform fits on the and SGM model by fixing the lighter singlet scalar $m_h$ to 95 GeV and the heavier singlet scalar $m_H$ to 125 GeV. We find that the observed excess in the SGM model would be expected to be correlated with a doubly charged Higgs with mass between 97 GeV to 160 GeV. 

\begin{figure}[hbt!]
 \centering
 \includegraphics[scale=0.8]{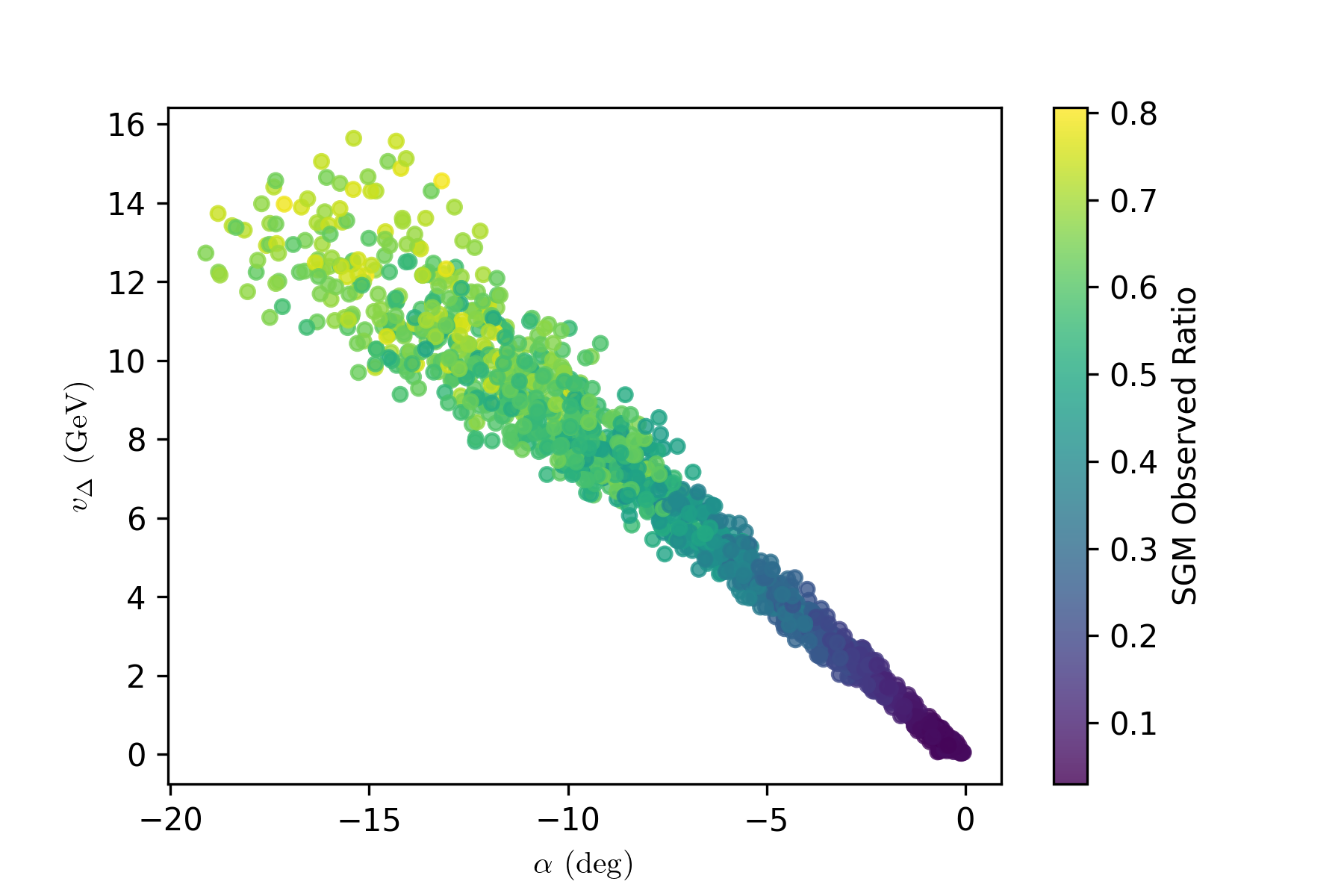} \caption[ Allowed regions at the $95\%$ confidence level in the $\alpha-v_\Delta$ plane when $m_h=95$ GeV in the SGM model.]
{Allowed regions at the $95\%$ confidence level in the $\alpha-v_\Delta$ plane when $m_h=95$ GeV in the SGM model.}
\label{fig:avdsgm}
\end{figure} 

\begin{figure}[hbt!]
 \centering
 \includegraphics[scale=0.8]{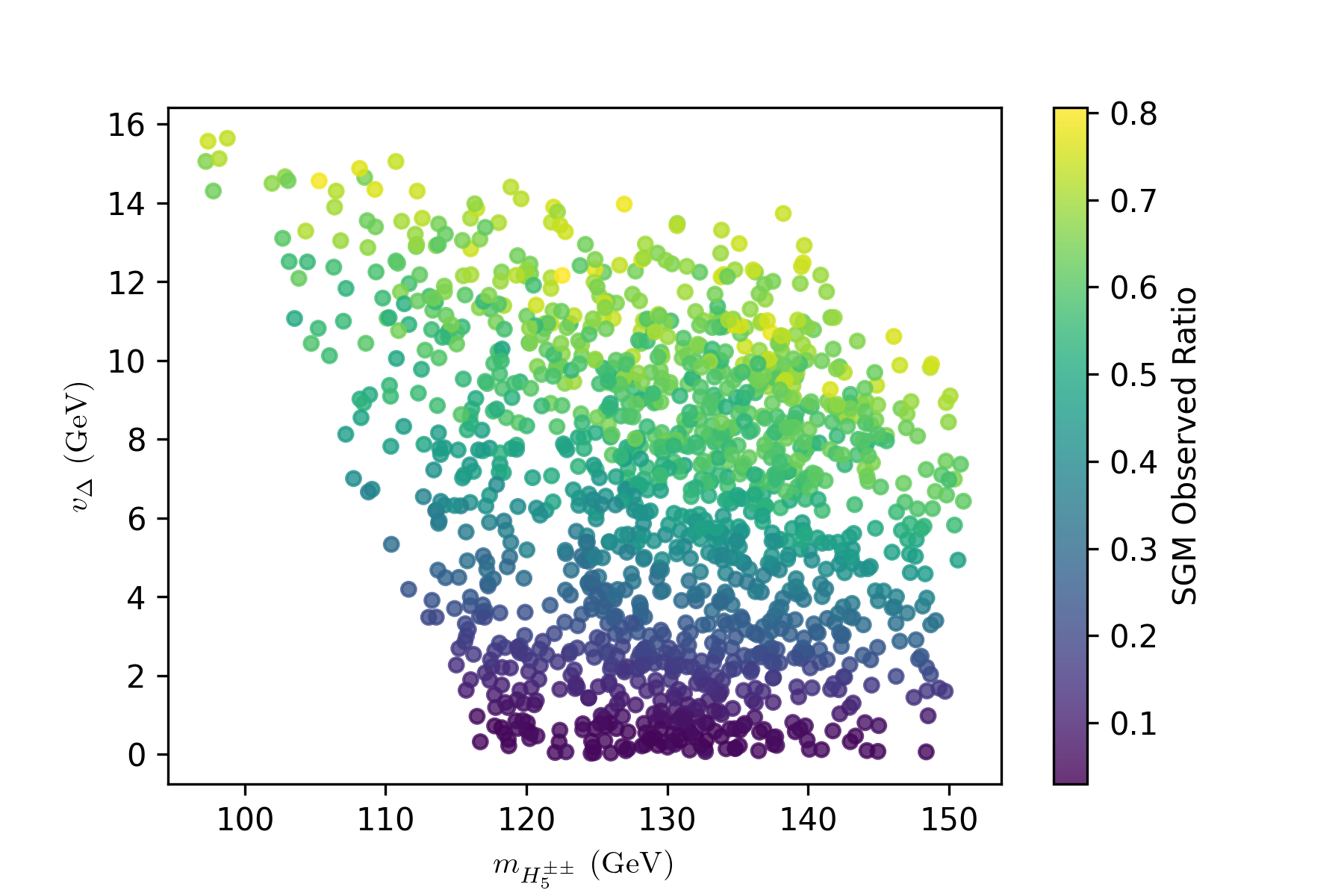} \caption[ Allowed regions at the $95\%$ confidence level in the $m_{H5}-v_\Delta$ plane when $m_h=95$ GeV in the SGM model.]
{Allowed regions at the $95\%$ confidence level in the $m_{H5}-v_\Delta$ plane when $m_h=95$ GeV in the SGM model.}
\label{fig:h5vdsgm}
\end{figure} 

In Figure~\ref{fig:avdsgm} and \ref{fig:h5vdsgm}, we present the allowed region on the $\alpha-v_\Delta$ plane and the $m_{H_5^{\pm\pm}}$ plane, at the $95\%$ confidence level for the SGM model. In the SGM model, the $95\%$ confidence limit on the triplet v.e.v's is further suppressed to $15.7$ GeV. The mixing angle $\alpha$ is bounded below by $-20^\circ$. We also find that the $95\%$ confidence limit mass of the quintuplet $H_5$ is between 97 GeV to 160 GeV. This implies that the $95$ GeV Higgs boson will be accompanied by a doubly charged scalar in this mass range in the SGM model, whose production cross section of pair productions varies from 1.1 pb to 200 fb in this mass range\cite{Ashanujjaman:2022ofg}. The confidence limits on other model parameters are also presented in Table~\ref{table:95pa}.

\begin{table}[hbt!]
\centering
\begin{tabular}{|c|c|c|c|}
\hline
Parameters  & $95\%$ CL in the SGM model & unit \\
\hline 
$\lambda_1$  & [0.03,0.034] & \\ 
\hline 
$\lambda_2$  & [0,0.045] & \\ 
\hline 
$\lambda_3$  & [-0.045,0] & \\ 
\hline 
$\lambda_4$  & [0.04,0.045] & \\ 
\hline 
$\lambda_5$  & [0.04,0.15] & \\ 
\hline 
$M_1$  & [0,11] &$\text{GeV}$  \\ 
\hline 
$M_2$ &  [-30,20] & $\text{GeV}$ \\ 
\hline 
$\alpha$  & [-19,0] & $\text{degree}$ \\ 
\hline
$v_\Delta$  & [0,15.7] &$\text{GeV}$ \\ 
\hline
\end{tabular}
 \caption[The $95\%$ confidence limits on model parameters in the SGM model when $m_h=95$ GeV.]{\centering
  Allowed parameter range at the $95\%$ confidence level in the SGM model when $m_h=95$ GeV.}
  \label{table:95pa}
\end{table}

In the SGM model, the doubly charged Higgs boson $H_5^{\pm\pm}$ can only decay to a pair of same sign $W$ bosons, with a decay width capped at $180$ keV. Since $m_{H_5^{\pm\pm}}$ in the allowed range at the $95\%$ confidence limits is less than twice the mass of $W$ bosons, one or both of the $W$ boson has to be off-shell. ATLAS and CMS have carried out several searches pertinent to this low mass range
\cite{ATLAS:2012hi,CMS:2012dun,ATLAS:2014kca,CMS:2014mra,ATLAS:2017xqs,CMS:2017fhs,ATLAS:2022pbd}. The non-observation of any significance over the SM expectations place stringent limit on the masses of doubly charged Higgs. However, these searches all focus on leptonic decay channels. For $v_\Delta\sim 10$ GeV, the couplings to like-sign leptons is suppressed, and thus these limits are not applicable to our model. Ref~\cite{ATLAS:2018ceg,ATLAS:2021jol} place lower limits on the mass of doubly charged Higgs bosons through the decay channel of same sign $W$ boson pairs but they only consider the on-shell case. A lower mass constraint of $84$ GeV is derived in \cite{Kanemura:2014ipa} through a diboson decay channel. In a nutshell, searches for a doubly charged Higgs decaying into same sign $W$ pairs in the mass range $[97,160]$ GeV are still open.

 \begin{singlespace}
\chapter{Summary and Outlook}\label{chapter:Summary}
\end{singlespace}

In this dissertation we have performed the global fits in the SGM model for the first time, using the $\texttt{GMcalc}$ and $\texttt{Higgstools}$ package and the latest experimental data. We also have performed global fits in the GM model for comparison purposes. By incorporating more recent experimental data\cite{ATLAS:2020tlo,ATLAS:2020fry}, we have been able to considerably shrink the triplet vacuum expectation value from 44 GeV in previous work\cite{Chiang:2018cgb} to 31 GeV for the GM model. We consider constraints from both theory and up-to-date experimental data from Run1 and Run2 of the LHC, primarily on direct searches sensitive to the neutral, singly charged and doubly charged heavy scalars. By considering both the theoretical constraints and experimental data from direct searches, we have constructed the allowed region at the $95\%$ confidence level for model parameters in the Lagrangians of the GM model and SGM model and several phenomenologial quantities. Current direct search results set significantly stronger limits on these model parameters than what is theoretically allowed. The LHC searches for heavy scalar states primarily constrain the vacuum expectation value of the triplet field $v_\Delta$. Though in theory $v_\Delta$ can be as large as 84 GeV, it is constrained to be less than 31 GeV by experimental data in the GM model. After the supersymmetric constraints being imposed, $v_\Delta$ is further constrained to 28 GeV. This translates to an upper bound on $\sin\theta_H=0.36$. $\alpha$ has to be purely negative and is bounded from below by $-44^\circ$ in the GM model. In the SGM model, this lower limit elevates to $-28^\circ$. We obtained the $95\%$ confidence limits on the differences of the masses between exotic Higgs particles in the to be less than 500 GeV in GM model and less than 200 GeV the SGM model. In particular, we have noted that the masses of exotic scalars are more degenerate in the SGM model. We also observed a mass hierarchy $m_H>m_{H3}>m_{H5}$ in the SGM model. This observation implied that some decay channels allowed in the GM model, such as $H_5^{++} \rightarrow H_3^{+} W^{+}$, $H_5^{+} \rightarrow H_3^{+} Z / H_3 W^{+}$, $H_5 \rightarrow H_3^{ \pm} / H_3 Z$, $H_3^{+} \rightarrow H_1 W^{+}$ and $H_3 \rightarrow H_1 Z$, are no longer kinematically allowed in the SGM model. 

In the SGM model, we have two singlets, either of which can be the SM-like Higgs boson. We have performed the global fits by assuming the lighter singlet to be the SM-like Higgs. We then consider the possibility of the heavier singlet being the SM-like Higgs boson. In this scenario, the mass of the lighter singlet is set to be 95 GeV, in order to address the excess reported by ATLAS and CMS with a combined local significant at $3.1\sigma$. We find, in this case, $v_\Delta$ is capped at 15.7 GeV and $\alpha$ is bounded from below by $-19^\circ$. Quartic couplings in the SGM model are restricted to be less than 0.05. Cubic couplings $M_{1,2}$ are also constrained to a fairly small range. The most interesting feature in the SGM model, is that a 95 GeV excess would be expected to be correlated with a doubly charged Higgs boson, with mass between 97 GeV to 160 GeV. This doubly charged scalar does not have fermion-anti-fermion couplings and can only decay to a pair of same sign $W$ bosons. Since the mass of the doubly charged Higgs boson is less than twice the mass of the $W$ boson, at least one of the $W$ boson in the decay channel has to be off-shell. All the current experimental limits in this mass range are not applicable to our model. The production cross section of pair productions varies from 1.1 pb to 200 fb in this mass range. We believe that the current LHC Run 2 data is enough to corroborate or falsify this possibility.

My contributions to this present work include modifying the \texttt{GMCalc} code to adapt it for the SGM model. \texttt{GMCalc} is written in \texttt{Fortran} and \texttt{HiggsTools} are written in \texttt{C++}. An interface that bridges these two packages is required. I modified and connected the two packages. In \texttt{HiggsTools}, I incorporated the interactions between exotic scalars and higgsinos. Adding such interactions and testing them against invisible decay data slightly lowered the decay branching ratios and relaxed experimental constraints in \texttt{HiggsTools}. I also constructed the allowed regions, at the $95\%$ confidence levels, for model parameters and physical observables, and extracted bounds on these quantities in the global fits. In addition, I ran the program for the 95 GeV boson case as well. 

In our global fits, we have only considered the theoretical constraints and the LHC direct search data. Observables from electroweak precision measurements can also be used to further constrain the model. Constraining the GM model from electroweak phase transition data has been performed in \cite{Chen:2022zsh}. A similar analysis could also be performed for the SGM model. Besides, we have a bountiful particle spectrum in the SCTM model, some of particles in which are stable and interact weakly with SM particles. These particles can function as weakly interacting massive particles (WIMP) that are potential dark matter candidates. The abundance evolution of these light stable particles after the Big Bang is also worth exploring in future work.

 \StartAppendix%
 \begin{singlespace}
\chapter{The Modified Frequentist Method}\label{appendix:A}
\end{singlespace}

In this Appendix we present the Modified Frequentist Method and the statistical details of the exclusion limits used by \texttt{HiggsTools}\cite{Bechtle:2008jh,Bechtle:2011sb,Read:2000ru,Bock:2004xz}. 

\section{Confidence Level}

The analysis of Higgs-like particle search results is formulated in terms of a hypothesis test. The null hypothesis is that a signal is absent while the alternative hypothesis is present. In other words, we want to make a distinction between two physical hypotheses, the data consists of physical signals and background signals and the data consists of background signals only. For that purpose, we introduce a discriminating function $Q$ to order observed events according to their signal-likeliness. Here by definition, $Q$ must be monotonic in signal-likeliness and is channel-dependent, i.e. $Q=Q(X_i)$, where $X_i$ is a decay channel. We also define the confidence limit $\operatorname{CL}$, which is the lower limit if the exclusion confidence is greater than the specified confidence level, for all values below the confidence limit\cite{Read:2000ru}. The confidence in the signal-plus-background hypothesis is therefore, the probability that $Q$ is less than or equal to the value observed in an experiment: 
\begin{equation}
\operatorname{CL}_{s+b}=P\left(Q \leq Q_{\mathrm{obs}} \mid \text{signal}+\text{background}\right)=P_{s+b}\left(Q \leq Q_{\text{obs}}\right),
\end{equation}
where \begin{equation}
P_{s+b}\left(Q \leq Q_{\text{obs}}\right)=\int_{-\infty}^{Q_{\text{obs}}} \frac{d P_{s+b}}{d Q} d Q.
\end{equation}
Here, the integrand is the probability distribution function (PDF) of the test statistics for the signal-plus-background experiments. Small values of $\mathrm{CL}_{s+b}$ indicates the analysis favors the null hypothesis. The confidence level $1-\operatorname{CL}_{s+b}$, however, is not a satisfying choice for exclusion purpose as it is prone to type II errors.

The confidence in the background hypothesis can be defined in a similar manner 
\begin{equation}
\operatorname{CL}_{b}=P\left(Q \leq Q_{\mathrm{obs}} \mid \text{background}\right)=P_b\left(Q \leq Q_{\text{obs}}\right),
\end{equation}
where \begin{equation}
P_{b}\left(Q \leq Q_{\text{obs}}\right)=\int_{-\infty}^{Q_{\text{obs}}} \frac{d P_{b}}{d Q} d Q,
\end{equation}
and the integrand is the PDF of the test statistics for the background-only experiments. Here, values $\mathrm{CL}_b\sim1$ favors the signal-plus-background hypothesis. 

Taking into account the background data, however, may result in a value of the estimator of a model parameter that is unphysical. To avoid such unacceptable situations, we adopt the modified frequentist approach by normalizing the confidence level for the signal-plus-background hypothesis to the confidence level for the background hypothesis proposed in \cite{Read:2000ru,Bock:2004xz},
\begin{equation}
\operatorname{CL}_s=\frac{\operatorname{CL}_{s+b}}{\operatorname{CL}_b}.
\end{equation}
It is impossible to determine this confidence level frequentistically as long as the background environment in an experiment is present. In this way, the ratio can be interpreted as an approximate confidence level for the signal-only hypothesis. The exclusion rule is $1-\operatorname{CL}_S\leq\operatorname{CL}= 0.95$. The experimentally observed and expected values of the quantity $\operatorname{CL}_s$ are determined by the above described process using the hypothesis of a Higgs production channel. $Q_{\text{obs}}$ and $Q_{\text{exp}}$ are obtained by numerically varying the production rate by a scaling factor until $\operatorname{CL}_s= 0.05$. This scaling factor is then interpreted as the observed limit and the expected limit\cite{Bechtle:2008jh}. In \texttt{HiggsTools}, the program evaluates the quantity $Q$ for the model, and then selects the most sensitive channel by maximizing the ratio $Q_{\text{model}}/Q_{\text{exp}}$. For the most sensitive channel, a point is excluded, at the $95\%$ confidence level, if $Q_{\text{model}}/Q_{\text{obs}}>1$.
\newpage

\section{Likelihood Ratio Test Statistics}

We have set the hypothesis and exclusion rules for a statistical test. We now need to define the test statistic. In the absence of background, the observation of a signal maximizes the signal likelihood $\mathcal{L}_s$. For a Poisson distribution, we have \begin{equation}
f(k ; \lambda)=P(X=k)=\frac{\lambda^k e^{-\lambda}}{k !}.
\end{equation}
The test statistic $Q$ for a signal event is defined as a likelihood ratio following from the definition of $\operatorname{CL}_s$ naturally,
\begin{equation}
Q=\frac{\mathcal{L}_{s+b}}{\mathcal{L}_b}=\frac{\prod_{i=1}^{N} \frac{e^{-\left(s_i+b_i\right)}\left(s_i+b_i\right)^{d_i}}{n_{i} !}}{\prod_{i=1}^{N} \frac{e^{-b_i b_i^{d_i}}}{d_{i} !}} \frac{\prod_{j=1}^{d_i} \frac{s_i S_i\left(x_{i j}\right)+b_i B_i\left(x_{i j}\right)}{s_i+b_i}}{\prod_{j=1}^{n_i} B_i\left(x_{i j}\right)},
\end{equation}
where $i$ is the channel index, $N$ is the total number of channels, $d_i$ is the observed events in channel $i$, $s_i$ is the expected signal events for a given hypothesis of the signal, and $b_i$ is the background events. $S_i$ and $B_i$ are PDFs of a discriminating variable $x$ for the signal and background, respectively. Here, we assume those two PDFs are identical. $Q$ can be simplified to
\begin{equation}
Q=\prod_{i=1}^N \frac{e^{-\left(s_i+b_i\right)}\left(s_i+b_i\right)^{d_i}}{d_{i} !} \Big/ \frac{e^{-b_i}b_i^{d_i}}{d_{i} !}=e^{-s_{\text{tot}}}\prod_i^N \frac{(s_i+b_i)^{d_i}}{(b_i)^{d_i}}
\end{equation}
This definition is equivalent to a weighted event counting\cite{Bock:2004xz}, where the weight for each channel is \begin{equation}
w_{i}=\ln \left(1+\frac{s_{ i}}{b_{ i}}\right).
\end{equation}
In the limit of large numbers, the negative log-likelihood $-2\ln Q$ converges to a $\chi^2$ distribution.

 \begin{singlespace}
\chapter{Higgs Search Experiments in HiggsTools}\label{appendix:exp}
\end{singlespace}

We present the experimental limits of Higgs searches relevant to GM and SGM scalars embedded in this Appendix. Table~\ref{t21}-\ref{t27} select the most recent experimental data from neutral scalar boson searches that are implemented in \texttt{HiggsTools}. Table~\ref{t28} presents invisible decay results implemented in \texttt{HiggsTools}. Table~\ref{t29} and \ref{t30} summarize the experimental searches of heavy Higgs bosons that may be sensitive to doubly charged and singly charged states in the GM and SGM model.

\begin{table}[H]
\centering
\begin{tabular}{ccccc}
\hline
Production channel  & Decay channel & $\sqrt{s}$TeV & $L[\mathrm{fb}^{-1}]$ & reference \\
\hline 
$pp\rightarrow HH$ & $HH\rightarrow WWWW$ & 13 & 36.1 & \cite{ATLAS:2018ili}\\
$pp\rightarrow HH$ & $HH\rightarrow \gamma\gamma WW$  & 8 & 20.3 & \cite{ATLAS:2015sxd}\\
$pp\rightarrow ggH$ & $HH\rightarrow \gamma\gamma WW$  & 13 & 36.1 & \cite{ATLAS:2018hqk}\\
$pp\rightarrow ggH$ & $HH\rightarrow \gamma\gamma\gamma\gamma$  & 8 & 20.3 & \cite{ATLAS:2015rsn}\\
$pp\rightarrow ggH$ & $HH\rightarrow \gamma\gamma\gamma\gamma$  & 13 & 36.7 & \cite{ATLAS:2018dfo}\\
$pp\rightarrow HH$ & $HH\rightarrow \gamma\gamma jj$ & 13 & 36.7 & \cite{ATLAS:2018jnf}\\
$pp\rightarrow HH$ & $HH\rightarrow \mu\mu\mu\mu$  & 8 & 20.7 & \cite{CMS:2015nay}\\
$pp\rightarrow HH$ & $HH\rightarrow \mu\mu\mu\mu$  & 13 & 139 & \cite{ATLAS:2021ldb}\\
$pp\rightarrow HH$, $\mathrm{vbf}H$, $HW$, $HZ$, $ttH$ & $HH\rightarrow \mu\mu\tau\tau$  & 8 & 19.7 & \cite{CMS:2017dmg}\\
$pp\rightarrow ggH$ & $HH\rightarrow \mu\mu\tau\tau$  & 8 & 20.3 & \cite{ATLAS:2015unc}\\
$pp\rightarrow HH$ & $HH\rightarrow \mu\mu\tau\tau$  & 13 & 35.9 & \cite{CMS:2018qvj}\\
$pp\rightarrow ggH$ & $HH\rightarrow \tau\tau\tau\tau$  & 8 & 19.7 & \cite{CMS:2015twz}\\
$pp\rightarrow HH$, $\mathrm{vbf}H$, $HW$, $HZ$ & $HH\rightarrow \tau\tau\tau\tau$  & 8 & 19.7 & \cite{CMS:2017dmg}\\
$pp\rightarrow HH$ & $HH\rightarrow \tau\tau\tau\tau$  & 13 & 35.9& \cite{CMS:2019spf}\\
\hline
\end{tabular}
 \caption[The production channel and decay channel for neutral Higgs pair production implemented in \texttt{HiggsTools}.]{\centering
  The production channel and decay channel for neutral Higgs pair production implemented in \texttt{HiggsTools}.}
  \label{t21}
\end{table}

\begin{table}[H]
\centering
\begin{tabular}{ccccc}
\hline
Production channel  & Decay channel & $\sqrt{s}$TeV & $L[\mathrm{fb}^{-1}]$ & reference \\
\hline 
$pp\rightarrow HH$ & $HH\rightarrow bbWW$  & 13 & 35.9 & \cite{CMS:2017rpp}\cite{CMS:2019noi}\\
$pp\rightarrow HH$ & $HH\rightarrow bbWW$  & 13 & 36.1 & \cite{ATLAS:2018fpd}\\
$pp\rightarrow HH$ & $HH\rightarrow bbWW$  & 13 & 138 & \cite{CMS:2021roc}\\
$pp\rightarrow HH$ & $HH\rightarrow bbZZ$  & 13 & 35.9 & \cite{CMS:2020jeo}\\
$pp\rightarrow HH$ & $HH\rightarrow bbbb$  & 8 & 17.9 & \cite{CMS:2015jal}\\
$pp\rightarrow HH$, $\mathrm{vbf}H$ & $HH\rightarrow bbbb$  & 8 & 19.5 & \cite{ATLAS:2015zug}\\
$pp\rightarrow HH$ & $HH\rightarrow bbbb$  & 13 & 35.9 & \cite{CMS:2018qmt}\\
$pp\rightarrow HH$ & $HH\rightarrow bbbb$  & 13 & 36.1 &\cite{ATLAS:2018pvw}\\
$pp\rightarrow HH$ & $HH\rightarrow bbbb$  & 13 & 126 &\cite{ATLAS:2020jgy}\\
$pp\rightarrow HH$ & $HH\rightarrow \gamma\gamma bb$  &8 & 19.7 &\cite{CMS:2016cma}\\
$pp\rightarrow HH$ & $HH\rightarrow \gamma\gamma bb$  &8 & 20 &\cite{ATLAS:2014pjm}\\
$pp\rightarrow HH$ & $HH\rightarrow \gamma\gamma bb$  &13 & 36.1 &\cite{ATLAS:2018dpp}\\
$pp\rightarrow HH$ & $HH\rightarrow \gamma\gamma bb$  &13 & 137 &\cite{CMS:2020tkr}\\
$pp\rightarrow HH$ & $HH\rightarrow bb\mu\mu$ & 8 & 19.7 & \cite{CMS:2017dmg}\\
$pp\rightarrow HH$ & $HH\rightarrow bb\mu\mu$ & 13 & 35.9 & \cite{CMS:2018nsh}\\
$pp\rightarrow HH$ & $HH\rightarrow bb\mu\mu$ & 13 & 139 & \cite{ATLAS:2021hbr}\\
$pp\rightarrow HH$ & $HH\rightarrow bb\tau\tau$  & 8 & 19.7 & \cite{ATLAS:2018uni}\\
$pp\rightarrow HH$ & $HH\rightarrow bb\tau\tau$  & 13 & 36.1 & \cite{ATLAS:2018uni}\\
$pp\rightarrow HH$ & $HH\rightarrow bb\tau\tau$  & 13 & 137 & \cite{CMS:2021yci}\\
$pp\rightarrow HH$ & $H\rightarrow bbbb$, $bb\tau\tau$, $bbZZ$, $bbWW$, $bb\gamma\gamma$  & 13 & 139 & \cite{ATLAS:2019vwv}\\
$pp\rightarrow HH$ & $H\rightarrow bb$, $\tau\tau$, $WW$, $\gamma\gamma$ 
$H\rightarrow bb$, $\tau\tau$, $WW$, $\gamma\gamma$  & 13 & 35.9 & \cite{CMS:2018ipl}\\
$pp\rightarrow HH$ & $H\rightarrow bb$, $\tau\tau$, $WW$, $\gamma\gamma$ $H\rightarrow bb$, $\tau\tau$, $WW$, $\gamma\gamma$ & 13 & 36.1 & \cite{ATLAS:2019qdc}\\
\hline
\end{tabular}
 \caption[The production channel and decay channel involving $b\bar{b}$ pair, for neutral Higgs pair production implemented in \texttt{HiggsTools}.]{\centering
  The production channel and decay channel involving $b\bar{b}$ pair, for neutral Higgs pair production implemented in \texttt{HiggsTools}.}
  \label{t22}
\end{table}

\begin{table}[H]
\centering
\begin{tabular}{ccccc}
\hline
Production channel  & Decay channel & $\sqrt{s}$TeV & $L[\mathrm{fb}^{-1}]$ & reference \\
\hline 
$p p \rightarrow HH$ & $H \rightarrow tt$ & 13 & 35.9 & \cite{CMS:2019pzc}\\
$p p \rightarrow \mathrm{vbf}H$ & $H \rightarrow bb$ & 8 & 19.8 & \cite{CMS:2015ebl}\\
$p p \rightarrow ZH$, $WH$ & $H \rightarrow bb$ & 8 & 24 & \cite{CMS:2013poe}\\
$p p \rightarrow ZH$, $WH$ & $H \rightarrow bb$ & 8 & 25 & \cite{ATLAS:2014vuz}\\
$p p \rightarrow bbH$ & $H \rightarrow bb$ & 8 & 19.7 & \cite{CMS:2015grx}\\
$p p \rightarrow bbH$ & $H \rightarrow bb$ & 13 & 27.8 & \cite{ATLAS:2019tpq}\\
$p p \rightarrow bbH$ & $H \rightarrow bb$ & 13 & 35.7 & \cite{CMS:2018hir}\\
$pp \rightarrow ggH$ & $H \rightarrow bb$ & 13 & 35.9 & \cite{CMS:2018pwl} \\
$p p \rightarrow ZH$ & $H \rightarrow cc$ & 13 & 36.1 & \cite{ATLAS:2018mgv}\\
$p p \rightarrow qq'H\gamma$ & $H\rightarrow qq'$ & 13 & 79.8 & \cite{ATLAS:2019itm}\\
$p p \rightarrow H$, $\mathrm{vbf}H$, $ZH$, $WH$ & $H \rightarrow ee$ & 8 & 19.7 & \cite{CMS:2014dqm}\\
$p p \rightarrow H$, $ZH$, $WH$ & $H \rightarrow ee$ & 13 & 139 & \cite{ATLAS:2019old}\\
$p p \rightarrow ggH$ & $H \rightarrow ee$, $\mu\mu$ & 8 & 19.3 & \cite{CMS:2015ooa}\\
$p p \rightarrow ggH$ & $H \rightarrow \mu\mu$ & 8 & 19.7 & \cite{CMS:2017nmj}\\
$p p \rightarrow ggH$ & $H \rightarrow \mu\mu$ & 13 & 36.1 & \cite{ATLAS:2019odt}\\
$p p \rightarrow ggH$, $bbH$ & $H \rightarrow \mu\mu$ & 13 & 35.9 & \cite{CMS:2019mij}\\
$p p \rightarrow ttH$ & $H \rightarrow ee$, $\mu\mu$ & 13 & 137 & \cite{CMS:2019lwf}\\
$p p \rightarrow H$, $\mathrm{vbf}H$, $ZH$, $WH$ & $H \rightarrow ee$, $\mu\mu$ & 13 & 60.7 & \cite{CMS:2018nak}\\
$p p \rightarrow ggH$ & $H \rightarrow ee$, $\mu\mu$ & 13 & 139 & \cite{ATLAS:2019erb}\\
$p p \rightarrow bbH$ & $H \rightarrow \tau\tau$ & 13 & 35.9 & \cite{CMS:2019hvr}\\
$p p \rightarrow ggH$, $bbH$ & $H \rightarrow \tau\tau$ & 13 & 35.9 & \cite{CMS:2018rmh}\\
$p p \rightarrow ggH$, $bbH$ & $H \rightarrow \tau\tau$ & 13 & 139 & \cite{CMS:2022goy}\cite{ATLAS:2020zms}\\
$p p \rightarrow H$, $ttH$, $\mathrm{vbf}H$, $ZH$, $WH$ & $H \rightarrow \gamma\gamma$ & 8 & 20.3 & \cite{ATLAS:2014jdv}\\
$p p \rightarrow H$, $ttH$, $\mathrm{vbf}H$, $ZH$, $WH$ & $H \rightarrow \gamma\gamma$ & 8 & 24.7 & \cite{CMS:2013uhr}\\
$p p \rightarrow H$, $ttH$, $\mathrm{vbf}H$, $ZH$, $WH$ & $H \rightarrow \gamma\gamma$ & 13 & 55.6 & \cite{CMS:2018cyk}\\
$p p \rightarrow H$, $ttH$, $\mathrm{vbf}H$, $ZH$, $WH$ & $H \rightarrow \gamma\gamma$ & 13 & 139 & \cite{ATLAS:2021uiz}\\
\hline
\end{tabular}
 \caption[ The production channel and decay channel for neutral Higgs bosons searches, where a neutral scalar decays into $\ell\ell$, $b\bar{b}$, $jj'$ and $\gamma\gamma$, implemented in \texttt{HiggsTools}.]{\centering
  The production channel and decay channel for neutral Higgs bosons searches, where a neutral scalar decays into $\ell\ell$, $b\bar{b}$, $jj'$ and $\gamma\gamma$, implemented in \texttt{HiggsTools}, where $\ell=e$, $\mu$, $\tau$ and $q$, $q'$ being jets.}
   \label{t23}
\end{table}

\begin{table}[H]
\centering
\begin{tabular}{ccccc}
\hline
Production channel  & Decay channel & $\sqrt{s}$TeV & $L[\mathrm{fb}^{-1}]$ & reference \\
\hline 
$pp \rightarrow HH$ & $H \rightarrow Z\gamma$ & 8 & 19.7 & \cite{CMS:2016ssv}\\
$pp \rightarrow HH$ & $H \rightarrow Z\gamma$ & 13 & 35.9 & \cite{CMS:2017dyb}\\
$pp \rightarrow HH$, $\mathrm{vbf}H$, $ZH$, $WH$, $ttH$ & $H \rightarrow Z\gamma$ & 8 & 24.8 & \cite{CMS:2013rmy}\\
$pp \rightarrow HH$, $\mathrm{vbf}H$, $ZH$, $WH$, $ttH$ & $H \rightarrow Z\gamma$ & 8 & 24.8 &\cite{ATLAS:2014fxe}\\
$pp \rightarrow ggH$, $\mathrm{vbf}H$, $ZH$, $WH$ & $H \rightarrow Z\gamma$ & 13 & 36.1 & \cite{ATLAS:2017zdf}\cite{ATLAS:2018sxj}\\
$pp \rightarrow HH$, $\mathrm{vbf}H$ & $H \rightarrow WW$, $ZZ$ & 13 & 36.1 & \cite{ATLAS:2018sbw}\\
$pp \rightarrow HH$, $\mathrm{vbf}H$ & $H \rightarrow WW$, $ZZ$ & 13 & 139 & \cite{ATLAS:2020fry}\\
$pp \rightarrow HH$, $\mathrm{vbf}H$ & $H \rightarrow ZZ$ & 8 & 20.3 & \cite{ATLAS:2015pre}\\
$pp \rightarrow ggH$, $\mathrm{vbf}H$ & $H \rightarrow ZZ$ & 13 & 35.9 & \cite{CMS:2018amk}\\
$pp \rightarrow ggH$, $\mathrm{vbf}H$ & $H \rightarrow ZZ$ & 13 & 36.1 & \cite{ATLAS:2017otj}\\
$pp \rightarrow ggH$, $\mathrm{vbf}H$ & $H \rightarrow ZZ$ & 13 & 137 & \cite{CMS:2021itu}\\
$pp \rightarrow ggH$, $\mathrm{vbf}H$ & $H \rightarrow ZZ$ & 13 & 139 & \cite{ATLAS:2020tlo}\\
$pp \rightarrow H$, $\mathrm{vbf}H$, $ZH$, $WH$, $ttH$ & $H \rightarrow ZZ$ & 8 & 24.8 &\cite{CMS:2013fjq}\\
$pp \rightarrow H$, $\mathrm{vbf}H$, $ZH$, $WH$, $ttH$ & $H \rightarrow ZZ$ & 8 & 25.3 &\cite{ATLAS:2013nma}\\
$pp \rightarrow\mathrm{vbf}H$ & $H \rightarrow WW$ & 8 & 24.4 & \cite{CMS:2013yea}\\
$pp \rightarrow ggH$, $\mathrm{vbf}H$ & $H \rightarrow WW$ & 8 & 24.3 & \cite{CMS:2012bea}\\
$pp \rightarrow ggH$, $\mathrm{vbf}H$ & $H \rightarrow WW$ & 8 & 24.3 & \cite{CMS:2012bea}\\
$pp \rightarrow ggH$, $\mathrm{vbf}H$ & $H \rightarrow WW$ & 13 & 35.9 & \cite{CMS:2019bnu}\\
$pp \rightarrow ggH$, $\mathrm{vbf}H$ & $H \rightarrow WW$ & 13 & 36.1 & \cite{ATLAS:2017uhp}\cite{ATLAS:2017jag}\\
$pp \rightarrow ggH$, $\mathrm{vbf}H$ & $H \rightarrow WW$ & 13 & 137 & \cite{CMS:2021klu}\\
$pp \rightarrow HH$ & $H \rightarrow WW$ & 8 & 24.4 & \cite{CMS:2013cul}\\
$pp \rightarrow HH$, $\mathrm{vbf}H$, $WH$, $ZH$ & $H \rightarrow WW$ & 8 & 24.3 & \cite{CMS:2013zmy}\\
$pp \rightarrow HH$, $\mathrm{vbf}H$ & $H \rightarrow WW$ & 8 & 20.3 & \cite{ATLAS:2015iie}\\
$pp \rightarrow HH$, $\mathrm{vbf}H$, $WH$, $ZH$ & $H \rightarrow WW$ & 8 & 25 & \cite{ATLAS:2014aga}\\
$pp \rightarrow HW$ & $H \rightarrow WW$, $\tau\tau$ & 8 & 24.4 & \cite{CMS:2013gel}\\
\hline
\end{tabular}
 \caption[The production channel and decay channel for neutral heavy Higgs searches, where a heavy scalar decays into two vector bosons, implemented in \texttt{HiggsTools}.]{\centering
  The production channel and decay channel for neutral heavy Higgs searches, where a heavy scalar decays into two vector bosons, implemented in \texttt{HiggsTools}.}
   \label{t24}
\end{table}

\begin{table}[H]
\centering
\begin{tabular}{ccccc}
\hline
Production channel  & Decay channel & $\sqrt{s}$TeV & $L[\mathrm{fb}^{-1}]$ & reference \\
\hline 
$pp \rightarrow ggH$ & $H\rightarrow ZH' \rightarrow ttZ$ & 13 & 140 & \cite{ATLAS:2023szc}\\
$pp \rightarrow HH$ & $H\rightarrow ZH' \rightarrow bbZ$ & 8 & 19.7 & \cite{CMS:2015flt}\\
$pp \rightarrow HH$ & $H\rightarrow ZH' \rightarrow bbZ$ & 8 & 20.3 & \cite{ATLAS:2015kpj}\\
$pp \rightarrow HH$ & $H\rightarrow ZH' \rightarrow bbZ$ & 13 & 35.9 & \cite{CMS:2018ljc}\cite{CMS:2019ogx}\\
$pp \rightarrow ggH$, $bbH$ & $H\rightarrow ZH' \rightarrow bbZ$ & 13 & 35.9 & \cite{CMS:2019qcx}\\
$pp \rightarrow ggH$, $bbH$ & $H\rightarrow ZH' \rightarrow bbZ$ & 13 & 139 & \cite{ATLAS:2020gxx}\\
$pp \rightarrow ggH$, $bbH$ & $H\rightarrow ZH' \rightarrow bbZ$ & 13 & 140 & \cite{ATLAS:2023szc}\\
$pp \rightarrow ggH$ & $H\rightarrow ZH' \rightarrow ggZ$, $ssZ$ & 13 & 139 & \cite{ATLAS:2020pcy}\\
$pp \rightarrow ggH$ & $H\rightarrow ZH' \rightarrow \mu\mu Z$ & 13 & 139 & \cite{ATLAS:2021ldb}\\
$pp \rightarrow ggH$ & $H\rightarrow ZH' \rightarrow \tau\tau Z$ & 8 & 19.7 & \cite{CMS:2015uzk}\\
$pp \rightarrow ggH$ & $H\rightarrow ZH' \rightarrow\tau\tau Z$ & 8 & 19.8 & \cite{CMS:2016xnc}\\
$pp \rightarrow HH$ & $H\rightarrow ZH' \rightarrow \tau\tau Z$ & 8 & 20.3 & \cite{ATLAS:2015kpj}\\
$pp \rightarrow HH$ & $H\rightarrow ZH' \rightarrow \ell\ell\tau\tau$ & 13 & 35.9 & \cite{CMS:2019kca}\\
\hline
\end{tabular}
 \caption[Selected neutral heavy Higgs searches for chain decays involving $Z$ boson implemented in \texttt{HiggsTools}.]{\centering
  The production channel and decay channel of chain decays involving one $Z$boson in neutral Higgs searchers implemented in \texttt{HiggsTools}, with $l$, $l'=e$, $\mu$, $\tau$.}
   \label{t25}
\end{table}

\begin{table}[H]
\centering
\begin{tabular}{ccccc}
\hline
Production channel  & Decay channel & $\sqrt{s}$TeV & $L[\mathrm{fb}^{-1}]$ & reference \\
\hline 
$pp \rightarrow HH$ & $H \rightarrow e\mu$, $e\tau$, $\mu\tau$ & 8 & 20.3 & \cite{ATLAS:2015dva}\\
$pp \rightarrow HH$ & $H \rightarrow e\mu$, $e\tau$, $\mu\tau$ & 13 & 36.1 & \cite{ATLAS:2018mrn}\\
$pp \rightarrow ggH$ & $H \rightarrow e\tau$, $\mu\tau$ & 13 & 35.9 & \cite{CMS:2019pex}\\
$pp \rightarrow ggH$, $\mathrm{vbf}H$ & $H \rightarrow e\mu$ & 13 & 35.9 & \cite{CMS:2018hnz}\\
$pp \rightarrow ggH$, $\mathrm{vbf}H$, $ZH$, $WH$ & $H \rightarrow e\mu$ & 13 & 139 & \cite{ATLAS:2019old}\\
$pp \rightarrow ggH$, $\mathrm{vbf}H$, $ZH$, $WH$ & $H \rightarrow e\tau$, $\mu\tau$ & 8 & 20.3 & \cite{ATLAS:2016joj}\\
$pp \rightarrow ggH$, $\mathrm{vbf}H$, $ZH$, $WH$ & $H \rightarrow \mu\tau$ & 8 & 19.7 & \cite{CMS:2015qee}\\
$pp \rightarrow ggH$, $\mathrm{vbf}H$, $ZH$, $WH$ & $H \rightarrow e\mu$, $e\tau$ & 8 & 19.7 & \cite{CMS:2016cvq}\\
$pp \rightarrow ggH$, $\mathrm{vbf}H$, $ZH$, $WH$ & $H \rightarrow e\mu$, $e\tau$ & 13 & 36.1 & \cite{ATLAS:2019pmk}\\
$pp \rightarrow ggH$, $\mathrm{vbf}H$ & $H \rightarrow e\tau$, $\mu\tau$ & 13 & 137 & \cite{CMS:2021rsq}\\
\hline
\end{tabular}
 \caption[Selected neutral heavy Higgs searches from flavor-changing process implemented in \texttt{HiggsTools}.]{\centering
  The production channel and decay channel for flavor-changing neutral heavy Higgs searches implemented in \texttt{HiggsTools}, with $l$, $l'=e$, $\mu$, $\tau$.}
   \label{t26}
\end{table}

\begin{table}[H]
\centering
\begin{tabular}{ccccc}
\hline
Production channel  & Decay channel & $\sqrt{s}$TeV & $L[\mathrm{fb}^{-1}]$ & reference \\
\hline 
$pp\rightarrow t\bar{t}$, $t\rightarrow Hu$,  & $H\rightarrow bb$, $\gamma\gamma$, $\tau\tau$, $WW$, $ZZ$  & 8 & 19.7 & \cite{CMS:2016obj}\\
$pp\rightarrow t\bar{t}$, $t\rightarrow Hc$ & $H\rightarrow bb$, $WW$, $\tau\tau$, $\gamma\gamma$  & 8 & 20.3 & \cite{ATLAS:2015ncl}\\
$pp\rightarrow t\bar{t}$, $t\rightarrow Hc$ & $H\rightarrow bb$  & 13 & 35.9 & \cite{CMS:2017bhz}\\
$pp\rightarrow t\bar{t}$, $t\rightarrow Hu$, $Hc$ & $H\rightarrow bb$, $\gamma\gamma$, $\tau\tau$, $WW$, $ZZ$  & 13 & 36.1 & \cite{ATLAS:2018jqi}\\
$pp\rightarrow t\bar{t}$, $t\rightarrow Hc$ & $H\rightarrow WW$, $ZZ$, $\tau\tau$  & 13 & 36.1 & \cite{ATLAS:2018xxe} \\
$pp\rightarrow t\bar{t}$, $t\rightarrow Hc$, $Hu$ & $H\rightarrow \gamma\gamma$  & 13 & 36.1 & \cite{ATLAS:2017tas}\\
\hline
\end{tabular}
 \caption[Selected neutral light Higgs searches produced from top-quark decay process implemented in \texttt{HiggsTools}.]{\centering
  The production channel and decay channel for light Higgs searches produced from top-quark decays implemented in \texttt{HiggsTools}.}
   \label{t27}
\end{table}

\begin{table}[H]
\centering
\begin{tabular}{ccccc}
\hline
Production channel  & Decay channel & $\sqrt{s}$TeV & $L[\mathrm{fb}^{-1}]$ & reference \\
\hline 
$pp\rightarrow HH$, $HZ$, $HW$ & \text{invisible} & 8 & 20.3 & \cite{ATLAS:2015qlt}\\
$pp\rightarrow HZ$, $\mathrm{vbf}H$ & \text{invisible} & 8 & 24.6 & \cite{CMS:2014gab}\\
$pp\rightarrow qqHZ$ & \text{invisible} & 13 & 36.1 & \cite{ATLAS:2017nyv}\\
$pp\rightarrow \mathrm{vbf}H$ & \text{invisible} & 13 & 36.1 & \cite{ATLAS:2018bnv}\\
\hline
\end{tabular}
 \caption[Selected neutral heavy Higgs searches of invisible decays implemented in \texttt{HiggsTools}.]{\centering
  The production channel and decay channel for neutral heavy Higgs searches of invisible decays implemented in \texttt{HiggsTools}.}
   \label{t28}
\end{table}

\begin{table}[H]
\centering
\begin{tabular}{ccccc}
\hline
Production channel  & Decay channel & $\sqrt{s}$TeV & $L[\mathrm{fb}^{-1}]$ & reference \\
\hline $p p \rightarrow H^{ \pm \pm} H^{\mp \mp}$ & $H^{ \pm \pm} \rightarrow \ell^{ \pm} \ell^{ \pm}$ & 8 & 19.7 & \cite{CMS:2016cpz}\\
$p p \rightarrow H^{ \pm \pm} H^{\mp}$ & $H^{ \pm \pm} \rightarrow \ell^{ \pm} \ell^{\prime \pm}, H^{\mp} \rightarrow \ell^{\mp} \nu_{\ell}$ & 8 & 19.7 & \cite{CMS:2016cpz}\\
$p p \rightarrow H^{ \pm \pm} H^{\mp \mp}$ & $H^{ \pm \pm} \rightarrow e^{ \pm} \tau^{ \pm}, \mu^{ \pm} \tau^{ \pm}$  & 8 & 20.3 & \cite{ATLAS:2014vih}\\
$p p \rightarrow H^{ \pm \pm} H^{\mp \mp}$ & $H^{ \pm \pm} \rightarrow e^{ \pm} e^{ \pm}, e^{ \pm} \mu^{ \pm}, \mu^{ \pm} \mu^{ \pm}$ & 8 & 20.3 & \cite{ATLAS:2014kca}\\
$p p \rightarrow H^{ \pm \pm} H^{\mp \mp}$ & $H^{ \pm \pm} \rightarrow \ell^{ \pm} \ell^{ \pm}$ & 13 & 12.9 & \cite{CMS:2017pet}\\
$p p \rightarrow H^{ \pm \pm} H^{\mp}$ & $H^{ \pm \pm} \rightarrow \ell^{ \pm} \ell^{\prime \pm}, H^{\mp} \rightarrow \ell^{\mp} \nu_{\ell}$ & 13 & 12.9 & \cite{CMS:2017pet}\\
$p p \rightarrow H^{ \pm \pm} H^{\mp \mp}$ & $H^{ \pm \pm} \rightarrow e^{ \pm} e^{ \pm}, e^{ \pm} \mu^{ \pm}, \mu^{ \pm} \mu^{ \pm}$ & 13 & 36 & \cite{ATLAS:2017xqs}\\
$p p \rightarrow H^{ \pm \pm} H^{\mp \mp}$ & $H^{ \pm \pm} \rightarrow W^{ \pm} W^{ \pm}$ & 13 & 139 & \cite{ATLAS:2021jol}\\
$p p \rightarrow H^{ \pm \pm} H^{\mp }$ & $H^{ \pm \pm} \rightarrow W^{ \pm} W^{ \pm}, H^{\mp} \rightarrow W^{\mp} Z$ & 13 & 139 & \cite{ATLAS:2021jol}\\
\hline
\end{tabular}
 \caption[Selected doubly charged Higgs searches implemented in \texttt{HiggsTools}.]{\centering
  The production channel and decay channel for doubly charged Higgs searches implemented in \texttt{HiggsTools}, with $\ell$, $\ell'=e$, $\mu$, $\tau$.}
   \label{t29}
\end{table}

\begin{table}[H]
\centering
\begin{tabular}{ccccc}
\hline
Production channel  & Decay channel & $\sqrt{s}$TeV & $L[\mathrm{fb}^{-1}]$ & reference \\
\hline 
$pp\rightarrow tb H^{\pm} $ & $H^{\pm} \rightarrow\tau^\pm\nu_\tau$ & 8 & 19.5 & \cite{ATLAS:2014otc}\\
$pp\rightarrow tbH^{\pm} $ & $H^{\pm} \rightarrow\tau^\pm\nu_\tau$ & 8 & 19.7 & \cite{CMS:2015lsf}\\
$pp\rightarrow tb H^{\pm} $ & $H^{\pm} \rightarrow\tau^\pm\nu_\tau$ & 13 & 35.9 & \cite{CMS:2019bfg}\\
$pp\rightarrow tt$, $t\rightarrow bH^{+}$ & $H^{\pm} \rightarrow\tau^\pm\nu_\tau$ & 13 & 35.9 & \cite{CMS:2019bfg}\\
$pp\rightarrow tb H^{\pm} $ & $H^{\pm} \rightarrow\tau^\pm\nu_\tau$ & 13 & 36.1 & \cite{ATLAS:2018gfm}\\
$pp\rightarrow tt$, $t\rightarrow bH^{+}$ & $H^{\pm} \rightarrow\tau^\pm\nu_\tau$ & 13 & 36.1 & \cite{ATLAS:2018gfm}\\
$pp\rightarrow tt$, $t\rightarrow bH^{+}$ & $H^{\pm} \rightarrow cb$ & 8 & 19.7 & \cite{CMS:2018dzl}\\
$pp\rightarrow tb H^{\pm}$, $qqH^\pm $ & $H^{\pm} \rightarrow tb$ & 8 & 20.3 & \cite{ATLAS:2015nkq}\\
$pp\rightarrow tb H^{\pm} $, $qqH^\pm $ & $H^{\pm} \rightarrow tb$ & 13 & 35.9 & \cite{CMS:2020imj}\\
$pp\rightarrow tb H^{\pm} $ & $H^{\pm} \rightarrow tb$ & 13 & 139 & \cite{ATLAS:2021upq}\\
$pp\rightarrow tt$, $t\rightarrow bH^{+}$ & $H^{+}\rightarrow c \bar{s} $, $H^{-}\rightarrow \bar{c} s $ & 8 & 19.7 & \cite{CMS:2015yvc}\\
$pp\rightarrow tt$, $t\rightarrow bH^{+}$ & $H^{+}\rightarrow c \bar{s} $, $H^{-}\rightarrow  \bar{c}s $ & 13 & 35.9 & \cite{CMS:2020osd}\\
$pp \rightarrow \text{vbf}H^{\pm}$  & $H^{ \pm} \rightarrow W^\pm Z$  & 8 & 20.3 & \cite{ATLAS:2015edr}\\
$pp \rightarrow \text{vbf}H^{\pm} $ & $H^{ \pm} \rightarrow W^\pm Z$  & 13 & 36.1 & \cite{ATLAS:2018iui}\\
$pp \rightarrow \text{vbf}H^{\pm}$ & $H^{ \pm} \rightarrow W^\pm Z$  & 13 & 137 & \cite{CMS:2021wlt}\\
$pp \rightarrow \text{vbf}H^{\pm}$ & $H^{ \pm} \rightarrow W^\pm Z$  & 13 & 139 & \cite{ATLAS:2022zuc}\\
\hline
\end{tabular}
 \caption[Selected singly charged Higgs searches implemented in \texttt{HiggsTools}.]{\centering
  The production channel and decay channel for singly charged Higgs searches implemented in \texttt{HiggsTools}.}
   \label{t30}
\end{table}

 \singlespacing%
 \setglossarystyle{list}
 \printglossary[title=GLOSSARY,toctitle=GLOSSARY]
 \doublespacing%


\nocite{*}
\bibliographystyle{bib/JHEP}
\raggedright
\bibliography{bib/citations}

\end{thesis}
\end{document}